\documentclass[onecolumn,seceqn,amsthm]{elsart3p}

\usepackage{amsmath}
\usepackage{amssymb}
\usepackage{amsbsy}
\usepackage{fixmath}
\usepackage{graphicx}
\usepackage{psfrag}
\usepackage{rotating}

\theoremstyle{plain} 
\newtheorem{proposition}{Proposition}

\allowdisplaybreaks

\newcommand{\cconj}[1]{{#1}^*}

\newcommand\sh{\ensuremath{^{1/2}}}

\newcommand\tti{\ensuremath{\to\infty}}
\newcommand\ttz{\ensuremath{\to 0}}

\newcommand\etc{etc.\ }
\newcommand\eg{e.g.\ }
\newcommand\ie{i.e.\ }
\newcommand\cf{cf.\ }

\providecommand\boldsymbol[1]{\mbox{\boldmath $##1$}}

\newcommand{\mat}[1]{\boldsymbol{#1}} % matrices
\newcommand{\colvec}[1]{\underline{#1}} % column vectors
\newcommand{\op}[1]{\mathcal{#1}} % operators
\newcommand{\form}[1]{\mathsf{#1}} % bilinear forms

\newcommand{\ord}{\operatorname{O}} % order of magnitude
\newcommand{\Real}{\operatorname{Re}} % real part 
\newcommand{\Imag}{\operatorname{Im}} % imaginary part
\newcommand{\spn}{\operatorname{span}} % span of sets of vectors
\newcommand{\sech}{\operatorname{sech}}
\newcommand\EE{\mathrm{E}}
\newcommand\ii{\mathrm{i}}
\newcommand\ee{\mathrm{e}}
\newcommand\DD{\operatorname{D}}
\newcommand\DDref{{\hat\DD}}
\newcommand\dd{\mathrm{d}}
\newcommand\ddz{\mathrm{d}z\,}
\newcommand{\tracemap}{\operatorname{S}}
\newcommand{\diag}{\operatorname{diag}}

\newcommand{\norm}[1]{|| #1 ||}
\newcommand{\normLtwo}[2]{\norm{ #2 }_{0, #1 }}
\newcommand{\seminorm}[1]{| #1 |}

\newcommand{\innerprod}[2]{( #1, #2 )}
\newcommand{\innerprodLtwo}[3]{\innerprod{#2}{#3}_{0, #1}}
\newcommand{\innerprodV}[3]{\innerprod{#2}{#3}_{V, #1}}

\newcommand\Rey{\mbox{\textit{Re}}}  % Reynolds number
\newcommand\Prm{\mbox{\textit{Pm}}} % Magnetic Prandtl number
\newcommand\Prg{\mbox{\textit{Pg}}} % Gravitational Prandtl number
\newcommand\Fro{\mbox{\textit{Fr}}} % Froude number
\newcommand\Ohn{\mbox{\textit{Oh}}} % Ohnesorge number
\newcommand\Har{\mbox{\textit{H}}} % Hartmann number
\newcommand\Harz{H_z}
\newcommand\Harx{H_x}
\newcommand\Harxi{H_\xi}
\newcommand\Reyc{\mbox{\textit{Rc}}}
\newcommand\Reym{\mbox{\textit{Rm}}}
\newcommand\Reyminv{\mbox{\Reym$^{-1}$}}
\newcommand\Reyinv{\mbox{\Rey$^{-1}$}}

\newcommand{\Alfx}{A_x} % streamwise Alfven number
\newcommand{\Alfxinv}{\Alfx^{-1}}
\newcommand{\Alfz}{A_z}
\newcommand{\Alfzinv}{\Alfz^{-1}}
\newcommand\Web{\mbox{\textit{We}}}
\newcommand\alphac{\mbox{$\alpha_\mathrm{c}$}}

\newcommand{\bux}{U}
\newcommand{\bbx}{B_x}
\newcommand{\bbz}{B_z}
\newcommand{\bbind}{B}

\newcommand{\duz}{u}
\newcommand{\dbz}{b}
\newcommand{\deh}{a}
 
\newcommand{ \utrial }{ u }
\newcommand{ \utest }{ \tilde{ u }}
\newcommand{ \btrial }{ b }
\newcommand{ \btest}{ \tilde{ b }}
\newcommand{ \htrial}{ a }
\newcommand{ \htest }{ \tilde{ a }}
\newcommand{ \vtrial }{ v }
\newcommand{ \vtest }{ \tilde{ v }}

\newcommand{\Omegac}{\Omega_\mathrm{ c }}
\newcommand{\Omegaf}{\Omega_\mathrm{ f } }
\newcommand{\Omegaref}{\hat{\Omega}}

\newcommand\tHOZ{\mat{T}_{H^1_0}}
\newcommand\tHTZ{\mat{T}_{H^2_0}}
\newcommand\tHO{\mat{T}_{H^1}}
\newcommand\tHTO{\mat{T}_{H^2_1}}
\newcommand\tHRZ{\mat{T}_{H^r_0}}
\newcommand\tHOHTZ{\mat{T}_{H^1 H^2_0}}
\newcommand\tHOHTO{\mat{T}_{H^1 H^2_1}}

\begin{document}

\begin{frontmatter}
\title{A spectral Galerkin method for the the coupled Orr--Sommerfeld and induction equations for free-surface MHD}
\author[label1]{Dimitrios Giannakis\corauthref{cor1}},
\corauth[cor1]{Corresponding author. Email: dg227@uchicago.edu.}
\author[label2]{Paul F. Fischer},
\author[label1,label2,label3]{Robert Rosner}
\address[label1]{Department of Physics, University of Chicago, Chicago, IL 60637, USA}
\address[label2]{Argonne National Laboratory, Argonne, IL 60439, USA}
\address[label3]{Department of Astronomy and Astrophysics,\\
University of Chicago, Chicago, IL 60637, USA}

\begin{abstract}
  We develop and test spectral Galerkin schemes to solve the coupled Orr--Sommerfeld (OS) and induction equations for parallel, incompressible MHD in free-surface and fixed-boundary geometries. The schemes' discrete bases consist of Legendre internal shape functions, supplemented with nodal shape functions for the weak imposition of the stress and insulating boundary conditions. The orthogonality properties of the basis polynomials solve the matrix-coefficient growth problem, and eigenvalue--eigenfunction pairs can be computed stably at spectral orders at least as large as $ p = \text{3,000} $ with $ p $-independent roundoff error. Accuracy is limited instead by roundoff sensitivity due to non-normality of the stability operators at large hydrodynamic and/or magnetic Reynolds numbers ($\text{\Rey, \Reym} \gtrsim 4 \times 10^4$). In problems with Hartmann velocity and magnetic-field profiles we employ suitable Gauss quadrature rules to evaluate the associated exponentially weighted sesquilinear forms without error. An alternative approach, which involves approximating the forms by means of Legendre--Gauss--Lobatto (LGL) quadrature at the $ 2 p - 1 $ precision level, is found to yield equal eigenvalues within roundoff error. As a consistency check, we compare modal growth rates to energy growth rates in nonlinear simulations and record relative discrepancy smaller than $ 10^{-5} $ for the least stable mode in free-surface flow at $ \Rey = 3 \times 10^4 $. Moreover, we confirm that the computed normal modes satisfy an energy conservation law for free-surface MHD with error smaller than $ 10^{-6} $. The critical Reynolds number in free-surface MHD is found to be sensitive to the magnetic Prandtl number $ \Prm $, even at the $ \Prm = \ord( 10^{-5} ) $ regime of liquid metals.  
\end{abstract}

\end{frontmatter}

\section{Introduction}

The Orr--Sommerfeld (OS) and induction equations, Eqs.~\eqref{eq:coupledOSInd} below, govern the linear stability of temporal normal modes in incompressible, parallel magnetohydrodynamics (MHD). These equations have mainly been applied to study the stability of flows with fixed domain boundaries in the presence of an external magnetic field (\cite{MullerBuhler01} and references therein). However, linear-stability analyses of free-surface flows have received comparatively little attention. Here the OS and induction equations, in conjunction with the kinematic boundary condition at the free surface~\eqref{eq:kinematic}, pose a coupled eigenvalue problem which must be solved for the complex growth rate $ \gamma $, the velocity and magnetic-field eigenfunctions, respectively $ u $ and $ b $, as well as the free-surface oscillatory amplitude $ a $.

Free-surface MHD arises in a variety of contexts, including cooling of fusion reactor walls by liquid-metal blankets \cite{AbdouEtAl01}, liquid-metal forced flow targets \cite{ShannonEtAl98}, and surface models of compact astrophysical objects \cite{AlexakisEtAl02}. In these and other cases of interest, hydrodynamic Reynolds numbers are large ($ \Rey \gtrsim 10^4 $), and the flow takes place in the presence of a strong background magnetic field ($ \Har \gtrsim 10^2$, where $ \Har $ is the Hartmann number). All terrestrial fluids have small magnetic Prandtl numbers (\eg for laboratory liquid metals $ \Prm \lesssim 10^{-5} $), suggesting that the magnetic field is well in the diffusive regime. On the other hand, $ \Prm = \ord( 1 ) $ flows have been conjectured to play a role in certain astrophysical accretion phenomena \cite{BalbusHenri07}.  

The main objective of the present work is to develop accurate and efficient spectral Galerkin schemes for linear-stability analyses of free-surface and fixed-boundary MHD. Our schemes build on the Galerkin method for plane Poiseuille flow by Kirchner \cite{Kirchner00}, and Melenk, Kirchner, and Schwab \cite{MelenkKirchnerSchwab00}, hereafter collectively referred to as KMS. A companion article \cite{GiannakisRosnerFischer07} discusses the operating physics in low-$ \Prm $ problems. A future objective is to test our linear models against wave-dispersion and critical-parameter data from a free-surface MHD experiment at Princeton Plasma Physics Laboratory (PPPL) by Ji and coworkers \cite{JiEtAl05,Katz04}.

\subsection{\label{sec:statusOfTheField}Background}

Since the pioneering work of Orszag \cite{Orszag71} in 1971, spectral methods have emerged as a powerful tool to solve hydrodynamic-stability problems. Orszag applied a Chebyshev tau technique to transform the OS equation for plane Poiseuille flow to a matrix generalized eigenproblem $ \mat{K} \colvec{u} = \gamma \mat{M} \colvec{u} $, which he solved at Reynolds numbers of order $ 10^4 $ using the QR algorithm. The superior performance of the Chebyshev tau method compared to existing finite-difference and spectral schemes led to its application to a diverse range of stability problems (\eg \cite{DongarraStraughanWalker96}). However, despite the widespread use of spectral techniques in flows with fixed domain boundaries, most numerical stability analyses of free-surface problems to date are based on finite-difference methods. Among these are the studies of gravity and shear-driven flows by De Bruin \cite{DeBruin74}, and Smith and Davis \cite{SmithDavis82}. To our knowledge, the only related work in the literature employing spectral techniques is contained in the PhD thesis by Ho \cite{Ho89}, where the OS equation for a vertically falling film is solved at small Reynolds numbers ($ \Rey \leq 10 $).

In MHD, numerical investigations on the stability of modified plane Poiseuille flow subject to a transverse magnetic field, also known as Hartmann flow, begin in 1973 with the work of Potter and Kutchey \cite{PotterKutchey73}, who used a Runge--Kutta technique to solve the coupled OS and induction equations at small Hartmann numbers ($ \Har \leq 6$). Lingwood and Alboussiere \cite{LingwoodAlboussiere99} also employed a Runge--Kutta method to study the stability of an unbounded Hartmann layer. An early application of spectral methods was performed by Dahlburg, Zang, and Montgomery \cite{DahlburgEtAl83} in 1983, who adopted Orszag's scheme to investigate the stability of a magnetostatic quasiequilibrium (\ie a state where the fluid is at rest but a slowly varying background magnetic field is present). A Chebyshev tau method for plane Poiseuille and plane Couette flows in the presence of a transverse magnetic field was later developed by Takashima \cite{Takashima96,Takashima98}. Takashima's calculations extend to high Reynolds and Hartmann numbers ($ \Rey \sim 10^7 $, $ \Har \sim 10^3 $) and over a range of magnetic Prandtl numbers up to $ \Prm = 0.1 $. In addition, he considers the limiting case of vanishing magnetic Prandtl number, where the OS and induction equations are replaced by a single equation~\eqref{eq:orrSommerfeldZeroPm}. However, his analysis does not take into account modes other than the least stable one (cf.~\cite{Kirchner00,Orszag71,DongarraStraughanWalker96}). 
            
A major challenge in hydrodynamic-stability problems at high Reynolds numbers is the existence of thin boundary layers, whose thickness scales as $ ( \alpha \Rey )^{-1/2} $ for a normal mode of wavenumber $ \alpha $ \cite{DrazinReid81}, which requires the use of large spectral orders $ p $ to achieve convergence. Specifically, Melenk \etal~\cite{MelenkKirchnerSchwab00} have shown that a necessary condition for accurate results is that the ratio $ \Rey / p^2 $ is small, implying that for problems of interest the required $ p $ can be in the thousands. At such high spectral orders the Chebyshev tau method can be problematic, since it gives rise to stiffness and mass matrices, respectively $ \mat{ K } $ and $ \mat{ M } $, that are (i) densely populated (the storage and computation cost therefore scale as $ p ^ 2 $ and $ p ^ 3 $, respectively), and (ii) ill conditioned (the matrix elements associated with a fourth-order differential operator, such as the OS one, grow as $ p^7 $). One way to alleviate the matrix-coefficient growth problem is to pass to a streamfunction--vorticity formulation \cite{McFaddenMurrayBoisvert90}, or, more generally, apply the $\DD^2$ method proposed by Dongarra, Straughan, and Walker~\cite{DongarraStraughanWalker96}. Here one achieves a $ p^3 $ coefficient scaling by casting the OS equation into two coupled second-order equations, but at the expense of doubling the problem size. 

Another drawback of the tau method is the occurrence of `spurious' eigenvalues, \ie eigenvalues with large magnitude (\eg $\ord( 10^{17} )$ \cite{StraughanWalker96}), and real-part oscillating between positive and negative values as $ p $ is varied. These numerical eigenvalues are not at all related to the spectrum of the OS operator, and in order to avoid drawing erroneous conclusions (\eg deciding that a flow is unstable when the unstable mode is spurious), the practitioner must either detect them and ignore them in the analysis (the non-spurious modes are computed correctly), or eliminate them by a suitable modification of the method (\eg \cite{DongarraStraughanWalker96,McFaddenMurrayBoisvert90}). Their origin has been elucidated by Dawkins, Dunbar, and Douglas \cite{DawkinsDunbarDouglass98}, who found that the large spurious eigenvalues in Chebyshev tau schemes are perturbations of infinite eigenvalues in nearby Legendre tau discretizations.     

Recently, KMS have developed a spectral Galerkin method that addresses some of the aforementioned shortcomings. Central to their scheme is the use of the so-called compact combinations of Legendre polynomials \cite{Shen94,Shen96}, or hierarchical shape functions \cite{Schwab98}, as a basis of the Sobolev space $ H^2_0$ (the trial and test space for velocity eigenfunctions). The resulting orthogonality properties solve the matrix-coefficient growth problem, and no reduction in the differential-equation order is required (in fact, the condition number of $ \mat{ K } $ has been found to be independent of $ p \gtrsim{ 100 }$ \cite{MelenkKirchnerSchwab00}). Moreover, the stiffness and mass matrices are sparse, \emph{provided that the basic velocity profile is polynomial}. In that case, memory requirements scale as $ p $, and iterative solvers can be used to compute the eigenvalues and eigenvectors efficiently. A further attractive feature of the method, which appears to be connected to the non-singularity of $ \mat{ M } $, is that it gives no rise to spurious eigenvalues.  

An additional, and perhaps more fundamental, challenge is related to the non-normality of hydrodynamic-stability operators, which becomes especially prominent at large Reynolds numbers. In that regime, even though the eigenfunctions may form a complete set (as has been proved for bounded domains \cite{DiPrimaHabetler69}), they are nearly linearly dependent. A key physical effect of the eigenfunctions' non-orthogonality is large transient growth of asymptotically stable perturbations, which suggests that eigenvalue analysis is of little physical significance \cite{TrefethenEtAl93}. An alternative method that aims to capture the effects of transient growth is pseudospectra \cite{TrefethenEmbree05}, but this will not be pursued here. We remark, however, that comparisons between spectra and pseudospectra are common in pseudospectral analyses, and stable and efficient schemes for eigenvalue computations are desirable even in that context. 

At the numerical level, non-normality is associated with high sensitivity of the spectrum to roundoff errors \cite{SchmidEtAl93}. This effect was noted by Orszag himself \cite{Orszag71}, who observed significant changes in the computed eigenvalues by artificially reducing numerical precision from $ 10^{-14} $ to $ 10^{-8} $. The eigenmodes that are most sensitive to perturbations of the OS operator and its matrix discrete analog are those lying close to the intersection point between the A, P, and S eigenvalue branches on the complex plane (see \cite{Mack76} for a description of the nomenclature). Reddy, Schmid, and Henningson~\cite{ReddySchmidHenningson93} have observed that in plane Poiseuille flow at $ \Rey \sim 10^4 $ perturbations of order $ 10^{ -6 } $ suffice to produce $ \ord( 1 ) $ changes in the eigenvalues near the branch point. Moreover, they found that roundoff sensitivity increases \emph{exponentially} with the Reynolds number. Qualitatively, this type of growth is attributed to the existence of solutions of the OS equation that satisfy the boundary conditions to within an exponentially small error. In consequence, double-precision arithmetic (typically 15 digits) rapidly becomes inadequate, and for $ \Rey \gtrsim 4 \times 10^4 $ one obtains a diamond shaped pattern of numerical eigenvalues instead of a well resolved branch point (\eg Fig.~4 in \cite{DongarraStraughanWalker96} and Figs.~\ref{fig:spectralInstabilityHydro}--\ref{fig:spectralInstabilityMhd2} below). Dongarra et al.~\cite{DongarraStraughanWalker96} have postulated that alleviation of this spectral instability requires the use of extended-precision arithmetic, and cannot be removed by improving the conditioning of the numerical scheme (\eg employing a $ \DD $ method instead of $ \DD^2 $ one). Melenk et al.~\cite{MelenkKirchnerSchwab00} note that their Galerkin method accurately resolves the eigenvalue branch point at $ \Rey = 2.7 \times 10^4 $ using 64-bit arithmetic, when the Chebyshev tau method already produces the diamond-shaped pattern. However, even a moderate Reynolds-number increase (\eg $ \Rey = 4 \times 10^4 $ in Fig.~\ref{fig:spectralInstabilityHydro}) results to the appearance of the pattern, despite the Galerkin scheme's superior conditioning. It therefore appears that, at least in these examples, the decisive factor in roundoff sensitivity is the non-normality of the OS operator rather than the details of the discretization scheme.
 
\subsection{Plan of the Present Work}           

The principal contribution of this article is twofold. First, we generalize the spectral Galerkin method of KMS for plane Poiseuille flow to free-surface and fixed-boundary MHD. Second, we present a number of test calculations aiming to assess our schemes' numerical performance, as well as to provide benchmark data. The calculations have been performed using a Matlab code, available upon request from the corresponding author.
  
As already stated, central to the stability and efficiency of the KMS scheme is the use of suitable linear combinations of Legendre polynomials as a basis of $ H^2_0 $. In the sequel, we employ similar constructions to treat the free-surface MHD problem. Here the velocity field obeys stress conditions at the free surface, which we enforce weakly (naturally) by supplementing the basis with nodal shape functions \cite{Schwab98}. Pertaining to the magnetic field, we assume throughout that the domain boundaries are electrically insulating, from which it follows that it obeys boundary conditions of Robin type, with extra contributions from the free-surface oscillation amplitude \cite{GiannakisRosnerFischer07}. We enforce naturally these boundary conditions as well, discretizing the magnetic-field by means of the internal and nodal shape functions for $ H^1 $. As we demonstrate in \S\ref{sec:pConvergence} and \S\ref{sec:hartmannProfile}, our choice of bases gives rise (and is essential) to a major advantage of our schemes, namely that roundoff error is independent of the spectral order $ p $.   
    
In problems with polynomial steady-state profiles the stiffness and mass matrices are sparse, and closed-form expressions exist for their evaluation (see Appendix~\ref{app:innerProducts}). On the other hand, in Hartmann flow $ \mat{ K } $ becomes full and must be computed numerically, since the discretization procedure introduces inner products of Legendre polynomials with exponential weight functions. We evaluate the required integrals stably and without error by means of the Gauss quadrature rules developed by Mach~\cite{Mach84}, who has studied a class of orthogonal polynomials with exponential weight functions on a finite interval. Following the standard practice in finite-element and spectral-element methods \cite{Ciarlet91,DevilleFischerMund02}, we also consider a method where the problem's weighted sesquilinear forms are replaced by approximate ones derived from Legendre--Gauss--Lobatto (LGL) quadrature rules. At an operational level, the latter approach has the advantage of being sufficiently general to treat arbitrary analytic profiles. However, it introduces quadrature errors, and one has to ensure that the stability and convergence of the scheme are not affected. As shown by Banerjee and Osborn \cite{BanerjeeOsborn90}, in finite-element schemes for \emph{elliptical} eigenvalue problems that is indeed the case, provided that the approximated eigenfunctions are smooth and the quadrature rule is exact for polynomial integrands of degree $ 2 p - 1 $. To our knowledge, however, no such result exists in the literature for the OS eigenproblems we study here, and is therefore not clear what (if any) quadrature precision would suffice. Even though we make no attempt to parallel Banerjee and Osborn's work, we nevertheless find that eigenvalues computed using approximate quadrature at the $ 2 p - 1 $ precision level converge, modulo roundoff error, to the same value as the corresponding ones from the exact-quadrature method.   

One of the advantages of the spectral Galerkin method is its flexibility. Our scheme for free-surface MHD can be straightforwardly adapted to treat MHD problems with fixed domain boundaries, problems in the limit of vanishing magnetic Prandtl number, as well as non-MHD problems. In \S\ref{sec:eigenvalueSpectra} we describe the basic properties of the eigenvalue spectra of these problems, leaving a discussion of the physical implications to Ref.~\cite{GiannakisRosnerFischer07}. We also present a series of critical-parameter calculations (see \S\ref{sec:criticalReynolds}), confirming that results obtained via the fixed-boundary variants of our schemes are in close agreement with the corresponding ones by Takashima \cite{Takashima96}. In free-surface problems, when $ \Prm $ is increased from $ 10^{-8} $ to $ 10^{-4} $ the critical Reynolds number is seen to drop by a factor of five, while the corresponding relative variation in fixed-boundary problems is less than $ 0.003 $. Due to the limited availability of eigenvalue data for free-surface flow (\cf fixed-boundary problems \cite{Kirchner00,MelenkKirchnerSchwab00,DongarraStraughanWalker96,Takashima96}), we were not able to directly compare our free-surface schemes to existing ones in the literature. Instead, we have carried out two other types of consistency checks (see \S\ref{sec:consistency}), one of which is based on energy-conservation laws in free-surface MHD, whereas the second involves growth-rate comparisons with fully nonlinear simulations.

A numerical caveat concerns the aforementioned roundoff sensitivity at high Reynolds numbers. In \S\ref{sec:nonNormality} we observe that as $ \Rey $ grows our schemes experience the spectral instability that has been widely encountered in Poiseuille flow \cite{Kirchner00,MelenkKirchnerSchwab00,Orszag71,DongarraStraughanWalker96,ReddySchmidHenningson93}. Most likely, this issue is caused by the physical parameters of the problem, rather than the properties of the discretization scheme, and can only be addressed by increasing the numerical precision. Unfortunately, since the latter option is (as of January 2008) not natively supported in Matlab, we merely acknowledge the existence of the instability, and work throughout in double-precision arithmetic. We remark, however, that only the eigenvalues near the branch-intersection points are affected. In particular, eigenvalues and eigenfunctions at the top end of the spectrum can be accurately computed at Reynolds numbers at least as high as $ 10^7 $. We also wish to note that the emphasis of our work is towards the numerical, rather than analytical, side, and even though techniques to study the stability and convergence of Galerkin methods for eigenvalue problems are well established in the literature (\cite{BabuskaOsborn91} and references therein), we do not pursue them here.
   
The plan of this paper is as follows. In \S\ref{sec:problemDescription} we specify the governing equations and boundary conditions of our models. In \S\ref{sec:weakFormulation} we develop their weak formulation. The associated Galerkin discretizations are described in \S\ref{sec:galerkinDiscretization}. We present our numerical results in \S\ref{sec:numericalResults}, and conclude in \S\ref{sec:conclusions}. Appendix~\ref{app:innerProducts} contains closed form expressions for the matrix representations of the sesquilinear forms used in the main text. Although some of these can also be found in \cite{Kirchner00}, we opted to reproduce them here because that paper contains a number of typographical errors. Finally, in Appendix~\ref{app:eigenvalueSpectra} we have collected tables of eigenvalues for the problems examined in \S\ref{sec:eigenvalueSpectra}.  

\section{\label{sec:problemDescription}Problem Description}

\subsection{\label{sec:governingEqs} Governing Equations}

Using $ x $ and $ z $ to denote the streamwise and flow-normal coordinates, and $ \DD $ to denote differentiation with respect to $ z $, we consider the coupled OS and induction equations,
\begin{subequations}
\label{eq:coupledOSInd}
\begin{equation}
\label{eq:orrSommerfeld}
\Reyinv ( \DD^2 - \alpha^2 )^2 \duz - ( \gamma + \ii \alpha \bux ) ( \DD^2 - \alpha^2 ) \duz + \ii \alpha ( \DD^2 \bux ) \duz
+ ( \ii \alpha \bbx + \bbz \DD ) ( \DD^2 - \alpha^2 ) \dbz - \ii \alpha ( \DD^2 \bbx ) \dbz = 0, 
\end{equation}
and
\begin{equation}
\label{eq:induction}
\Reyminv ( \DD^2 - \alpha^2 ) \dbz - ( \gamma + \ii \alpha \bux )  \dbz + ( \ii \alpha \bbx + \bbz \DD ) \duz = 0,
\end{equation}
\end{subequations}
defined over an interval $ \Omega = ( z_1, z_2 ) \in \mathbb{ R } $. Here $ \duz \in C^4( \bar \Omega ) $ and $ \dbz \in C^3( \bar \Omega ) $ are respectively the velocity and magnetic-field eigenfunctions corresponding to the eigenvalue $ \gamma \in \mathbb{ C } $. Also, $ \alpha > 0 $ is the wavenumber, and $ \Rey > 0 $ and $ \Reym > 0 $ are the hydrodynamic and magnetic Reynolds numbers. The functions $ \bux \in C^2( \bar \Omega ) $ and $ \bbx \in C^2( \bar \Omega ) $ are the steady-state velocity and streamwise magnetic field. The flow-normal, steady-state magnetic field $ \bbz $ is constant, since $ (B_x, B_z )$, where $( \cdot, \cdot )$ stands for $ (x,z)$ vector components, is divergence free and streamwise invariant. The two-dimensional velocity and magnetic fields associated to $ u $ and $ b $ are given by $ \Real( ( \ii \DD \duz / \alpha, \duz ) \ee^{\ii \alpha x + \gamma t } )$ and $ \Real( ( \ii \DD \dbz / \alpha, \dbz ) \ee^{\ii \alpha x + \gamma t } ) $.  

A physical derivation of~\eqref{eq:coupledOSInd}, as well as~\eqref{eq:orrSommerfeldZeroPm} and the boundary conditions~\eqref{eq:noSlipChannel}--\eqref{eq:normalStressZeroPm} ahead, can be found in Refs.~\cite{GiannakisRosnerFischer07,Takashima96}. Here we note that the magnetic-field variables $ b $, $ \bbx $, and $ \bbz $ have been rendered to non-dimensional form using the characteristic magnetic-field  $ B_0 := ( \mu_0 \rho )\sh U_0 $, where $ \mu_0 $, $ \rho $, and $ U_0 $ are the permeability of free space, the fluid density, and the characteristic velocity, respectively. With this choice of magnetic-field scale, $ u $ and $ b $ are naturally additive. Another option (employed \eg by Takashima~\cite{Takashima96}) is to set $ B_0 = B' $, where $ B' $ is the typical steady-state magnetic field. The resulting non-dimensional eigenfunction $ b' $ is related to the one used here according to $ b' = A b $, where $ A := ( \mu_0 \rho )\sh U_0 / B' $ is the Alfv\'en number. We also remark that we have adopted the eigenvalue convention used by Ho~\cite{Ho89}, under which $ \Real( \gamma ) =: \Gamma $ corresponds to the modal growth rate (\ie a mode is unstable if $ \Gamma > 0 $), while $ C := - \Imag( \gamma ) / \alpha $ is the phase velocity. The complex phase velocity $ c = \ii \gamma / \alpha $, where $ \Real( c ) = C $ and $ \Imag( c ) \alpha = \Gamma $, is frequently employed in the literature (\eg \cite{Kirchner00,MelenkKirchnerSchwab00,DongarraStraughanWalker96,DeBruin74,SmithDavis82,Takashima96}) in place of $ \gamma $.  
 
Let $ \Prm := \Reym / \Rey $ denote the magnetic Prandtl number; the ratio of magnetic to viscous diffusivity (see \eg \cite{Shercliff65} for an overview of the dimensionless parameters in MHD). A limit of physical interest, hereafter referred to as the \emph{inductionless limit} \cite{MullerBuhler01}, is $ \Prm \ttz $ with $ \bbx $ independent of $ z $, and the Hartmann numbers $ \Harx := \bbx \Rey \Prm\sh $ and $ \Harz := \bbz \Rey \Prm\sh $, measuring the square root of the ratio of Lorentz to viscous forces, non-negligible. This situation corresponds to a fluid of sufficiently high magnetic diffusivity so that magnetic-field perturbations are small ($ \norm{ b } \ll \norm{ u } $ in some suitable norm), but  Lorentz forces due to currents induced by the perturbed fluid motions $ \duz $ within the steady-state field $( \bbx, \bbz )$ are nonetheless present. It is then customary to make the approximation $ ( \DD^2 - \alpha^2 ) b = - \Reym ( \ii \alpha \bbx + \bbz \DD ) u $ \cite{Stuart54} and replace~\eqref{eq:coupledOSInd} by the single equation
\begin{equation}
\label{eq:orrSommerfeldZeroPm}
( \DD^2 - \alpha^2 )^2 \duz - ( \ii \alpha \Harx + \Harz \DD )^2 \duz - \Rey ( \gamma + \ii \alpha \bux ) ( \DD^2 - \alpha^2 ) \duz + \ii \alpha \Rey ( \DD^2 \bux ) \duz = 0.  
\end{equation}

\subsection{\label{sec:boundaryConditions}Boundary Conditions}

\begin{figure}
\psfrag{z}{\small{$z$}}
\psfrag{0}{\small{$0$}}
\psfrag{x}{\small{$x$}}
\psfrag{U(z)}{\small{$U(z)$}}
\psfrag{B(z)}{\small{$B(z)$}}
\psfrag{z=l}[][r]{\small{$z=z_2=1$}}
\psfrag{z=-l}[][r]{\small{$z=z_1=-1$}}
\psfrag{z=0}[][r]{\small{$z=z_2=0$}}
\psfrag{a}{\small{$a$}}
\psfrag{theta}{\small{$\theta$}}
\begin{center}
\includegraphics{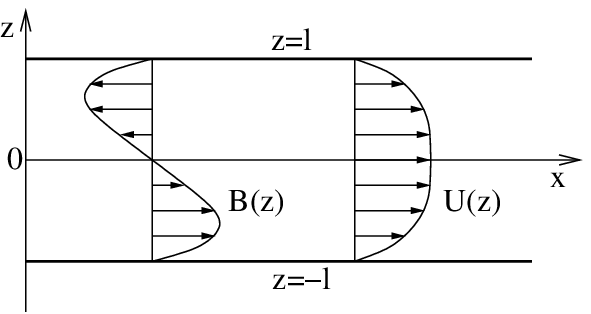}
\qquad
\includegraphics{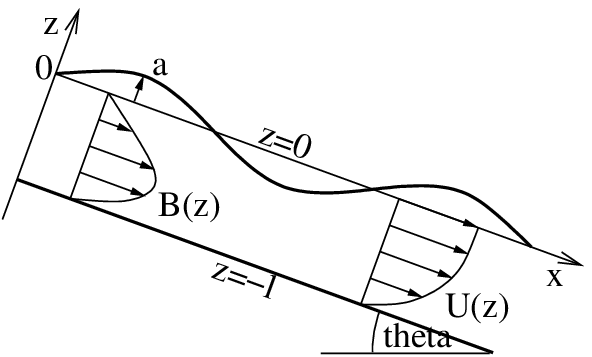}
% rotation is 20 degrees 
\end{center}
\caption{\label{fig:modelGeometry}Geometry of channel (left) and film (right) problems. $ \bux( z ) $ and $ \bbind( z ) $ are the steady-state velocity and induced magnetic-field profiles, respectively (see \S\ref{sec:steadyState}).}
\end{figure}

We study two types of problems, which we refer to as \emph{channel} and \emph{film} problems according to their geometrical configuration (see Fig.~\ref{fig:modelGeometry} for an illustration). Within each category we further distinguish among MHD problems, their counterparts in the inductionless limit, and non-MHD problems, where all electromagnetic effects are neglected.   

In channel problems the flow takes place between two fixed, parallel plates. As is customary, we make the domain choice $ \Omega = \Omega_\mathrm{ c } := ( -1, 1 ) $, and enforce the no-slip boundary conditions
\begin{equation}
\label{eq:noSlipChannel}
\duz( \pm 1 ) = \DD \duz( \pm 1 ) = 0.
\end{equation} 
Moreover, we assume that the plates and the region exterior to the flow are perfect insulators, which leads to the Robin boundary conditions 
\begin{equation}
\label{eq:insulatingChannel}
\DD \dbz( \pm 1 ) \pm \alpha \dbz( \pm 1 ) = 0
\end{equation} 
for the magnetic field. 

In film problems we set $ \Omega = \Omega_\mathrm{ f } := ( -1, 0 ) $, and consider that the domain boundary $ z_2 = 0 $ corresponds to a free surface, whose oscillation amplitude $ \deh \in \mathbb{ C } $ obeys the kinematic boundary condition 
\begin{equation}
\label{eq:kinematic}
\duz( 0 ) - ( \gamma + \ii \alpha \bux( 0 ) ) \deh = 0.
\end{equation}
We assume that the free surface is subject to surface tension and gravity (see \eg \cite{Batchelor67} for a discussion of free-surface dynamics). The non-dimensional stress due to surface tension is given by $ a\alpha^2/(\Ohn\Rey)^2 $, where the Ohnesorge number $ \Ohn $ measures the  ratio of viscous to capillary velocity scales. $ \Ohn $ is related to the Weber number $ \Web $ (the ratio of surface tension to inertial stresses) via $ \Ohn = 1 / ( \Rey \Web\sh ) $. Moreover, we express the $ z $ component of the gravitational force as $ - ( \Prg \Rey )^{-2} $, where $ \Prg $ is a parameter that we refer to as the \emph{gravitational Prandtl number}. $ \Prg $ is equal to the ratio between the kinematic viscosity $ \nu $ and the diffusion constant $ ( g \cos(\theta) l^3 )\sh $, formed by the flow-normal gravitational-field strength, $ g \cos( \theta ) $, and the characteristic length scale $ l $ (\cf the magnetic Prandtl number $ \Prm = \nu / \eta $, where $ \eta $ is the magnetic diffusivity), and is related to the Froude number $ \Fro $ (the ratio of convective to gravity velocity scales) according to $ \Prg = \Fro / \Rey $ \cite{GiannakisRosnerFischer07}. Our use of the parameters $ \Ohn $ and $ \Prg $, rather than the more familiar $ \Web $ and $ \Fro $, is motivated by the fact that they do not depend on the characteristic flow velocity, and are thus likely to remain fixed in situations where a single working fluid is driven at different flow speeds. Typical values for a laboratory liquid-metal film of thickness $ \simeq 1~\mbox{cm} $ are $ \Ohn \simeq 10^{-4} $ and $ \Prg \simeq 10^{-4} $ (\eg \cite{HofEtAl04,Har85,Kol99}). 

Balancing the forces acting on the free surface leads to the normal-stress condition
\begin{multline}
\label{eq:normalStress}
( ( ( \DD^2 - 3 \alpha ^ 2 ) \DD - \Rey ( \gamma + \ii \alpha \bux ) \DD + \ii \alpha \Rey ( \DD \bux ) ) \duz )|_{z=0} + \Rey ( \bbz ( \DD^2 - \alpha^2 ) - \ii \alpha ( \DD \bbx ) ) \dbz |_{z=0} \\
- \alpha^2 \left( \frac{ 1 }{ \Prg^2 \Rey } + \frac{ \alpha^2 }{ \Ohn^2 \Rey } + \Rey \bbx( 0 ) \DD \bbx( 0 ) - 2 \ii \alpha \DD \bux( 0 ) \right) \deh
 = 0,
\end{multline}
and the shear-stress condition
\begin{equation}
\label{eq:shearStress}
\DD^2 \duz( 0 ) + \alpha^2 \duz( 0 ) - \ii \alpha \DD^2 \bux( 0 ) \deh = 0.
\end{equation}
The no-slip boundary conditions are again 
\begin{equation}
\label{eq:noSlipFilm}
\duz( -1 ) = \DD \duz( -1 ) = 0,
\end{equation}
but the insulating boundary conditions
\begin{equation}
\label{eq:insulatingFilm}
\DD \dbz( -1 ) - \alpha \dbz( -1 ) = \DD \dbz( 0 ) + \alpha \dbz( 0 ) - \ii \alpha \DD \bbx( 0 ) a = 0
\end{equation}
now involve the free-surface oscillation amplitude (\cf \eqref{eq:insulatingChannel}). In the inductionless limit, the boundary conditions for $ b $ are not required. Furthermore, \eqref{eq:normalStress} reduces to
\begin{multline}
\label{eq:normalStressZeroPm} ( ( \DD^2 - 3 \alpha^2 ) \DD - \Rey( \gamma + \ii \alpha U )\DD + \ii \alpha \Rey ( \DD U ) - \Harz ( \ii \alpha \Harx + \Harz \DD ) ) \duz|_{z=0} \\
- \alpha^2 \left( \frac{ 1 }{ \Prg^2 \Rey } + \frac{ \alpha^2 }{  \Ohn^2 \Rey } - 2 \ii \alpha \DD U( 0 ) \right) \deh = 0.
\end{multline} 
  
To summarize, we refer to all problems involving a free surface as film problems and those that take place within fixed boundaries as channel problems. Within the film category, we call \emph{film MHD problems} those governed by~\eqref{eq:coupledOSInd}, subject to the boundary conditions~\eqref{eq:kinematic}--\eqref{eq:insulatingFilm}. These are to be distinguished from \emph{inductionless film problems}, where the coupled OS and induction equations are replaced by~\eqref{eq:orrSommerfeldZeroPm}, and the boundary conditions are \eqref{eq:kinematic}, \eqref{eq:shearStress}, \eqref{eq:noSlipFilm}, and~\eqref{eq:normalStressZeroPm}. Similarly, we differentiate between \emph{channel MHD problems}, specified by~\eqref{eq:coupledOSInd}, \eqref{eq:noSlipChannel}, and~\eqref{eq:insulatingChannel}, and their inductionless variants, where the differential equation and boundary conditions are respectively~\eqref{eq:orrSommerfeldZeroPm} and~\eqref{eq:noSlipChannel}. Finally, for what we refer to as \emph{non-MHD film problems} and \emph{non-MHD channel problems} we set $ \Harx = \Harz = 0 $ in~\eqref{eq:orrSommerfeldZeroPm} and~\eqref{eq:normalStressZeroPm}. We mention in passing that one can treat in a similar manner `jet' problems, where free-surface boundary conditions are enforced at $ z = \pm 1 $, although problems of this type will not be considered here.

\subsection{\label{sec:steadyState}Steady-State Configuration}

\begin{figure}
\begin{center}
\includegraphics{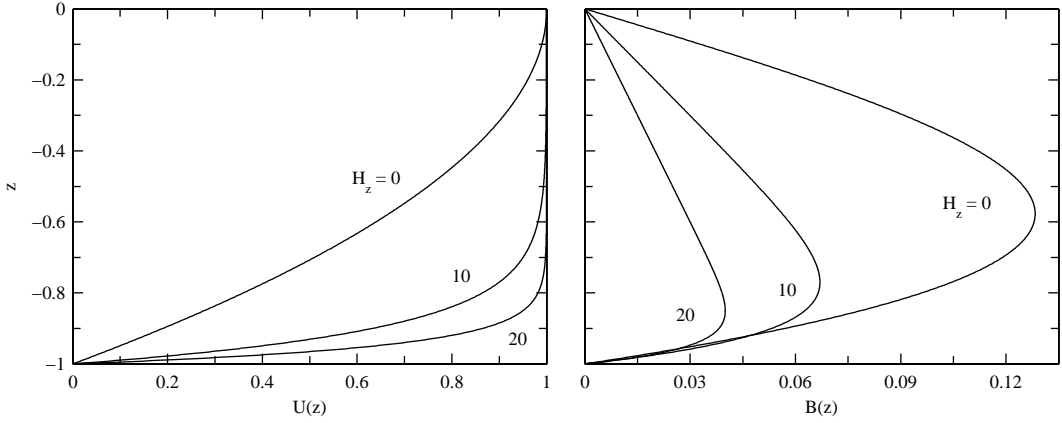}
\end{center}
\caption{\label{fig:baseProfile}Steady-state velocity $ U $ (left) and magnetic field $ B $ (right) for Hartmann flow \eqref{eq:baseHartmann} at $ \Harz = \text{0, 10, 20}$}
\end{figure}

In what follows we consider the magnetic-field configuration
\begin{equation}
\label{eq:bbx}
( \bbx( z ), \bbz ) = ( \Alfxinv, \Alfzinv ) + ( \Alfzinv \Reym \bbind( z ), 0 ),
\end{equation}
where $ ( \Alfxinv, \Alfzinv ) $ is a uniform, externally imposed magnetic field, quantified in terms of the streamwise and flow-normal Alfv\'en numbers $ \Alfx $ and $ \Alfz $, and $ \bbind \in C^2( \bar \Omega ) $ is a function representing the magnetic field induced by the fluid motion $ \bux( z ) $ within the background field  ($ B $ is equal to the corresponding function $ \bar B $ in~\cite{Takashima96}). For the test calculations presented in \S\ref{sec:numericalResults} we employ the Hartmann profiles \cite{MullerBuhler01}
\begin{equation}
\label{eq:baseHartmann}
\bux( z ) = ( \cosh( \Harz ) - \cosh( \Harz z ) ) / X,
\quad
\Harz \bbind( z ) = ( \sinh( \Harz z ) - \sinh( \Harz ) z ) / X,
\end{equation}
where $ \Harz = ( \Rey \Reym )\sh \Alfzinv $, $ X = \cosh( \Harz ) - 1 $, and $ z \in [ -1, 1] $. Note that the expressions above are valid for both channel and film problems. In the latter case, one restricts $ z $ to the interval $ \bar\Omegaf $ to obtain `half' of the corresponding channel profile.  A further useful quantity is the mean velocity,
\begin{equation}
\label{eq:baseUAverage}
\langle U \rangle := \int_{z_1}^{z_2} \ddz \, \frac{ U( z ) }{ z_2 - z_1 } = ( \cosh( \Harz ) - \sinh( \Harz ) / \Harz ) / X,
\end{equation} 
which grows monotonically from $ 2/3 $ to 1 as $ \Harz $ increases from zero to infinity. The steady-state configuration described by~\eqref{eq:bbx} and \eqref{eq:baseHartmann} is a solution of the unperturbed Navier--Stokes and induction equations \cite{GiannakisRosnerFischer07}. In the limit $ \Harz \to 0 $ we have,
\begin{equation}
\label{eq:baseNonMhd}
\bux( z )  = 1 - z^2,
\quad
\bbind( z ) = - z ( 1 - z^2 ) / 3, 
\end{equation}
indicating that the velocity profile reduces to the usual Poiseuille one. Even though $ B $ is nonzero in the limit, the streamwise induced magnetic field $ \Alfzinv \Reym B = \Prm\sh \Harz B $ vanishes. For $ \Harz > 0 $ the velocity and magnetic-field profiles develop exponential tails of thickness $ 1 / \Harz $, where the vorticity and current are concentrated. These so-called Hartmann layers form near the no-slip walls, as shown in Fig.~\ref{fig:baseProfile}.

\subsection{\label{sec:energyBalance}Energy Balance}
In \S\ref{sec:eigenvalueSpectra} and \S\ref{sec:consistency} ahead we shall make use of energy conservation laws for the normal modes, which follow from the linearized Navier--Stokes and induction equations governing the evolution of linear perturbations in MHD. Leaving the details of the derivation to \cite{GiannakisRosnerFischer07}, to each $ ( u, b, a ) $ satisfying \eqref{eq:coupledOSInd} and the boundary conditions~\eqref{eq:kinematic}--\eqref{eq:insulatingFilm} we assign an energy $ E := E_u + E_b + E_a $, consisting of kinetic, magnetic, and surface contributions
\begin{subequations}
\label{eq:energySum}
\begin{align}
\label{eq:eU}
E_u & := \int_{-1}^{0} \ddz \, ( | D u( z ) |^2 + \alpha^2 | u( z ) |^2 ), \\
E_b & := \int_{-1}^{0} \ddz \, ( | D b( z ) |^2 + \alpha^2 | b( z ) |^2 ) + 2 \alpha ( | b( -1 ) |^2 + | b( 0 ) |^2 ), \\
\label{eq:eA}
E_a & := \alpha^2 \left( \Fro^{-2} + \bbx( 0 ) \DD \bbx( 0 ) + \Web \alpha^2 \right) | a |^2.
\end{align}
\end{subequations}
Here the kinetic energy $ E_u $ is (up to a proportionality constant) the energy norm of the 2D velocity field associated with the velocity eigenfunction $ u $, while the magnetic energy contains, in addition to the energy norm of the magnetic field within the fluid, boundary terms representing the energy of the field penetrating through the insulating boundaries. The surface energy consists of potential energy due to gravitational and magnetic stresses, plus a contribution from surface tension. In inductionless problems the modal energy is $ E = E_u + E_a $, where $ E_u $ is given by~\eqref{eq:eU}, and $ E_a $ follows from~\eqref{eq:eA} with $ \bbx $ set to zero. 

Aside from $ E $, to each $ ( u, b, a ) $ correspond power-transfer terms
\begin{subequations}
\label{eq:gammaSum}
\begin{align}
\label{eq:gammaR}
\Gamma_\mathrm{R} & := \frac{\alpha}{E} \int_{-1}^{0} \ddz \, ( \DD U( z ) )\Imag( \cconj{ u( z ) } \DD u( z ) ), \\
\label{eq:gammaM}
\Gamma_\mathrm{M} & := - \frac{ \alpha }{ E } \int_{-1}^{0} \ddz \, ( \DD U( z ) )\Imag( \cconj{ b( z ) } \DD b( z ) ), \\
\label{eq:gammaJ}
\Gamma_J & := \frac{ \alpha }{ E } \int_{-1}^{0} \ddz \, ( \DD \bbx( z ) ) \Imag( \cconj{ u( z ) } \DD b( z ) - \cconj{ b( z ) } \DD u( z ) ), \\ 
\Gamma_\nu & := - \frac{ 1 }{ E \Rey } \int_{-1}^{0} \ddz \, ( | \DD^2 u( z ) |^2 - 2 \alpha^2 \Real( \cconj{ u( z ) } \DD^2 u( z ) ) + \alpha^4 | u( z ) |^2 ), \\
\label{eq:gammaEta}
\Gamma_\eta & := - \frac{ 1 }{ E \Reym } \int_{-1}^{0} \ddz \, ( | \DD^2 b( z ) |^2 - 2 \alpha^2 \Real( \cconj{ b( z ) } \DD^2 b( z ) ) + \alpha^4 | b( z ) |^2), \\
\Gamma_{a \nu } & := - \frac{ \alpha }{ E \Rey } ( \DD U( 0 ) ) \Imag( \DD u( 0 ) \cconj{ a } ), \\
\label{eq:gammaAEta}
\Gamma_{ a \eta } & := \frac{ \alpha }{ E \Reym } ( \DD \bbx( 0 ) )\left( \Imag( ( \DD^2b( 0 ) - \alpha^2 b( 0 ) ) \cconj{ a } )  + \bbz \Imag( \DD u( 0 ) \cconj{ a } )  + \alpha \bbx( 0 ) \Real( u( 0 ) \cconj{ a }) \right),
\end{align}
\end{subequations}
each of which has a physical interpretation. $ \Gamma_\mathrm{ R } $ and $ \Gamma_\mathrm{ M } $ are the Reynolds and Maxwell stresses, \ie the power transferred from the steady-state velocity field $ U $ to the velocity and magnetic-field perturbations. $ \Gamma_J $ is the so-called current interaction; the power transfer from the steady-state current (represented by the $ \DD \bbx $ term in~\eqref{eq:gammaJ}) to the perturbations. The non-positive quantities $ \Gamma_\nu $ and $ \Gamma_\eta $ are respectively the viscous and resistive dissipation (the subscripts $ \nu $ and $ \eta $ stand for the viscous and magnetic diffusivities). Finally, the surface terms $ \Gamma_{ a \nu } $ and $ \Gamma_{a\eta} $ represent the power transferred to the free surface by viscous and electromagnetic forces, respectively. It can be shown that the sum of the terms in~\eqref{eq:gammaSum} is equal to the modal growth rate. That is, the real part $ \Gamma $ of the eigenvalue $ \gamma $ corresponding to $ ( u, b, a ) $ is expressible as
\begin{equation}
\label{eq:energyBalance}
\Gamma = \Real( \gamma ) = \Gamma_\mathrm{ R } + \Gamma_\mathrm{ M } + \Gamma_J + \Gamma_\nu + \Gamma_\eta + \Gamma_{ a \nu } + \Gamma_{ a \eta }.    
\end{equation}
Similar energy-balance relations can be derived for channel and inductionless problems, but we do not require them here. 

\section{\label{sec:weakFormulation}Weak Formulation}

We now cast the eigenvalue problems specified in~\S\ref{sec:problemDescription} into weak (variational) form, suitable for the Galerkin schemes developed in~\S\ref{sec:galerkinDiscretization}. With a slight abuse of notation we use the symbol $ \Omega $ to denote the domain of both film and channel problems, where it is understood that $ \Omega $ stands for either $ \Omegaf $ (film problems) or $ \Omegac $ (channel problems), depending on the context. Also, we collectively denote the vector spaces of admissible solutions for the velocity and magnetic-field eigenfunctions by $ V_u $ and $ V_b $, respectively, even though different versions of these spaces will be constructed for film and channel problems. In what follows, we describe the procedure of obtaining the weak formulation of film MHD problems. Channel MHD problems, as well as the inductionless variants of film and channel problems, can be treated in an analogous manner, and, in the interests of brevity, we shall merely state the results.

Given an interval $ \Omega = ( z_1, z_2 ) \in \mathbb{ R } $, we denote by $ L^2( \Omega ) $ the Hilbert space of square-integrable complex-valued functions on $ \Omega $, equipped with the inner product $ \innerprodLtwo{ \Omega }{ v_1 }{v _2 } := \int_{z_1}^{z_2} \ddz v_1( z ) \cconj{ v_2( z ) }$ and induced norm $ \normLtwo{ \Omega}{ v }^2 := \innerprodLtwo{ \Omega }{ v }{ v } $. We then introduce as usual (\eg \cite{AdamsFournier03}) the Sobolev spaces $ H^k( \Omega ) $, $ k \in \mathbb{ N } $, consisting of elements $ v \in L^2( \Omega ) $, whose weak derivatives $ \DD^d v $ for $ | d | \leq k $ are also in $ L^2( \Omega ) $. Moreover, $ H^k_0( \Omega ) $ are the closures in $ H^k( \Omega ) $ of $ C_0^\infty( \Omega ) $, the space of smooth, compactly supported functions on $ \Omega $. The associated semi-norms and norms are given by $ \seminorm{ v }_{k, \Omega}^2 := \normLtwo{ \Omega }{ \DD^k v }^2 $ and $ \norm{ v }_{k, \Omega}^2 := \sum_{ n = 0 }^k \seminorm{ v }_k^2 $, where $ \seminorm{ \cdot }_{k, \Omega } $ and $ \norm{ \cdot }_{k, \Omega } $ are equivalent norms on $ H_0^k( \Omega ) $. Using the symbol $ \hookrightarrow $ to denote embedding, it is a consequence of the Sobolev embedding theorem that $ H^2( \Omega ) \hookrightarrow C^1( \bar \Omega ) $ \cite{AdamsFournier03}. That is, each $ v \in H^2( \Omega ) $ is equal to a unique function $ \tilde v \in C^1( \bar \Omega ) $, except on a measure-zero subset of $ \Omega $. This allows us to define the boundary-value maps $ \tracemap_i^j : H^2( \Omega ) \mapsto \mathbb{ C } $ for $ i \in \{ 1, 2 \} $ and $ j \in \{ 0, 1 \} $, where $ \tracemap_i^j( v ) = \DD^j( \tilde v( z_i ) ) $. We then construct the space
\begin{equation}
H^2_1( \Omega ) :=  \{ v \in H^2( \Omega ); \, \tracemap_1^0( v )  = \tracemap_1^1( v ) = 0 \},
\end{equation}  
which will be used as trial and test space of velocity eigenfunctions in film problems. Using the embedding $ H^1( \Omega ) \hookrightarrow C^0( \bar \Omega ) $, we also introduce the boundary-value maps $ \tracemap_i^0( v ) = \tilde v( z_i ) $ for $ H^1( \Omega ) $, where now $ v \in H^1( \Omega ) $ and $ \tilde v $ is its image in $ C^0( \bar \Omega ) $. The latter two maps will be used to (weakly) enforce the insulating boundary conditions obeyed by the magnetic field. 

In the strong (classical) formulation of film MHD problems we express Eqs.~\eqref{eq:coupledOSInd} and~\eqref{eq:kinematic} in the form
\begin{equation}
\label{eq:strongForm}
\op{ K }( \duz, \dbz, \deh ) = \gamma \op{ M }( \duz, \dbz, \deh ),
\end{equation}
where $ \op{ K } $ and $ \op{ M } $ are matrix differential operators. These so-called `stiffness' and `mass' operators, respectively with domain $ \mathcal{ D }_\op{ K } = C^4( \bar \Omega ) \times C^2( \bar \Omega ) \times \mathbb{ C } $ and $ \mathcal{ D }_\mathcal{ M } = C^2( \bar \Omega ) \times C^1( \bar \Omega ) \times \mathbb{ C } \supset \mathcal{ D }_\op{ K } $, are given by
\begin{equation}
\op{ K }( \duz, \dbz, \deh ) = 
\left( 
\begin{array}{lll}
\op{ K }_{uu} & \op{ K }_{ub } & 0 \\
\op{ K }_{bu} & \op{ K }_{bb} & 0 \\
\tracemap_1^0 & 0 & - \ii \alpha \bux( 0 ) 
\end{array}
\right)
\left(
\begin{array}{l}
\duz \\
\dbz \\
\deh
\end{array}
\right),
\quad
\op{ M }( \duz, \dbz, \deh ) = 
\left( 
\begin{array}{lll}
- \Rey ( \DD^2 - \alpha^2 ) & 0 & 0\\
0 & \Reym & 0\\
0 & 0 & 1 
\end{array}
\right)
\left(
\begin{array}{l}
\duz \\
\dbz \\
\deh
\end{array}
\right),
\end{equation}
where
\begin{subequations}
\begin{gather}
\op{ K }_{uu} = - ( \DD^2 - \alpha^2 )^2 + \ii \alpha \Rey ( \bux  ( \DD^2 - \alpha^2 ) -  ( \DD^2 \bux ) ), \quad \op{ K }_{bb}  = \DD^2 - \alpha^2 - \ii \alpha \Reym \bux, \\
\op{ K }_{ub} = - \Rey ( \ii \alpha \bbx + \bbz \DD ) ( \DD^2 - \alpha^2 ) + \ii \alpha \Rey ( \DD^2 \bbx ), \quad
\op{ K }_{bu} = \Reym ( \ii \alpha \bbx + \bbz \DD ).
\end{gather}
\end{subequations} 
Note that in~\eqref{eq:strongForm} we have multiplied the OS equation~\eqref{eq:orrSommerfeld} by $ - 1 $. This is a conventional manipulation, with no influence on the scheme's numerical behavior, made in order to obtain a positive-definite mass form in~\eqref{eq:weakForm} below. The strong version of the problem may then be stated as follows: Find $ \gamma \in \mathbb{ C } $ and $ ( \duz, \dbz, \deh ) \in \mathcal{ D }_\op{ K } \setminus \{ ( 0, 0, 0 ) \} $, such that the governing equations~\eqref{eq:strongForm}, and the boundary conditions~\eqref{eq:kinematic}--\eqref{eq:insulatingFilm} are satisfied.

In order to pass from the strong to the weak (variational) formulation, one begins by identifying the spaces of admissible solutions $ V_u $ and $ V_b $ for the velocity and magnetic-field eigenfunctions, respectively. In film problems we set $ V_u = H^2_1( \Omega ) $ and $ V_b = H^1( \Omega ) $, so that the no-slip boundary conditions~\eqref{eq:noSlipFilm} are enforced strongly, whereas the stress and insulating boundary conditions, \eqref{eq:normalStress}, \eqref{eq:shearStress}, and \eqref{eq:insulatingFilm}, must be imposed in a natural (weak) sense (\eg \cite{Ho89,HoPatera91}). Taking the free-surface amplitude into account, the full solution space is therefore $ V = V_u \times V_b \times \mathbb{ C } $, which we equip with the direct-sum inner product $ \innerprodV{\Omega}{ v_1 }{ v_2 } = \innerprodLtwo{ \Omega }{ u_1 }{ u_2 } + \innerprodLtwo{\Omega}{ b_1 }{ b_2 } + a_1 \cconj{ a_2} $, where $ u_j \in V_u $, $ b_j \in V_b $, $ a_j \in \mathbb{ C } $, and $ v_j = ( u_j, b_j, a_j ) \in V $ for $j \in \{1, 2 \}$.

We now proceed to construct sesquilinear forms $ \form{ K } $ and $ \form{ M } $ associated to $ \op{ K } $ and $ \op{ M } $, respectively. Introducing a test element $ \vtest = ( \utest, \btest, \htest ) \in V $, we form the $ \innerprodV{ \Omega }{\cdot}{\cdot} $ inner product of~\eqref{eq:strongForm} with $ \vtest $, namely $ \innerprodV{ \Omega }{ \op{ K }( \duz, \dbz, \deh ) }{ \vtest } = \gamma \innerprodV{ \Omega }{ \op{ M }( \duz, \dbz, \deh ) }{ \vtest } $. Upon integration by parts this leads to
\begin{equation}
\label{eq:weakForm}
\form{ K }( \vtrial, \vtest ) = \gamma \form{ M }( \vtrial, \vtest ),
\end{equation}
where now $ \vtrial = ( \utrial, \btrial, \htrial ) \in V \supset \mathcal{D}(\op{ K } )$. Also, $ \form{ K } : V \times V \mapsto \mathbb{ C } $ and $ \form{ M } : V \times V \mapsto \mathbb{ C } $ are sesquilinear forms associated with the mass and stiffness operators $ \op{ K } $ and $ \op{ M } $, respectively. 
We make the decompositions
\begin{subequations}
\label{eq:stiffMassFilmMhd}
\begin{align}
\nonumber
\form{ K }( \vtrial, \vtest ) & = \form{ K }_{uu}( \utrial, \utest ) + \form{ K }_{ub}( \btrial, \utest ) + \form{ K }_{ua}( \htrial, \utest ) + \form{ K }_{bu}( \utrial, \btest ) + \form{ K }_{bb}( \btrial, \btest ) + \form{ K }_{ba}( \btrial, \htest ) \\
\label{eq:stiffnessFilmMHD} & \quad + \form{ K }_{au}( \utrial, \htest ) + \form{ K }_{aa}( \htest, \htrial ), \\
\label{eq:massFilmMHD}
\form{ M }( \vtrial, \vtest ) & = \form{ M }_{uu}( \utrial, \utest ) + \form{ M }_{bb}( \btrial, \btest ) + \form{ M }_{aa} ( \htrial, \htest ),
\end{align}
\end{subequations}
which consist of the following objects: In~\eqref{eq:stiffnessFilmMHD}, $ \form{ K }_{uu}: V_u \times V_u \mapsto \mathbb{ C } $, $ \form{ K }_{bb}: V_b \times V_b \mapsto \mathbb{ C } $, and $ \form{ K }_{aa} : \mathbb{ C } \times \mathbb{ C } \mapsto \mathbb{ C } $ are sesquilinear forms given by
\begin{subequations}
\begin{align}
\label{eq:stiffWeakUU}
\form{ K }_{uu}( \utrial, \utest ) & = \form{ K }_{uu}^{[0]}( \utrial, \utest ) + \form{ K }_{uu}^{[U]}( \utrial, \utest ) + \form{ K }_{uu}^{[\mathrm{S}]}( \utrial, \utest ), \\
\label{eq:stiffWeakBB}
\form{ K }_{bb}( \btrial, \btest ) & = \form{ K }_{bb}^{[0]}( \btrial, \btest ) + \form{ K }_{bb}^{[U]}( \btrial, \btest ) + \form{ K }_{bb}^{[\mathrm{I}]}( \btrial, \btest ), \\
\label{eq:stiffWeakHH} 
\form{ K }_{aa}( \htrial, \htest ) & = - \ii \alpha \bux( 0 ) \htrial \cconj{ \htest }, 
\end{align}
\end{subequations}
where we have split~\eqref{eq:stiffWeakUU} and~\eqref{eq:stiffWeakBB} into free-stream terms,
\begin{subequations} 
\begin{align}
\label{eq:stiffWeakUUFS}
\form{ K }_{uu}^{[0]}( \utrial, \utest ) &:= - ( \innerprodLtwo{ \Omega }{ \DD^2 \utrial }{ \DD^2 \utest } + 2 \alpha^2 \innerprodLtwo{ \Omega }{ \DD \utrial }{ \DD \utest } + \alpha^4 \innerprodLtwo{ \Omega }{ \utrial }{ \utest } ), \\
\label{eq:stiffWeakBBFS}
\form{ K }_{bb}^{[0]}( \btrial, \btest ) &:= - ( \innerprodLtwo{ \Omega }{ \DD \btrial }{ \DD \btest } + \alpha^2 \innerprodLtwo{ \Omega }{ \btrial }{ \btest } ),
\end{align}
\end{subequations}
contributions from the velocity profile $ U $, 
\begin{subequations}
\label{eq:stiffWeakUUBBBaseU}
\begin{align}
\label{eq:stiffWeakUUBaseU}
\form{ K }_{uu}^{[U]}( \utrial, \utest ) & := - \ii \alpha \Rey ( \innerprodLtwo{ \Omega}{ \bux \DD \utrial }{ \DD \utest } + \alpha^2 \innerprodLtwo{ \Omega }{ \bux \utrial }{ \utest } - \innerprodLtwo{ \Omega }{ ( \DD \bux ) \utrial }{ \DD \utest }), \\
\label{eq:stiffWeakBBBaseU}
\form{ K }_{bb}^{[U]}( \btrial, \btest ) & :=  - \ii \alpha \Reym \innerprodLtwo{ \Omega }{ \bux \btrial }{\btest },
\end{align}
\end{subequations}
free-surface terms
\begin{equation}
\label{eq:stiffWeakBoundaryUU}
\form{ K }^{[\mathrm{S}]}_{uu}( \utrial, \utest ) := - \alpha^2 ( \tracemap_2^0( \utrial ) \cconj{ \tracemap_2^1( \utest ) } + \tracemap_2^1( \utrial ) \cconj{ \tracemap_2^0( \utest ) } ),
\end{equation}
and contributions from the insulating boundary conditions
\begin{equation}
\label{eq:stiffWeakBoundaryBB}
\form{ K }^{[\mathrm{I}]}_{bb}( \btrial, \btest ) := - \alpha ( \tracemap_1^0( \btrial ) \cconj{ \tracemap_1^0( \btest ) } + \tracemap_2^0( \btrial ) \cconj{ \tracemap_2^0( \btest ) } ).
\end{equation}
Moreover, $ \form{ K }_{ ub } : V_b \times V_u \mapsto \mathbb{ C } $ and $ \form{ K }_{ bu } : V_u \times V_b \mapsto \mathbb{ C } $ are maps defined by
\begin{subequations}
\label{eq:stiffWeakUBBU}
\begin{align}
\nonumber
\form{ K }_{ub}( \btrial, \utest ) & = \ii \alpha \Rey ( \innerprodLtwo{ \Omega }{ \bbx \DD \btrial }{ \DD \utest } + \alpha^2 \innerprodLtwo{ \Omega }{ \bbx \btrial }{ \utest } - \innerprodLtwo{ \Omega }{ ( \DD \bbx ) \btrial }{ \DD \utest } ) \\
\label{eq:stiffWeakUB} & \quad - \Rey \bbz ( \innerprodLtwo{ \Omega }{ \DD \btrial }{ \DD^2 \utest } + \alpha^2 \innerprodLtwo{ \Omega }{ \btrial }{ \DD \utest } ) - \alpha \Rey \tracemap_2^0( \btrial ) \cconj{ ( \ii \alpha \bbx( 0 ) \tracemap_2^0( \utest ) + \bbz \tracemap_2^1( \utest ) ) },\\
\label{eq:stiffWeakBU}
\form{ K }_{bu}( \utrial, \btest ) & = \Reym ( \ii \alpha \innerprodLtwo{ \Omega }{ \bbx \utrial }{ \btest } + \bbz \innerprodLtwo{ \Omega }{ \DD \utrial }{ \btest }).
\end{align}
\end{subequations}
For the parameterization~\eqref{eq:bbx} of the magnetic field we have
\begin{equation}
\form{ K }_{ub}( \btrial, \utest ) = \form{ K }_{ub}^{[0]}( \btrial, \utest ) + \form{ K }_{ub}^{[\bbind]}( \btrial, \utest ) + \form{ K }_{ub}^{[\mathrm{S}]}( \btrial, \utest ), \quad
\form{ K }_{bu}( \utrial, \btest ) = \form{ K }_{bu}^{[0]}( \utrial, \btest ) + \form{ K }_{bu}^{[\bbind]}( \utrial, \btest ),
\end{equation}
where, employing a similar notation as above, 
\begin{subequations}
\label{eq:stiffWeakUBBUFS}
\begin{align}
\label{eq:stiffWeakUBFS}
\form{ K }_{ub}^{[0]}( \btrial, \utest ) & := \ii \alpha \Rey \Alfxinv ( \innerprodLtwo{ \Omega }{ \DD \btrial }{ \DD \utest } + \alpha^2 \innerprodLtwo{ \Omega }{ \btrial }{ \utest } ) - \Rey \Alfzinv ( \innerprodLtwo{ \Omega }{ \DD \btrial }{ \DD^2 \utest } + \alpha^2 \innerprodLtwo{ \Omega }{ \btrial }{ \DD \utest } ), \\
\label{eq:stiffWeakBUFS}
\form{ K }_{bu}^{[0]}( \utrial, \btest ) & := \ii \alpha \Reym \Alfxinv \innerprodLtwo{ \Omega }{ \utrial }{ \btest } + \Reym \Alfzinv \innerprodLtwo{ \Omega }{ \DD \utrial }{ \btest }
\end{align}
\end{subequations}
are free-stream terms,
\begin{subequations}
\label{eq:stiffWeakUBBUIndB}
\begin{align}
\label{eq:stiffWeakUBIndB}
\form{ K }_{ub}^{[B]}( \btrial, \utest ) & := \ii \alpha \Rey \Harz \Prm\sh ( \innerprodLtwo{ \Omega }{ \bbind \DD \btrial }{ \DD \utest } + \alpha^2 \innerprodLtwo{ \Omega }{ \bbind \btrial }{ \utest } - \innerprodLtwo{ \Omega }{ ( \DD \bbind ) \btrial }{ \DD \utest } ), \\
\label{eq:stiffWeakBUIndB}
\form{ K }_{bu}^{[B]}( \btrial, \utest ) & := \ii \alpha \Reym \Harz \Prm\sh \innerprodLtwo{ \Omega }{ \bbind \utrial }{ \btest }
\end{align}
are the contributions from the induced magnetic field $ \bbind $, and
\end{subequations}
\begin{equation}
\label{eq:stiffWeakBoundaryUB}
\form{ K }_{ub}^{[\mathrm{S}]}( \btrial, \utest ) := - \alpha \Rey \tracemap_2^0( \btrial ) \cconj{ ( \ii \alpha( \Alfxinv + \Alfzinv \Reym \bbind( 0 ) ) \tracemap_2^0( \utest ) + \Alfzinv \tracemap_2^1( \utest ) ) }
\end{equation}
are free-surface terms. The maps $ \form{ K }_{ ua } : \mathbb{ C } \times V_u \mapsto \mathbb{ C } $, $ \form{ K }_{ ba } : \mathbb{ C } \times V_b \mapsto \mathbb{ C } $, and $ \form{ K }_{au} : V_u \times \mathbb{ C } \mapsto \mathbb{ C } $, where
\begin{subequations}
\label{eq:boundaryFormsA}
\begin{align}
\label{eq:stiffWeakUH}
\form{ K }_{ua}( \htrial, \utest ) & := - \alpha^2 \left( \frac{ 1 }{ \Prg^2 \Rey } + \frac{ \alpha^2 }{ \Ohn^2 \Rey }  - 2 \ii \alpha \DD \bux( 0 ) \right) \htrial \cconj{ \tracemap_2^0(\utest) } + \ii \alpha ( \DD^2\bux( 0 ) + \Harz^2  \DD B( 0 ) ) \htrial \cconj{ \tracemap_2^1( \utest ) }, \\
\label{eq:stiffWeakBH}
\form{ K }_{ba}( \htrial, \btest ) & := \ii \alpha \Alfzinv \Reym \DD B( 0 ) \htrial \cconj{ \tracemap_2^0( \btest ) }, \\
\label{eq:stiffWeakHU}
\form{ K }_{au}( \utrial, \htest ) & := \tracemap_2^0( \utrial ) \cconj{ \htest },
\end{align}
\end{subequations}
represent the coupling of the velocity and magnetic field to the free-surface amplitude. Finally, Eq.~\eqref{eq:massFilmMHD} contains the forms $ \form{ M }_{uu}: V_u \times V_u \mapsto \mathbb{ C } $, $ \form{ M }_{ bb} : V_b \times V_b \mapsto \mathbb{ C } $, and $ \form{ M }_{aa} : \mathbb{ C } \times \mathbb{ C } \mapsto \mathbb{ C } $, where
\begin{subequations}
\label{eq:massWeak}
\begin{align}
\label{eq:massWeakUU}
\form{ M }_{uu}( \utrial, \utest ) & :=  \Rey ( \innerprodLtwo{ \Omega }{ \DD \utrial }{ \DD \utest } + \alpha^2\innerprodLtwo{ \Omega }{ \DD \utrial }{ \DD \utest } ), \\
\label{eq:massWeakBB}
\form{ M }_{bb}( \btrial, \btest  ) & := \Reym \innerprodLtwo{ \Omega }{ \btrial }{ \btest }, \\
\label{eq:massWeakHH}
\form{ M }_{aa}( \htrial, \htest ) & := \htrial \cconj{ \htest }.
\end{align}
\end{subequations}
 We are now ready to state the weak formulation of film MHD problems:
\begin{defn}[Film MHD problem] 
\label{def:filmMhd}
Let $ \Omega = \Omegaf $, $ V_u = H^2_1( \Omega ) $, $ V_b = H^1( \Omega ) $, and $ V = V_u \times V_b \times \mathbb{ C } $. Then, find $ ( \gamma, \vtrial ) \in \mathbb{  C } \times V \setminus \{ 0 \} $, such that for all $ \vtest \in V $ Eq.~\eqref{eq:weakForm}, with $ \form{ K } $ and $ \form{ M }$ given by~\eqref{eq:stiffMassFilmMhd}, is satisfied.
\end{defn}

In a similar manner, one can construct weak formulations of the form~\eqref{eq:weakForm} for channel MHD problems, as well as film and channel problems in the inductionless limit. In what follows, we will always use $ V $ to denote the full (direct sum) solution space. Also, we shall employ throughout the notation $ \form{ K } $ and $ \form{ M } $ for the stiffness and mass forms, and $ \form{ K }_{uu} $, $ \form{ M }_{uu} $, \etc for their constituent sub-maps. It is understood that the maps act on pairs of elements from the appropriate vector space, and their definitions are restricted to the problem type under consideration. In channel MHD problems, we select the solution spaces $ V_u =  H^2_0( \Omega ) $,  $ V_b = H^1( \Omega ) $, and $ V = V_u \times V_b $, where now $ \Omega = \Omegac $. The stiffness and mass forms in~\eqref{eq:weakForm} read
\begin{subequations}
\label{eq:stiffnessMassChannelMhd}
\begin{align}
\label{eq:stiffnessChannelMHD}
\form{ K }( \vtrial, \vtest ) & = \form{ K }_{uu}( \utrial, \utest ) + \form{ K }_{ub}( \btrial, \utest ) + \form{ K }_{bu}( \utrial, \btest ) + \form{ K }_{bb}( \btrial, \btest ),\\
\label{eq:massChannelMHD}
\form{ M }( \vtrial, \vtest ) & = \form{ M }_{uu}( \utrial, \utest ) + \form{ M }_{bb}( \btrial, \btest ),
\end{align}
\end{subequations}
where $ \form{ K }_{bu} $, $ \form{ K }_{bb} $, $ \form{ M }_{uu} $, and $ \form{ M }_{bb} $ are defined as the corresponding maps for the film problem, \ie \eqref{eq:stiffWeakBU}, \eqref{eq:stiffWeakBB}, \eqref{eq:massWeakUU}, and~\eqref{eq:massWeakBB}. However, $ \form{ K }_{uu } $ and $ \form{ K }_{ ub } $ are now given by
\begin{equation}
\label{eq:stiffWeakUUBBChannelMHD}
\form{ K }_{uu}( \utrial, \utest ) = \form{ K }_{uu}^{[0]}( \utrial, \utest ) + \form{ K }_{uu}^{[U]}( \utrial, \utest ), \quad
\form{ K }_{ub}( \btrial, \utest ) = \form{ K }_{ub}^{[0]}( \btrial, \utest ) + \form{ K }_{ub}^{[\bbind]}( \btrial, \utest ),
\end{equation} 
where $ \form{ K }_{uu}^{[0]} $, $ \form{ K }_{uu}^{[U]} $, $ \form{ K }_{ub}^{[0]} $, and $ \form{ K }_{ub}^{[\bbind]} $ have the same form as~\eqref{eq:stiffWeakUUFS}, \eqref{eq:stiffWeakUUBaseU}, \eqref{eq:stiffWeakUBFS} and \eqref{eq:stiffWeakUBIndB}, respectively. The absence of the boundary terms in~\eqref{eq:stiffWeakUUBBChannelMHD} is due to the essential imposition of the velocity boundary conditions~\eqref{eq:noSlipChannel}. With these specifications, the variational formulations of channel MHD problems is as follows: 
\begin{defn}[Channel MHD problem]
\label{def:channelMhd}
Set $ \Omega = \Omegac $, and $ V_u = H^2_0( \Omega ) $, $ V_b = H^1( \Omega ) $, $ V = V_u \times V_b $. Then, find $ ( \gamma, \vtrial ) \in \mathbb{  C } \times V \setminus \{ 0 \} $, such that for all $ \vtest \in V $ Eq.~\eqref{eq:weakForm}, with $ \form{ K } $ and $ \form{ M }$ given by~\eqref{eq:stiffnessMassChannelMhd}, holds.
\end{defn}
Film problems in the inductionless limit are governed by~\eqref{eq:orrSommerfeldZeroPm} subject to the boundary conditions~\eqref{eq:kinematic}, \eqref{eq:shearStress}, \eqref{eq:noSlipFilm} and~\eqref{eq:normalStressZeroPm}. Like in full MHD problems we set $ V_u = H^2_1( \Omega ) $, but now $ V = V_u \times \mathbb{ C } $. The stiffness and mass forms then become
\begin{equation}
\label{eq:stiffnessMassFilmZeroPm}
\form{K}( \vtrial, \vtest ) = \form{ K }_{uu}( \utrial, \utest ) + \form{ K }_{ ua }( \htrial, \utest ) + \form{ K }_{au}( \utrial, \htest ) + \form{ K }_{aa}( \htrial, \htest ), \quad
\form{M}( \vtrial, \vtest ) = \form{ M }_{uu}( \utrial, \utest ) + \form{ M }_{aa}( \htrial, \htest ),
\end{equation}
where
\begin{equation}
\label{eq:stiffWeakUUFilmPrm0}
\form{ K }_{uu}( \utrial, \utest ) = \form{ K }_{uu}^{[0]}( \utrial, \utest ) + \form{ K }_{uu}^{[U]}( \utrial, \utest ) + \form{ K }_{uu}^{[\mathrm{S}]}( \utrial, \utest )
+ \form{ K }_{uu}^{[\mathrm{L}]}( \utrial, \utest ).
\end{equation} 
Here the form $ \form{ K }_{uu}^{[\mathrm{L}]} : V_u \times V_u \mapsto \mathbb{ C } $, defined by
\begin{align}
\label{eq:stiffWeakUULorentz}
\form{ K }_{uu}^{[\mathrm{L}]}( \utrial, \utest ) & := - \alpha^2 \Harx^2 \innerprodLtwo{ \Omega }{ \utrial }{ \utest } + \ii \alpha \Harx \Harz ( \innerprodLtwo{ \Omega }{ \DD \utrial }{ \utest } - \innerprodLtwo{ \Omega }{ \utrial }{ \DD \utest } ) - \Harz^2 \innerprodLtwo{ \Omega }{ \DD \utrial }{ \DD \utest },
\end{align}
represents the contributions from Lorentz forces, and $ \form{ K }_{uu}^{[0]} $, $ \form{ K }_{uu}^{[U]} $, and $ \form{ K }_{uu}^{[\mathrm{S}]} $ are given by~\eqref{eq:stiffWeakUUFS}, \eqref{eq:stiffWeakUUBaseU} and~\eqref{eq:stiffWeakBoundaryUU}. Moreover,
\begin{align}
\label{eq:stiffWeakUHZeroPm}
\form{ K }_{ua}( \htrial, \utest ) & := - \alpha^2 \left( \frac{ 1 }{ \Prg^2 \Rey }  + \frac{ \alpha^2 }{ \Ohn^2 \Rey } - 2 \ii \alpha  \DD \bux( 0 ) \right) \htrial \cconj{ \tracemap_2^0(\utest ) } - \ii \alpha  \DD^2 \bux( 0 ) \htrial \cconj{ \tracemap_2^1( \utest ) }  
\end{align}
is the analog of~\eqref{eq:stiffWeakUH} in the inductionless limit. The maps $ \form{ K }_{ au } $, $ \form{ K }_{aa} $, $ \form{ M }_{uu} $, and $ \form{ M }_{aa} $ are defined in~\eqref{eq:stiffWeakHU}, \eqref{eq:stiffWeakHH}, \eqref{eq:massWeakUU}, and~\eqref{eq:massWeakHH}, respectively, and therefore we can now state the weak formulation of inductionless film problems:  
\begin{defn}[Inductionless film problem]
\label{def:filmZeroPm}
Let $ \Omega = \Omegaf $, $ V_u = H^2_1( \Omega ) $, and $ V = V_u \times \mathbb{ C } $. Find $ ( \gamma, \vtrial ) \in \mathbb{ C } \times V \setminus \{ 0 \} $, such that~\eqref{eq:weakForm}, with $ \form{ K } $ and $ \form{ M } $ given by~\eqref{eq:stiffnessMassFilmZeroPm}, is satisfied for all $ \vtest \in V $.
\end{defn}
The trial and test space for inductionless channel problems is simply $ V = V_u = H^2_0( \Omega ) $. Moreover, the stiffness and mass forms reduce to
\begin{equation}
\label{eq:stiffnessMassChannelZeroPm}
\form{ K }( \vtrial, \vtest ) = \form{ K }_{uu}( \utrial, \utest ), \quad
\form{ M }( \vtrial, \vtest ) = \form{ M }_{uu}( \utrial, \utest ).
\end{equation}
where 
\begin{equation}
\form{ K }_{uu} = \form{ K }_{uu}^{[0]}( \utrial, \utest ) + \form{ K }_{uu}^{[U]}( \utrial, \utest ) + \form{ K }_{uu}^{[\mathrm{L}]}( \utrial, \utest ),
\end{equation}
and, as usual, $ \form{ K }^{[0]}_{uu} $, $ \form { K }^{[U]}_{uu} $ and $ \form{ K }^{[\mathrm{L}]}_{uu} $ are given by~\eqref{eq:stiffWeakUUFS}, \eqref{eq:stiffWeakUUBaseU} and \eqref{eq:stiffWeakUULorentz}, and $ \form{ M }_{uu} $ by~\eqref{eq:massWeakUU}. Inductionless channel problems then have the following weak formulation:
\begin{defn}[Inductionless channel problem]
\label{def:channelZeroPm}
Let $ V = H^2_0( \Omega ) $, where $ \Omega = \Omegac $. Let also $ \form{ K } $ and $ \form{ M } $ be the stiffness and mass forms given by~\eqref{eq:stiffnessMassChannelZeroPm}. Then, find $ \vtrial \in V $ such that the relation~\eqref{eq:weakForm} is satisfied for all $ \vtest $ in $ V $.
\end{defn}

We note that the weak formulation of non-MHD problems, in both film and channel geometries, follows by setting the Hartmann numbers Defs.~\ref{def:filmZeroPm} and~\ref{def:channelZeroPm} equal to zero.

\section{\label{sec:galerkinDiscretization} Galerkin Discretization}

The Galerkin discretization of the variational problems formulated in \S\ref{sec:weakFormulation}, collectively represented by equations of the form~\eqref{eq:weakForm}, involves replacing the spaces $ V_u $ and, where applicable, $ V_b $ by finite-dimensional spaces $ V_u^{N_u } \subset V_u $ and $ V_b^{N_b } \subset V_b $, respectively of dimension $ N_u $ and $ N_b $. Denoting the set of polynomials of degree $ p $ on $ \Omega $ by $ \mathcal{ P }^p( \Omega ) $, we define
\begin{equation}
\label{eq:discreteVuVb}
V_u^{ N_u } := V_u \cap \mathcal{ P }^{p_u}( \Omega ), \quad
V_b^{ N_b } := V_b \cap \mathcal{ P }^{p_b}( \Omega ),
\end{equation}
where it is understood that $ \Omega $ stands for $ \Omegaf $  $ ( \Omegac $) when the problem under consideration is of film (channel) type. The subspaces $ V_u^{N_u} $ and $ V_b^{N_b} $  provide a dense coverage of $ V_u $ and $ V_b $ in the limit $ \mbox{$ N_u $, $ N_b $} \tti $. Introducing the  multi-index $ \boldsymbol{N} $, where $ \boldsymbol{ N }= ( N_u, N_b ) $ for MHD problems, and $ \boldsymbol{ N } = N_u $ for their inductionless counterparts, finite-dimensional spaces $ V^{\boldsymbol{N}} \in V $, where $ \dim{ V^{\boldsymbol{N}} } =: N $ can be constructed by substituting $ V_u^{N_u} $ for $ V_u $, and $ V_b^{N_b} $ for $ V_b $ in the definitions for $ V $. Then, the Galerkin discretization of the variational problems~\eqref{eq:weakForm} can be stated as follows: Find $ ( \gamma, \vtrial ) \in \mathbb{ C } \times V^{\boldsymbol{ N } } \setminus \{ 0 \} $ such that for all $ \vtest \in V^{\boldsymbol{N}} $ the relation
\begin{equation}
\label{eq:discreteWeakForm}
\tilde{\form{ K }}( \vtrial, \vtest ) = \gamma \form{ M }( \vtrial, \vtest )
\end{equation} 
is satisfied. Here $ \tilde{\form{ K }} : V^{ \boldsymbol{N} } \times V^{ \boldsymbol{N} } \mapsto \mathbb{ C }$ is an approximation of $ \form{ K } $ (the details of which will be specified in \S\ref{sec:structureDiscrete}), oftentimes introduced to perform numerically the quadratures associated with the velocity and/or magnetic field profiles, $ \bux $ and $ \bbind $. However, in a number of cases, including the Poiseuille and Hartmann profiles considered below, the quadratures can be performed exactly and $ \tilde{\form{ K }} = \form{ K } $ for all elements of $ V^{ \boldsymbol{N} } $. Given a basis $ \{ \psi_i \}_{ i = 1 }^N $ of $ V^{\boldsymbol{N}} $, Eq.~\eqref{eq:discreteWeakForm} is equivalent to the matrix generalized eigenproblem
\begin{equation}
\label{eq:matrixWeakForm}
\mat{ K } \colvec{ v } = \gamma \mat{ M } \colvec{ v },
\end{equation}
where the stiffness and mass matrices, $ \mat{ K } \in \mathbb{ C }^{ N \times N } $ and $ \mat{ M } \in \mathbb{ C }^{ N \times N } $, have elements
\begin{equation}
\label{eq:matrixWeakForm2}
K_{m n} = \tilde{\form{ K }}( \psi_n, \psi_m ), \quad M_{ m n } = \form{ M }( \psi_n, \psi_m ),
\end{equation}
and $ \colvec{ v } = ( v_1, \ldots, v_N )^\mathrm{T} \in \mathbb{ C }^N $ is a column vector of the components of $ \vtrial $ in the $ \{ \psi_i \}_{i=1}^{N} $ basis. 

The matrix eigenproblem~\eqref{eq:matrixWeakForm} can be solved using \eg the QZ algorithm \cite{MolerStewart73,AndersonEtAl99}, or implicitly restarted Arnoldi methods \cite{LehoucqSorensenYang98}. Its numerical properties, such as roundoff sensitivity and memory requirements, depend strongly on the choice of basis for $ V^{\boldsymbol{N}} $. Following the approach of KMS, in the sequel we use basis functions that are linear combinations of Legendre polynomials, constructed according to the smoothness of the underlying infinite-dimensional solution space (\ie its Sobolev order), as well as the boundary conditions. In these bases, the matrices $ \mat{ K } $ and $ \mat{ M } $ are well conditioned at large spectral orders. Moreover, if the velocity and magnetic-field profiles are polynomial, they are banded and sparse, and stable, closed-form expressions exist for their nonzero elements. In problems with Hartmann steady-state profiles the stiffness matrix ceases to be sparse, and the contributions from $ U $ and $ B $ must be computed using numerical quadrature. Yet, $ \mat{ K } $  remains well-conditioned even at high spectral orders (see \S\ref{sec:hartmannProfile} below).  

\subsection{Choice of Basis} 

In order to construct our bases for $ V_u^{ N_u } $ and $ V_b^{ N_b } $ we first introduce the reference interval $ \Omegaref := ( -1, 1 ) $ and the linear map $ Q : \Omegaref \mapsto \Omega = ( z_1,z_2) $, where 
\begin{equation}
\label{eq:mapQ}
Q( \xi ) =  z_0 + j \xi, \quad z_0 := ( z_1 + z_2 ) / 2, \quad j := h / 2 := ( z_2 - z_1 ) / 2.
\end{equation} 
In film problems we have $ z_2 = 0 $, $ z_1 = -1 $, $ z_0 = 1 / 2 $, and $ h = 1 $, whereas in channel problems $ Q $ becomes the identity map ($ z_2 = 1 $, $ z_1 = - 1$, $ z_0 = 0 $, and $ h = 2 $). Uniformly continuous functions on $ \Omega $ can be transported to $ \Omegaref $ via the pullback map $ Q^* : C^0( \Omega ) \mapsto C^0( \Omegaref ) $, where $ ( Q^* f )( \xi ) = f( Q( \xi ) ) $ for any $ f \in C^0( \Omega ) $. The pushforward map $ Q_* : C^0( \Omegaref ) \mapsto C^0( \Omega ) $, where $ ( Q_* \hat f )( z ) = \hat f ( Q^{-1}( z ) ) $ and $ \hat f \in C^0( \Omegaref )$, carries out the reverse operation. Moreover, a straightforward application of the chain rule leads to the relations
\begin{subequations}
\label{eq:formScalings}
\begin{align}
\label{eq:formScalingBoundary}
\DD^{ d_1 }(Q_* \hat f_1 )( Q( \pm 1 ) ) & = j^{-d_1} \DDref^{d_1}\hat f_1(\pm 1) , \\
\label{eq:formScaling}
\innerprodLtwo{ \Omega }{ ( \DD^{d_g} g ) \DD^{ d_1 } Q_* \hat f_1 }{ \DD^{ d_2} Q_* \hat f_2 } & = j^{ 1 - d_g - d_1 - d_2 }
\innerprodLtwo{ \Omegaref }{ \DDref^{ d_g } ( Q^* g ) \DDref^{d_1} \hat f_1}{ \DDref^{d_2} f_2 },
\end{align}
\end{subequations}
where $ \DDref $ is the derivative operator on $ \Omegaref $, and $ \hat f_i $ and $ g $ are sufficiently smooth functions, respectively on $ \Omegaref $ and $ \Omega $. 

Let $ L_n $, where $ n = 0, 1, 2, \ldots $, denote the $ n $-th Legendre polynomial defined on $ \Omegaref $ and normalized such that $ L_n( 1 ) = 1 $ (see \eg \cite{AbramowitzStegun71} for various properties of the Legendre polynomials). The Legendre polynomials obey the orthogonality relation
\begin{subequations}
\label{eq:legendreOrthogonality}
\begin{equation}
\label{eq:legendreOrthogonality0}
\innerprodLtwo{ \Omegaref }{ L_n }{ L_m } = 2 \delta_{m n } / ( 2 n + 1 ),
\end{equation}
where $ \delta_{mn} $ is the Kronecker delta. In addition, the inner-product relations
\begin{align} 
\label{eq:legendreOrthogonality1}
\frac{\innerprodLtwo{ \Omegaref }{ w_1 L_n }{ L_m }}{2} & = \frac{ ( n + 1 ) \delta_{ m, n + 1 }}{ ( 2 n + 1 )( 2 n + 3 ) } + \frac{ n \delta_{m, n-1}}{ ( 2 n - 1 )( 2 n + 1 ) }, \\
\label{eq:legendreOrthogonality2}
\frac{\innerprodLtwo{ \Omegaref }{ w_2 L_n }{ L_m }}{2} & = \frac{ ( n + 1 )( n + 2 ) \delta_{ m, n + 2 }}{ ( 2 n + 1 )( 2 n + 3 )( 2 n + 5 ) } + \frac{ ( n - 1 )( n + 1 ) + n ( n + 2 )  }{ ( 2 n - 1 )( 2 n + 1 )( 2 n + 3 ) } \delta_{mn} + \frac{ n ( n - 1 ) \delta_{m, n - 2 }}{ ( 2 n - 3 )( 2 n - 1 )( 2 n + 1 ) } 
\end{align}
\end{subequations}
hold, where $ w_k $ denote the weight functions $ w_k( \xi ) = \xi^k $. The values of the Legendre polynomials and their first derivatives at the domain boundaries are given by
\begin{equation}
\label{eq:legendreBoundaries}
L_n( - 1 ) = ( - 1 )^n, \quad \DDref L_n( 1 ) =  n ( n + 1 ) / 2, \quad \DDref L_n( - 1 ) = ( - 1 )^{ n + 1 } n ( n + 1 ) /2.
\end{equation}
Moreover, the property
\begin{equation}
\label{eq:legendreIntegration}
( 2 n + 1 ) L_n = \DDref L_n - \DDref L_{ n - 1 }
\end{equation}
is useful for evaluating integrals of the $ L_n $.

We introduce the following linear combinations of Legendre polynomials, which will be used as bases of the vector spaces $ H^k_0( \Omegaref ) \cap \mathcal{P}^p( \Omegaref ) $ (for $ k \in \{ 0, 1, 2 \} $), $ H^1( \Omegaref ) \cap \mathcal{P}^p( \Omegaref ) $, and $ H^2_1( \Omegaref ) \cap \mathcal{P}^p( \Omegaref ) $:

\begin{proposition}
\label{prop:lambda}The polynomials
\begin{subequations}
\label{eq:lambda}
\begin{align}
\label{eq:lambda0}
\lambda_n^{[ 0 ]}( \xi ) & := \sqrt{ \frac{ 2 n - 1 }{ 2 } } L_{ n - 1 }( \xi ), \\
\label{eq:lambda1}
\lambda_n^{[ 1 ]}( \xi ) & : = \int_{ - 1 }^{ \xi } \dd\eta \, \lambda^{[ 0 ]}_{ n + 1 }( \eta ) = \frac{ L_{ n + 1 }( \xi ) - L_{ n - 1 }( \xi ) }{ \sqrt{ 2 ( 2 n + 1 ) } }, \\
\label{eq:lambda2}
\lambda_n^{[ 2 ]}( \xi ) & : = \int_{ - 1 }^{ \xi } \dd\eta \, \lambda^{[ 1 ]}_{ n + 1 }( \eta ) = \frac{ 1 }{ \sqrt{ 2 ( 2 n + 3 ) } } \left( \frac{ L_{ n + 3 }( \xi ) - L_{ n + 1 }( \xi ) }{ 2 n + 5 } - \frac{ L_{ n + 1 }( \xi ) - L_{ n - 1 }( \xi ) }{ 2 n + 1 } \right),
\end{align}
\end{subequations}
where $ 1 \leq n \leq N $, span the spaces  $ L^2( \Omegaref ) \cap \mathcal{P}^{ N - 1 }( \Omegaref ) $, $ H^1_0( \Omegaref ) \cap \mathcal{ P }^{ N + 1 }( \Omegaref ) $, and $ H^2_0( \Omegaref ) \cap \mathcal{ P }^{ N + 3 }( \Omegaref ) $, respectively. In addition, they satisfy the orthogonality relations
\begin{equation}
\label{eq:orthogonalityLambda}
\innerprodLtwo{ \Omegaref }{ \lambda_n^{[0]} }{ \lambda_m^{[0]} } = \innerprodLtwo{ \Omegaref }{ \DDref \lambda_n^{[1]} }{ \DDref \lambda_m^{[1]}} = \innerprodLtwo{ \Omegaref }{ \DDref^2 \lambda_n^{[2]} }{ \DDref^2 \lambda_m^{[2]} } = \delta_{mn}.
\end{equation} 
\end{proposition}    

\begin{proposition}
\label{prop:mu}Let
\begin{equation}
\label{eq:muNodal}
\mu_n( \xi ) := 
\begin{cases}
( 1 - \xi ) / 2, & n = 1, \\
( 1 + \xi ) / 2, & n = 2, \\
\lambda^{[ 1 ]}_{ n - 2 }( \xi ), & n \geq 3. 
\end{cases}
\end{equation}
Then, $ \{ \mu_n \}_{ n = 1 }^N $ is a basis of $ H^1( \Omegaref ) \cap \mathcal{ P }^{ N - 1 }( \Omegaref ) $. The values of the basis functions at the domain boundaries are
\begin{subequations}
\label{eq:muBoundary}
\begin{gather}
\label{eq:muBoundary1}
\mu_1( - 1 ) = \mu_2( 1 ) = 1, \quad \mu_1( 1 ) = \mu_2( - 1 ) = 0, \\
\label{eq:muBoundary2}
\mu_n( \pm 1 ) = 0, \quad n \geq 3 
\end{gather}
\end{subequations}
Furthermore, the inner-product relations
\begin{subequations}
\label{eq:orthogonalityMu}
\begin{gather}
\label{eq:orthogonalityMu1}
\innerprodLtwo{ \Omegaref }{ \DDref \mu_1 }{ \DDref \mu_1 } = \innerprodLtwo{ \Omegaref }{ \DDref \mu_2 }{ \DDref \mu_2 } = - \innerprodLtwo{ \Omegaref }{ \DDref \mu_1 }{ \DDref \mu_2 } = - 1 / 2,\\
\label{eq:orthogonalityMu2}
\innerprodLtwo{ \Omegaref }{ \DDref \mu_n }{ \DDref \mu_m } = 0, \quad \mbox{$ n \in \{ 1, 2 \} $ and $ m \geq 3 $}, \\
\label{eq:orthogonalityMu3}
\innerprodLtwo{ \Omegaref }{ \DDref \mu_n }{ \DDref \mu_m } = \delta_{ mn }, \quad \mbox{$ n \geq 3 $ and $ m \geq 3 $}
\end{gather}
\end{subequations}
hold.
\end{proposition}

\begin{proposition}
\label{prop:nu}The polynomials $ \nu_n( \xi ) $, where 
\begin{equation}
\label{eq:nuNodal}
\nu_n( \xi ) := 
\begin{cases}
- ( 1 + \xi )^2( \xi - 2 ) / 4, & n = 1, \\
( 1 + \xi )^2 ( \xi - 1 ) / 4, & n = 2, \\
\lambda^{[2]}_{ n - 2 }( \xi ), & 3 \leq n \leq N,
\end{cases}
\end{equation}
span the space $ H^2_1( \Omegaref ) \cap \mathcal{ P }^{ N + 1 }( \Omegaref ) $. They have the properties
\begin{subequations}
\label{eq:nuBoundary}
\begin{gather}
\label{eq:nuBoundary1}
\nu_1( 1 ) = \DDref \nu_2( 1 ) = 1, \quad \DDref \nu_1( 1 ) = \nu_2( 1 ) = 0,\\
\label{eq:nuBoundary2} 
\nu_n( -1 ) = \DDref \nu_n( -1 ) = 0, \quad n \in \{ 1 , 2 \}, \\
\label{eq:nuBoundary3}
\nu_n( \pm 1 ) = \DDref \nu_n( \pm 1 ) = 0, \quad n \geq 3, 
\end{gather}
\end{subequations}
and also satisfy the orthogonality relations
\begin{subequations}
\label{eq:orthogonalityNu}
\begin{gather}
\label{eq:orthogonalityNu1}
\innerprodLtwo{ \Omegaref }{ \DDref^2 \nu_1 }{ \DDref^2 \nu_1 } = - \innerprodLtwo{ \Omegaref }{ \DDref^2 \nu_1 }{ \DDref^2 \nu_2 } = 3 / 2,
\quad
\innerprodLtwo{ \Omegaref }{ \DDref^2 \nu_2 }{ \DDref^2 \nu_2 } = 2, \\
\label{eq:orthogonalityNu2}
\innerprodLtwo{ \Omegaref }{ \DDref^2 \nu_n }{ \DDref^2 \nu_m } = 0, \quad \mbox{$ n \in \{ 1, 2 \} $ and $ m \geq 3 $}, \\
\label{eq:orthogonalityNu3}
\innerprodLtwo{ \Omegaref }{ \DDref^2 \nu_n }{ \DDref^2 \nu_m } = \delta_{mn}, \quad \mbox{$ n \geq 3 $ and $ m \geq 3 $}.
\end{gather}
\end{subequations}
\end{proposition}  

One can check that within each of the sets of polynomials defined in Propositions~\ref{prop:lambda}--\ref{prop:nu} the elements are linearly independent and, as follows from~\eqref{eq:legendreBoundaries}, satisfy the appropriate boundary conditions. In particular, the polynomials $ \lambda^{[k]}_n $ have the properties 
\begin{equation}
\label{eq:internalShapeFunctions}
\DDref^j \lambda^{[k]}_n( \pm 1 ) = 0, 
\end{equation}
where $ 0 \leq j \leq k - 1 $. In the context of FEMs they are referred to as \emph{internal shape functions of order $ k $ }. On the other hand, $ \mu_1 $ and $ \mu_2 $, and $ \nu_1 $ and $ \nu_2 $ are called \emph{nodal shape functions} because they satisfy all but one of the conditions~\eqref{eq:internalShapeFunctions}, respectively for $ k = 1 $ and $ k = 2 $. Separating the basis functions into internal and nodal ones facilitates the application of the natural boundary conditions at the free surface. For example, the forms~\eqref{eq:boundaryFormsA} contribute only one nonzero matrix element, while~\eqref{eq:stiffWeakBoundaryUU}, \eqref{eq:stiffWeakBoundaryBB}, and~\eqref{eq:stiffWeakBoundaryUB} contribute two.

\begin{rem}\label{rem:orthogonality}The $ \lambda_n^{[k]} $ polynomials embody $ H^k $-regularity in the sense that they are, by construction, $k$-th antiderivatives of $ L^2 $-orthonormal polynomials. As a result, the principal forms of the continuous spaces (\ie $ \innerprodLtwo{ \Omegaref }{ \DDref^k v_1 }{ \DDref^k v_2 } $ for $ v_1 $, $ v_2 \in H^k_0( \Omegaref ) $) are, in accordance with~\eqref{eq:orthogonalityLambda}, represented by identity matrices, and do not exhibit an element-growth problem with $ p $. By virtue of \eqref{eq:orthogonalityMu} and~\eqref{eq:orthogonalityNu}, the corresponding matrices in the $ \{ \mu_n \} $ and $\{ \nu_n \}$ bases are, in each case, the direct sum of a $ 2 \times 2 $ matrix and an identity matrix, and therefore are also well behaved.  
\end{rem}   

We obtain our basis polynomials for the discrete spaces  $ V_u^{ N_u } $ and $ V_b^{ N_b } $~\eqref{eq:discreteVuVb} by transporting $ \lambda_n^{[2]} $, $ \mu_n $, and $ \nu_n $ from the reference interval $ \Omegaref $ to the problem domain $ \Omega $ by means of the pushforward map $ Q_* $. Introducing
\begin{equation}
\phi_n := 
\begin{cases}
Q_* \nu_n & \mbox{film problems}, \\
Q_* \lambda_n^{[2]} & \mbox{channel problems},
\end{cases}
\end{equation}
and $ \chi_n := Q_* \mu_n$, it follows from~\eqref{eq:discreteVuVb}, in conjunction with Defs.~\ref{def:filmMhd}--\ref{def:channelZeroPm}, that $ V_u^{N_u} = \spn\{ \phi_n \}_{n=1}^{N_u} $ and $ V_b^{N_b} = \spn\{ \chi_n \}_{n=1}^{N_b} $. Then, our bases $ \{ \psi_n \}_{n=1}^N $ are constructed as follows:
\begin{defn}[Bases of the discrete solution spaces $ V^{\boldsymbol{N} } $]
\label{def:discreteBases}In film and channel MHD problems we respectively set 
\begin{subequations}
\begin{equation}
\label{eq:psiFilmMhd}
\psi_n := 
\begin{cases}
( \phi_n, 0, 0 ), & 1 \leq n \leq N_u,\\
( 0, \chi_n, 0 ), & N_u + 1 \leq n \leq N_u + N_b,\\
( 0, 0, 1 ), & n + N_u + N_b + 1,
\end{cases}
\quad
\psi_n := 
\begin{cases}
( \phi_n, 0 ), & 1 \leq n \leq N_u,\\
( 0, \chi_n ), & N_u + 1 \leq n \leq N_u + N_b.
\end{cases}
\end{equation}
Moreover, our basis vectors for inductionless film problems are 
\begin{equation}
\psi_n := 
\begin{cases}
( \phi_n, 0 ), & 1 \leq n \leq N_u,\\
( 0, 1 ), & n = N_u + 1,
\end{cases}
\end{equation}
\end{subequations} 
whereas for inductionless channel problems we simply have $ \psi_n := \phi_n$, where $ 1 \leq n \leq N_u $. Thus, for all $ v \in V^{\boldsymbol{N}} $ one can write $ v = \sum_{n=1}^N [ \colvec{v}]_n \psi_n $, where
\begin{equation}
\label{eq:colvec}
\colvec{ v }^\mathrm{T} = 
\begin{cases}
( \colvec{ u }^\mathrm{T}, \colvec{ b }^\mathrm{T}, a ), & \mbox{film MHD problems}, \\
( \colvec{ u }^\mathrm{T}, \colvec{ b }^\mathrm{T} ), & \mbox{channel MHD problems}, \\
( \colvec{ u }^\mathrm{T}, a ), & \mbox{inductionless film problems}, \\
\colvec{ u }^\mathrm{T}, & \mbox{inductionless channel problems},
\end{cases}
\end{equation}
with $ \colvec{ u } \in \mathbb{ C }^{N_u} $ and $ \colvec{ b } \in \mathbb{ C }^{N_b} $.
\end{defn}

We note here that the procedure of constructing finite-dimensional solution spaces by transporting polynomial functions from the reference element to the problem domain is extensively applied in $ hp $-finite-element methods (FEMs), with the difference that $ \Omegaref $ is mapped to the \emph{mesh elements} rather than the full domain $ \Omega $ (\eg \cite{Schwab98}). Our method can thus be viewed as a single-element $hp$-FEM, with $ h = 1 $ (film problems) or $ h = 2 $ (channel problems). One of the benefits of working with affine families of finite elements is that the action of sesquilinear forms on basis-function pairs only needs to be computed on $ \Omegaref $, as the corresponding values on the mesh elements follow by scalings of the form~\eqref{eq:formScalings}. Even though this type of computational gain is not relevant to our single mesh-element scheme, working with $ \Omegaref $ leads to a more unified treatment of channel and film problems, and also allows for extensions of the method to problems defined over multiple domains (\eg vertically-stacked layers of fluids). The salient properties of the discrete solution spaces and their bases are displayed in Table~\ref{table:discreteSolutionSpaces}.

\begin{table}
\centering
\caption{\label{table:discreteSolutionSpaces}Properties of the discrete spaces $V_u^{N_u}$ and $ V_b^{N_b} $}
\begin{tabular*}{\linewidth}{@{\extracolsep{\fill}}lllllll}
\hline
& \multicolumn{3}{c}{$V_u^{N_u}$} & \multicolumn{3}{c}{$V_b^{N_b}$}\\
\cline{2-4} \cline{5-7}
Problem & Definition & Basis & $ p $ & Definition & Basis & $ p $ \\
\hline
Film & $ H^2_1( \Omegaf ) \cap \mathcal{P}^p( \Omegaf ) $ & $ \{ Q_* \nu_n \}_{n=1}^{N_u} $ & $ N_u + 1 $ & $ H^1( \Omegaf ) \cap \mathcal{P}^p( \Omegaf ) $ & $ \{ Q_* \mu_n \}_{n=1}^{N_b} $ & $ N_b - 1 $ \\ 
Channel & $ H^2_0( \Omegac ) \cap \mathcal{P}^p( \Omegac ) $ & $ \{ Q_* \lambda_n^{[2]} \}_{n=1}^{N_u} $ & $ N_u + 3 $ & $ H^1( \Omegac ) \cap \mathcal{P}^p( \Omegac ) $ & $ \{ Q_* \mu_n \}_{n=1}^{N_b} $ & $ N_b - 1 $\\
\hline
\end{tabular*}
\end{table}

\begin{rem}In channel MHD problems it is also possible to apply the procedure described by Shen \cite{Shen94,Shen96} to construct linear combinations of Legendre polynomials that satisfy strongly (essentially) the Robin boundary conditions~\eqref{eq:insulatingChannel} for the magnetic field. That approach would lead to well-conditioned and (for polynomial $ U $ and $ B $) sparse stiffness and mass matrices as well. However, since the boundary-value maps $ S_i^1( b ) = \DD( \tilde b( z_i ) ) $ cannot be defined for all elements of $ H^1 $, the trial and test space for $ b $ would have to be an $ H^2( \Omega ) $ subspace, such as $ H^2_\alpha( \Omega ) := \{ b \in H^2( \Omega ); \, S_1^1( b ) - \alpha S_1^0( b ) = S_2^1( b ) + \alpha S_2^0( b ) = 0 \} $. In our treatment of channel MHD problems, we opted to consider that $ b $ is an element of $ H^1( \Omega ) $ and enforce the boundary conditions weakly in the interests of commonality with our film-problem formulation. 
\end{rem}

\subsection{\label{sec:structureDiscrete}Structure of the Discrete Problems}  

We are now ready to write down explicit expressions for the stiffness and mass matrices in~\eqref{eq:matrixWeakForm}. With the choice of basis functions in Def.~\ref{def:discreteBases}, the free-stream contributions can be evaluated in closed form by means of the properties of the Legendre polynomials. This is also the case for the $ U$-dependent forms in problems with the Poiseuille velocity profile, since~\eqref{eq:legendreOrthogonality1} and~\eqref{eq:legendreOrthogonality2} can be used to evaluate terms that are, respectively, linear and quadratic in the reference coordinate $ \xi $. On the other hand, the exponential terms in Hartmann profiles~\eqref{eq:baseHartmann} preclude the derivation of closed-form expressions for the integrals, and one has to resort to numerical quadrature instead. Here we pursue two alternative approaches, either of which can be used to obtain highly accurate solutions of our stability problems. 
 
The first approach is based on specialized Gauss quadrature rules, by means of which the exponentially weighted sesquilinear forms are computed exactly (modulo roundoff error). Numerical methods for orthogonal polynomials with exponential weight function over a finite interval, and the associated Gauss quadrature knots and weights, have been developed by Mach \cite{Mach84}. As with many classes of orthogonal polynomials, the challenge is to compute the coefficients of the three-term recurrence relation in a manner that is stable with the polynomial degree $ p $. In the context of a study on optical scattering (a problem of seemingly little relevance to spectral methods), Mach presents an iterative algorithm that yields the required coefficients and, importantly, is stable at large $ p $. By computing the eigenvalues and eigenvectors of the resulting Jacobian matrix (\eg \cite{DavisRabinowitz07}), it is therefore possible to obtain quadrature knots and weights suitable for the evaluation of polynomial inner products weighted by $ \exp( \pm \Harz  z ) $.  

We also propose an alternative approach, which, following the widely used practice in spectral methods (\cite{Ciarlet91} and references therein), involves replacing the weighted sesquilinear forms by approximate ones derived from numerical quadrature rules (in the present case, LGL quadrature). Banerjee and Osborn \cite{BanerjeeOsborn90} have shown that in elliptical eigenvalue problems the incurred integration error does not affect the exponential $ p $-convergence of the discrete solution, provided that the eigenfunction being approximated is smooth, and the quadrature method is exact for polynomial integrands of degree $ 2 p - 1 $. To our knowledge, however, no corresponding theorem is available in the literature for the OS eigenvalue problems studied here, and, although surely an interesting direction for future research, an investigation along those lines lies beyond the scope of our work. Instead, in \S\ref{sec:hartmannProfile} below we contend ourselves with a series of comparisons with the exact-quadrature method supporting the adequacy of the $ 2 p - 1 $ precision level in our schemes for free-surface MHD as well.

\subsubsection{\label{sec:freeStreamMatrices}Free-Stream Matrices}

For the matrix representations of the $ U $ and $ B $-independent forms it is convenient to introduce the square matrices  $ \tHRZ^{ [ k d_1 d_2 ] } $, $ \tHTO^{ [ k d_1 d_2 ] } $, and $ \tHO^{ [ k d_1 d_2 ] } $, whose size is equal to the number of basis polynomials (\ie $N_u $ and/or $N_b$). Using, as in~\eqref{eq:legendreOrthogonality}, $ w_k $ to denote the power-law weight functions $ w_k( \xi ) = \xi^k $, we set
\begin{subequations}
\label{eq:matricesT}
\begin{gather}
\label{eq:tHr0}
\left[ \tHRZ^{ [ k d_1 d_2 ] } \right]_{ m n } := \innerprodLtwo{ \Omegaref }{ w_k \DDref^{d_2 } \lambda_n^{[r]} }{ \DDref^{d_1} \lambda_m^{[r]} }, \\
\label{eq:tH21}
\left[ \tHTO^{ [ k d_1 d_2 ] } \right]_{ m n } := \innerprodLtwo{ \Omegaref }{ w_k \DDref^{d_2} \nu_n }{ \DDref^{d_1} \nu_m }, \quad
\left[ \tHO^{ [ k d_1 d_2 ] } \right]_{ m n } := \innerprodLtwo{ \Omegaref }{ w_k \DDref^{d_2} \mu_n}{ \DDref^{d_1} \mu_m }.
\end{gather}
\end{subequations}
For our purposes it suffices to restrict attention to the cases where all of $ k $, $ r $, $ d_1 $, and $ d_2 $ are non-negative integers smaller than three. Then, closed-form expressions for~\eqref{eq:matricesT}, which we quote in Appendices~\ref{app:tInternal} and \ref{app:tNodal}, can be evaluated with the help of the orthogonality relations \eqref{eq:orthogonalityLambda}, \eqref{eq:orthogonalityMu}, and~\eqref{eq:orthogonalityNu}. We note that several of the calculations can be performed in a hierarchical manner. Specifically, the property $ \DDref \lambda^{[r]}_n = \lambda^{[r-1]}_{n+1} $ (see Proposition~\ref{prop:lambda}) carries over to the corresponding matrices, where the relation
\begin{equation}
\label{eq:hierarchicalT}
\left[  \tHRZ^{ [ k d_1 d_2 ] } \right ]_{ m n } = \left[ \mat{ T }_{ H^{ r- 1 }_0}^{[k, d_1 -1, d_2 - 1]} \right]_{ m + 1, n + 1 }
\end{equation}
applies for $\text{$ r $, $ d_1 $, $ d_2 $} \geq 1 $. Moreover, by construction of the $ \mu_n $ and $ \nu_n $ polynomials (Propositions~\ref{prop:mu} and~\ref{prop:nu}), we have 
\begin{equation}
\label{eq:tH2MinusSubmatrix}
\left[ \tHO^{ [ k d_1 d_2 ] } \right]_{ m n } = \left[ \mat{ T }_{ H^1_0 }^{ [ k d_1 d_2 ] } \right]_{ m - 2, n -2 }, \quad
\left[ \tHTO^{ [ k d_1 d_2 ] } \right]_{ m n } = \left[ \mat{ T }_{ H^2_0 }^{ [ k d_1 d_2 ] } \right]_{ m - 2, n -2 },
\end{equation}
where $ \text{$ m $, $ n $} \geq 3 $. That is, every $ N \times N $ matrix $ \tHTO^{ [ k d_1 d_2 ] } $ contains a $ \tHTZ^{ [ k d_1 d_2 ] } $ submatrix of size $ ( N - 2 ) \times ( N - 2 ) $, and similarly a $ \tHOZ^{ [ k d_1 d_2 ] } $ submatrix of size $ ( N - 2 ) \times ( N - 2 ) $ is contained in every $ N \times N $ matrix $ \tHO^{ [ k d_1 d_2 ] } $.  

\begin{rem}
It follows from~\eqref{eq:legendreOrthogonality} that the matrices $ \tHRZ^{ [ k d_1 d_2 ] } $ are banded and sparse (see Table~\ref{table:tHr0}). Moreover, their bands are not fully populated, as every other diagonal consists of zeros. The bandwidth of $ \mat{ T }_{ H^r_0 }^{ [ k d_1 d_2 ] } $ is equal to $ 2 r + k - d_1 - d_2 $.   
\end{rem}

\begin{rem}
\label{rem:leakage}
Let $ m $ and $ n $ respectively denote the row and column indices of $ \tHO^{ [ k d_1 d_2 ] } $ and $ \tHTO^{ [ k d_1 d_2 ] } $. Then, elements with $ m > 2 $ and $ n \leq 2 $, or $ m \leq 2 $ and $ n > 2 $, are the results of (weighted) inner products between nodal shape functions, respectively $ \mu_1 $, $ \mu_2 $ and $ \nu_1 $, $ \nu_2 $, and the internal shape functions $ \mu_3 $, $\mu_4, $ \ldots and $ \nu_3 $, $ \nu_4 $, \ldots. It can be checked by explicit calculation (see Appendix~\ref{app:tNodal}) that the spectral leakage between the nodal and internal shape functions is small. Specifically, the quantities
\begin{equation}
l_{ H^2_1 } := \max_m \left\{ \left[ \tHTO^{ [ k d_1 d_2 ] } \right]_{ m n } \neq 0; \, n \in \{ 1, 2 \} \right\}, \quad
l_{ H^1 } :=  \max_m \left\{ \left[ \mat{ T }_{ H^1 }^{ [ k d_1 d_2 ] } \right]_{ m n } \neq 0; \, n \in \{ 1, 2 \} \right\}
\end{equation}
are found to obey the relation
\begin{equation}
\label{eq:spectralLeakage}
l_{  H^2_1 } = l_{ H^1 } = 4 + k - d_1 - d_2.
\end{equation}
Note that defining $ l_{ H^2_1 } $ and $ l_{  H^1 } $ as maxima over the matrix columns $ n $ leads to the same expression as~\eqref{eq:spectralLeakage}.    
\end{rem}

In order to compute the matrix representations $ \mat{ K }_{uu}^{[0]} $, $ \mat{ K }_{bb}^{[0]} $,  and $ \mat{ K }_{uu}^{[\mathrm{L}]} $ of the free-stream forms $ \form{ K }_{uu}^{[0]} $~\eqref{eq:stiffWeakUUFS}, $ \form{ K }_{bb}^{[0]} $~\eqref{eq:stiffWeakBBFS}, and $ \form{ K }_{uu}^{[\mathrm{ L }]} $~\eqref{eq:stiffWeakUULorentz}, we employ the collective notation
\begin{equation}
\label{eq:tVu}
\mat{ T }_{ uu }^{ [ k d_1 d_2 ] } := 
\begin{cases}
\tHTO^{[ k d_1 d_2 ]}, & \mbox{film problems}, \\
\tHTZ^{[ k d_1 d_2 ]}, & \mbox{channel problems},
\end{cases}
\end{equation}
where $ \mat{ T }_{ uu }^{ [ k d_1 d_2 ] } \in \mathbb{ R }^{ N_u \times N_u} $, and also write $ \mat{ T }_{ bb }^{ [ k d_1 d_2 ] } :=  \tHO^{ [ k d_1 d_2 ] } \in \mathbb{ R }^{ N_b \times N_b} $ in both channel and film problems. As above, we denote matrix rows and columns respectively by $ m $ and $ n $. Then, making use of~\eqref{eq:formScaling}, we obtain 
\begin{subequations}
\label{eq:matKuubb0}
\begin{align}
\label{eq:matKuu0}
\mat{ K }_{uu}^{[0]} & := \left[ \form{ K }_{uu}^{[0]}( \phi_n, \phi_m ) \right] = - j \left( j^{ - 4 } \mat{ T }^{[022]}_{uu} + 2 \alpha^2 j^{ - 2 } \mat{ T}^{[011]}_{uu} + \alpha^4 \mat{ T }_{uu}^{[000]} \right), \\
\label{eq:matKbb0}
\mat{ K }_{bb}^{[0]} & := \left[ \form{ K }_{bb}^{[0]}( \chi_n, \chi_m ) \right] = - j \left( j^{ - 2 } \mat{ T }^{[011]}_{bb} + \alpha ^ 2 \mat{ T }^{[000]}_{bb} \right),
 \end{align}
\end{subequations}
and
\begin{align}
\label{eq:matKuuL}
\mat{ K }_{uu}^{[\mathrm{L}]} & := \left[\form{ K }_{uu}^{[\mathrm{L}]}( \phi_n, \phi_m )\right] = - \alpha^2 \Harx^2 j \mat{ T }^{[000]}_{uu} + \ii \alpha \Harx \Harz \left( \mat{ T }^{[001]}_{uu} - \mat{ T }^{[010]}_{uu} \right) - \Harz^2 j^{-1} \mat{ T }^{[011]}_{uu}.
\end{align} 
As for the mass forms $ \form{ M }_{uu} $~\eqref{eq:massWeakUU} and $ \form{ M }_{bb} $~\eqref{eq:massWeakBB}, these are represented by the matrices
\begin{subequations}
\label{eq:matM}
\begin{align}
\label{eq:matMuu0}
\mat{ M }_{uu} & := [ \form{ M }_{uu }( \phi_n, \phi_m ) ] = \Rey \, j \left( j^{ - 2 } \mat{ T }_{uu}^{[011]} + \alpha^2 \mat{ T }_{uu}^{[000]} \right), \\
\label{eq:matMbb0}
\mat{ M }_{bb} & := [ \form{ M }_{bb}( \chi_n, \chi_m ) ] = \Reym \, j \mat{ T }_{ bb }^{[000]}.
\end{align}
\end{subequations}

The maps~\eqref{eq:stiffWeakUBBUFS}, coupling the velocity and magnetic fields, can be treated by introducing $ \mat{ T }_{ H^1 H^2_0 }^{ [ k d_1 d_2 ] } \in \mathbb{ R }^{ N_b \times N_u } $ and $  \mat{ T }_{ H^1 H^2_1 }^{ [ k d_1 d_2 ] } \in \mathbb{ R }^{ N_b \times N_u } $, where
\begin{equation}
\label{eq:matricesTCross}
\left[ \mat{ T }_{ H^1 H^2_0  }^{ [ k d_1 d_2 ] } \right]_{ m n} := \innerprodLtwo{ \Omegaref }{ w_k \DDref^{d_2} \lambda_n^{[2]} }{ \DDref^{d_1} \mu_m  }, \quad
\left[ \mat{ T }_{ H^1 H^2_1 }^{ [ k d_1 d_2 ] } \right]_{ m n } := \innerprodLtwo{ \Omegaref }{ w_k \DDref^{d_2} \nu_n }{ \DDref^{d_1} \mu_m },
\end{equation}
and also $  \mat{ T }_{ b u }^{ [ k d_1 d_2 ] } \in \mathbb{ R }^{ N_b \times N_u } $, with
\begin{equation}
\mat{ T }_{ b u }^{ [ k d_1 d_2 ] } := 
\begin{cases}
\tHOHTO^{[ k d_1 d_2 ]}, & \mbox{film problems},\\
\tHOHTZ^{[ k d_1 d_2 ]}, & \mbox{channel problems}.
\end{cases}
\end{equation} 
Then, the matrices associated with $ \form{ K }_{ub}^{[0]} $ and $ \form{ K }_{bu}^{[0]} $~\eqref{eq:stiffWeakUBBUFS} are
\begin{subequations}
\label{eq:matKubbu0}
\begin{align}
\label{eq:matKub0}
\mat{ K }_{ub}^{[0]} := \left[ \form{ K }_{ub}^{[0]}( \chi_n, \phi_m ) \right] & = \ii \alpha \Rey \Alfxinv j \left( j^{-2} \mat{ T }_{ u b }^{ [ 0 1 1 ] } + \alpha^2 \mat{ T }_{ u b }^{[ 0 0 0 ]} \right) - \Rey \Alfzinv \left( j^{-2} \mat{ T }_{ u b }^{[ 0 2 1 ]} + \alpha^2 \mat{ T }_{ u b }^{[ 0 1 0 ]} \right),\\
\label{eq:matKbu0}
\mat{ K }_{bu}^{[0]} := \left[ \form{ K }_{bu}^{[0]}( \phi_n, \chi_m ) \right] & = \ii \alpha \Reym \Alfxinv j \mat{ T }_{ b u }^{[0 0 0 ]} + \Reym \Alfzinv \mat{ T }_{ b u }^{[0 0 1 ]},
\end{align}
where $ \mat{ T }_{ u b}^{[ k d_1 d_2 ]} := \left( \mat{ T }_{ b u }^{[k d_2 d_1 ]} \right)^\mathrm{T} $. 
\end{subequations}

\subsubsection{\label{sec:baseUBMatrices}$U$ and $B$-Dependent Matrices}

We now examine the matrix representations of the forms $ \form{ K }_{uu}^{[U]} $ and $ \form{ K }_{bb}^{[U]} $ \eqref{eq:stiffWeakUUBBBaseU}, and the maps $ \form{ K }_{ub}^{[B]} $ and $ \form{ K }_{bu}^{[B]} $ \eqref{eq:stiffWeakUBBUIndB}, all of which involve inner products of Legendre polynomials with non-trivial weight functions.

Problems with the Poiseuille profile~\eqref{eq:baseNonMhd} can be treated using the matrices $ \mat{ T }_{uu}^{[ k d_1 d_2]} $ and $ \mat{ T }_{bb}^{[ k d_1 d_2 ]} $ established in \S\ref{sec:freeStreamMatrices}. First, we compute the action of the pullback map $ Q^* $ (defined below~\eqref{eq:mapQ}) on $ U $, 
\begin{equation}
( Q^* U )( \xi ) = 1 - ( Q( \xi ) )^ 2 =: \hat U_0 + \hat U_1 \xi + \hat U_2 \xi^2, 
\end{equation}
where $ \hat U_0 = 1 - z_0^2 $, $ \hat U_1 = - z_0 h $, and $ \hat U_2 = - h^2 / 4 $. Specifically, in film problems ($ z_1 = -1 $, $ z_2 = 0 $) we have $ \hat U_0 = 3 / 4 $, $ \hat U_1 = - 1 / 2 $, and $ \hat U_2 = - 1 / 4 $, whereas in channel problems ($ z_1 = -1 $, $ z_2 = 1 $) the trivial result $ \hat U_0  = 1 $, $ \hat U_1 = 0 $, and $ \hat U_2 = - 1 $ applies. We then set
\begin{subequations}
\label{eq:matKuubbUPoiseuille}
\begin{align}
\nonumber
\mat{ K }_{uu}^{[ U ]} := \left[ \form{ K }_{uu}^{[ U ]}( \phi_n, \phi_m ) \right]  & = - \ii \alpha \Rey j \left( \hat U_0 \left( j^{ - 2 }  \mat{ T }_{uu}^{ [ 0 1 1 ] } + \alpha^2 \mat{ T }_{ uu }^{ [ 0 0 0 ]} \right) + \hat U_1 \left( j^{ -2 } \mat{ T }_{ uu }^{[ 1 1 1 ]} + \alpha^2 \mat{ T }_{ uu }^{[ 1 0 0 ]} - j^{ - 2 } \mat{ T }_{ uu }^{[ 0 1 0 ]} \right) \right. \\
\label{eq:matKuuUPoiseuille}
& \quad \left. + \hat U_2 \left( j^{ -2 } \mat{ T }_{ uu }^{ [ 2 1 1 ]} + \alpha^2 \mat{ T }_{ uu }^{ [ 2 0 0 ]} - 2 j^{ - 2 } \mat{ T }_{ uu }^{[ 1 1 0 ]} \right) \right) , \\ 
\label{eq:matKbbUPoiseuille}
\mat{ K }_{ bb }^{[ U ]} := \left[ \form{ K }_{bb}^{[ B ]}( \phi_n, \phi_m ) \right]  & = - \ii \alpha \Reym j \left( \hat U_0 \mat{ T }_{ bb }^{[ 0 0 0 ]} + \hat U_1 \mat{ T }_{ bb }^{[ 1 0 0 ]} + \hat U_2 \mat{ T }_{ bb }^{[ 2 0 0 ]} \right),
\end{align} 
\end{subequations}
where $ \mat{ K }_{uu}^{[U]} \in \mathbb{ C }^{ N_u \times N_u } $ and $ \mat{ K }_{bb}^{[U]} \in \mathbb{ C }^{N_b \times N_b } $ respectively represent $ \form{ K }_{uu}^{[U]} $ and $ \form{ K }_{bb}^{[U]} $.

Turning to problems with Hartmann profiles~\eqref{eq:baseHartmann}, it is convenient to introduce the shorthand notation $ \Harxi = \Harz  h / 2 $, $ s_{ \Harxi }( \xi ) := \sinh( \Harxi \xi )  $, and $ c_{ \Harxi }( \xi ) := \cosh( \Harxi \xi ) $, which leads to the relations
\begin{equation}
(Q^*U)( \xi ) = \hat U_0 - \hat U_s( \xi ) - \hat U_c( \xi ), \quad
(Q^*B)( \xi ) = - \hat B_0 - \hat B_1 \xi + \hat B_s( \xi )  + \hat B_c( \xi ),
\end{equation}   
with $ \hat U_0 = \cosh( \Harz ) / X $, $ \hat B_0 = \sinh( \Harz ) z_0 / (\Harz X) $, $ \hat B_1 = \sinh( \Harz ) h / ( 2 \Harz X ) $, and 
\begin{equation}
\label{eq:weightHartmannUB} 
\hat U_s = \frac{ \sinh( \Harz  z_0 ) s_{ \Harxi }}{X}, \quad
\hat U_c = \frac{\cosh( \Harz  z_0 ) c_{ \Harxi }}{X}, \quad
\hat B_s = \frac{ \cosh( \Harz  z_0 ) s_{ \Harxi }}{\Harz X }, \quad
\hat B_c = \frac{ \sinh( \Harz  z_0 ) c_{ \Harxi}}{\Harz X }.
\end{equation}
We also use  $ \xi_{G,k}^{[H]} \in ( -1, 1 ) $ and $ \hat\rho_{G,k}^{[H]} $, where $ H \geq 0 $ and $ k \in \{ 1, 2, \ldots, G \} $, to denote the quadrature knots and weights computed via Mach's algorithm \cite{Mach84}, such that
\begin{equation}
\label{eq:expQuadrature}
\int_{ - 1 }^1 \dd \xi \, e^{ H \xi } f( \xi )  = \sum_{ k = 1 }^G \hat \rho_{G,k}^{[H]} f( \xi_{G,k}^{[H]} ) 
\end{equation}
holds for any polynomial $ f \in \mathcal{ P }^{2 G -1 }( \Omegaref )$. Following the procedure outlined in Appendix~\ref{app:hartmann}, Eq.~\eqref{eq:expQuadrature} can be used to evaluate the matrices $ \mat{ S }_{uu}^{[d d_1 d_2]} \in \mathbb{ R }^{ N_u \times N_u } $, $ \mat{ S }_{ bb }^{[d d_1 d_2]} \in \mathbb{ R }^{ N_b \times N_b } $, and $ \mat{ S }_{ b u }^{[d d_1 d_2]} \in \mathbb{ R}^{ N_b \times N_u } $, where
\begin{subequations}
\label{eq:matS}
\begin{align}
\label{eq:matSVu}
\left[ \mat{ S }_{uu}^{[ d d_1 d_2 ]} \right]_{ mn } & := 
\begin{cases}
 \innerprodLtwo{ \Omegaref }{ ( \DDref^d \hat U_s ) \DDref^{ d_2 } \lambda_n^{[2]} }{ \DDref^{d_1} \lambda_m^{[2]} }, & \mbox{channel problems}, \\
\innerprodLtwo{ \Omegaref }{ ( \DDref^d \hat U_s ) \DDref^{ d_2 } \nu_n }{ \DDref^{d_1} \nu_m}, & \mbox{film problems},
\end{cases} \\
\label{eq:matSVb}
\left[ \mat{ S }_{ bb}^{[d d_1 d_2 ]} \right]_{ mn } & :=
\innerprodLtwo{ \Omegaref }{ (\DDref^d \hat U_s ) \DDref^{d_2} \mu_n }{ \DDref^{d_1} \mu_m }, \\
\label{eq:matSVuVb}
\left[ \mat{ S }_{b u}^{[d d_1 d_2]} \right]_{ mn } & := 
\begin{cases}
\innerprodLtwo{ \Omegaref }{ ( \DDref^d \hat B_s) \DDref^{d_2} \lambda_n^{[2]} }{ \DDref^{d_1} \mu_m }, & \mbox{channel problems},\\
\innerprodLtwo{ \Omegaref}{ ( \DDref^d \hat B_s ) \DDref^{d_2} \nu_n }{ \DDref^{d_1 } \mu_m }, & \mbox{film problems}. 
\end{cases}
\end{align}
\end{subequations}
Similarly, one can compute the matrices $ \mat{ C }_{uu}^{[d d_1 d_2]} \in \mathbb{ R }^{ N_u \times N_u } $, $ \mat{ C }_{ bb }^{[d d_1 d_2]} \in \mathbb{ R }^{ N_b \times N_b } $, and $ \mat{ C }_{ u b }^{[d d_1 d_2]} \in \mathbb{ R}^{ N_u \times N_b } $, whose elements are given by expressions analogous to~\eqref{eq:matS}, but with $ \hat U_s $ and $ \hat B_s $ respectively replaced by $ \hat U_c $ and $ \hat B_c $. Then, $ \mat{ K }_{ uu}^{[U]} $ and $ \mat{ K }_{bb}^{[U]} $ become
\begin{subequations}  
\label{eq:matKuubbUHartmannExact}
\begin{align}
\nonumber
\mat{ K }_{ uu}^{[U]}  & = - \ii \alpha \Rey j \left( \hat U_0 \left( j^{ -2 } \mat{ T }_{ V_u }^{[011]} + \alpha^2 \mat{ T }^{[0 0 0 ]}_{ V_u } \right)  - \left( j^{ -2 } \mat{ S }_{ V_u }^{[011]} + \alpha^2 \mat{ S }_{ V_u }^{[00 0]} - j^{-2} \mat{ S }_{V_u}^{[110]} \right) \right. \\
\label{eq:matKuuUHartmannExact}
& \quad - \left.\left( j^{ - 2 }\mat{ C }_{V_u}^{[011]} + \alpha^2 \mat{ C }_{ V_u }^{[000]} - j^{-2} \mat{ C }_{V_u}^{[110]} \right) \right), \\
\label{eq:matKbbUHartmannExact}
\mat{ K }_{ bb }^{[U]} & = - \ii \alpha \Reym j \left( \hat U_0 \mat{ T }_{ V_b }^{[000]} - \mat{ S }_{ V_b}^{[000]} - \mat{ S }_{ V_b }^{[000]} \right).
\end{align}
\end{subequations}
Also, the maps $ \form{ K }_{ u b}^{[B]} $ and $ \form{ K }_{ b u }^{[B]} $~\eqref{eq:stiffWeakUBBUIndB} induce the matrices $ \mat{ K }_{ub}^{[B]} \in \mathbb{ C }^{N_u \times N_b} $ and $ \mat{ K }_{bu}^{[B]} \in \mathbb{ C }^{ N_b \times N_u } $ given by
\begin{subequations}
\label{eq:matKubbuBHartmannExact}  
\begin{align}
\nonumber
\mat{ K }_{ u b }^{[B]} := \left[ \form{ K }_{ ub }^{[B]}( \chi_n, \phi_m ) \right ]  & = \ii \alpha \Rey \Harz \Prm\sh j \left( - \hat B_0 \left( j^{ -2 } \mat{ T }_{u b}^{[011]} + \alpha^2 \mat{ T }_{u b}^{[000]} \right) \right. \\
\nonumber
& \quad \left. + j^{ -2 } \mat{ S }^{[011]}_{u b} + \alpha^2 \mat{ S }^{[000]}_{u b} - j^{ -2 }  \mat{ S }_{u b}^{[110]} +  j^{ -2 } \mat{ C }^{[011]}_{u b} + \alpha^2 \mat{ C }^{[000]}_{u b} - j^{-2} \mat{ C }_{u b}^{[110]} \right. \\
\label{eq:matKubBHartmannExact}
& \quad \left. - \hat B_1 \left( j^{ -2 } \mat{ T }^{[111]}_{ u b } + \alpha^2  \mat{ T }^{[100]}_{ u b } - j^{ -2 } \mat{ T }^{[010]}_{ u b} \right) \right),\\
\label{eq:matKbuBHartmannExact}
\mat{ K }_{ b u }^{[B]} := \left[ \form{ K }_{ bu}^{[B]}( \phi_n, \chi_m ) \right] 
 & = \ii \alpha \Reym \Harz \Prm\sh j \left( - \hat B_0 \mat{ T }^{[000]}_{b u} - \hat B_1 \mat{ T }^{[100]}_{b u} + \mat{ S }_{b u}^{[000] } + \mat{ C }_{b u}^{[000]} \right),
\end{align}
\end{subequations}
where $ \mat{ S }_{u b}^{[dd_1d_2]} := \left( \mat{ S }_{b u }^{[dd_2d_1]} \right)^\mathrm{ T} $ and $ \mat{ C }_{u b}^{[dd_1d_2]} := \left( \mat{ C }_{b u}^{[dd_1d_2]} \right)^\mathrm{ T }$.

\begin{rem}
Due to the non-polynomial nature of the Hartmann profiles~\eqref{eq:baseHartmann}, the matrices in~\eqref{eq:matKuubbUHartmannExact} and~\eqref{eq:matKubbuBHartmannExact} are fully populated, and no simple closed form expressions exist for their evaluation (\cf \eqref{eq:matKuubbUPoiseuille}). However, by virtue of~\eqref{eq:expQuadrature} no quadrature errors are made in the computation of their elements.
\end{rem}

The expressions presented thus far are restricted to the specific cases of the Poiseuille and Hartmann profiles. Oftentimes, however, one is faced with the task of studying the stability properties of arbitrary steady-state profiles, and, although in principle possible, deriving each time specialized quadrature schemes would be a laborious task. An alternative approach is to replace the weighted forms and maps by approximate ones defined on the discrete solution spaces by means of the following procedure: 

Let $ \zeta_{G,k} \in [ z_1, z_2 ] $, where $ k = 0, 1, \ldots, G + 1 $, $ \zeta_0 = z_1 $, and $ \zeta_{G+1} = z_2 $, be the abssicas of LGL quadrature with $ G $ interior points on the interval $ \bar\Omega = [ z_1, z_2 ] $, and let $ \varrho_{G,k }$ be the corresponding weights (this type of quadrature is exact for polynomial integrands of degree up to $ 2 G + 1 $ \cite{DavisRabinowitz07}). Also, consider inner products of the form $ \innerprodLtwo{ \Omega }{ W f_1 }{ f_2 } $, where $ W $ stands for either $ U $ or $ B $, or their derivatives, and $ f_1 $ and $ f_2 $ are polynomials of degree $ p_1 $ and $ p_2 $, respectively. For all such inner products appearing in $ \form{ K }^{[U]}_{uu} $, $ \form{ K }_{bb}^{[U]} $, $ \form{ K }_{ub}^{[B]} $, and $ \form{ K }_{bu}^{[B]} $ set $ G \geq ( p_1 + p_2 )  / 2 - 1 $ and make the substitution $\innerprodLtwo{ \Omega }{ W f_1 }{ f_2 } \mapsto \sum_{ k = 0 }^{ G + 1 } \varrho_{G,k} W( \zeta_{G,k} )  f_1( \zeta_{G,k} ) \cconj{ f_2( \zeta_{G,k} ) }$. The resulting forms and maps, respectively denoted by $ \tilde{ \form{ K } }_{uu}^{[U]} : V_u^{N_u} \times V_u^{N_u} \mapsto \mathbb{ C } $, $ \tilde{ \form{ K } }_{bb}^{[U]} : V_b^{N_b} \times V_b^{N_b} \mapsto \mathbb{ C }$, $ \tilde{ \form{ K } }_{ub}^{[B]} : V_b^{N_b} \times V_u^{N_u} \mapsto \mathbb{ C }$, and $ \tilde{ \form{ K } }_{bu}^{[B]} : V_u^{N_u } \times V_b^{N_b} \mapsto \mathbb{ C }$, are
\begin{subequations}
\label{eq:matKuubbUHartmannLgl}
\begin{align}
\nonumber
\tilde{\form{ K }}_{uu}^{[U]}( \utrial, \utest ) & := - \ii \alpha \Rey \sum_{ k = 0 }^{G_u+1} \varrho_{G,k} \left( U( \zeta_{G,k} ) ( \DD \utrial( \zeta_{G,k} ) \cconj{ \DD \utest( \zeta_{G,k} ) } + \alpha^2 \utrial( \zeta_{G,k} ) \cconj \utest( \zeta_{G,k} ) ) \right.  \\
\label{eq:matKuuUHartmannLgl}
& \left. \quad - ( \DD U( \zeta_{G,k} ) ) \utrial( \zeta_{G,k} ) \cconj{ \DD \utest( \zeta_{G,k} ) } \right),\\
\label{eq:matKbbUHartmannLgl}
\tilde{\form{ K }}_{bb}^{[U]}( \btrial, \btest ) & := - \ii \alpha \Reym \sum_{ k = 0 }^{G_b+1} \varrho_{G,k} U( \zeta_{G,k} ) \btrial( \zeta_{G,k} ) \cconj{ \btest( \zeta_{G,k} ) },  
\end{align}
\end{subequations}
and
\begin{subequations}
\label{eq:matKubbuBHartmannLgl}
\begin{align}
\nonumber
\tilde{\form{ K }}_{ub}^{[B]}( \btrial, \utest ) & := \ii \alpha \Rey  \Harz  \Prm\sh \sum_{ k = 0 }^{G_{ub}+1} \varrho_{G_{ub},k} \left( \bbind( \zeta_{G_{ub},k} )(  \DD\btrial( \zeta_{G_{ub},k} ) \cconj{ \DD\utest( \zeta_{G_{ub},k} ) } \right.\\
\label{eq:matKubBHartmannLgl} & \quad \left.
+ \alpha^2 \btrial( \zeta_{G_{ub},k} ) \cconj{ \utest( \zeta_{G_{ub},k} ) } ) - ( \DD \bbind( \zeta_{G_{ub},k} ) ) \btrial( \zeta_{G_{ub},k} ) \cconj{ \DD \utest( \zeta_{G_{ub},k} ) } \right), \\
\label{eq:matKbuBHartmannLgl}
\tilde{\form{ K }}_{bu}^{[ B ]}( \utrial, \btest ) & := \ii \alpha \Reym  \Harz  \Prm\sh \sum_{ k = 0 }^{ G_{ub}+1 } \varrho_{G_{ub},k} \bbind( \zeta_{G_{ub},k} ) \utrial( \zeta_{G_{ub},k} ) \cconj{ \btest( \zeta_{G_{ub},k} ) },   
\end{align}
\end{subequations} 
where
\begin{equation}
\label{eq:lglPrecision}
G_u \geq p_u - 1, \quad G_b \geq p_b - 1, \quad G_{ ub } \geq ( p_u + p_b ) / 2- 1,
\end{equation}
and, as usual,  $ p_u $ and $ p_b $ are the polynomial degrees of the velocity and magnetic-field bases (see Table~\ref{table:discreteSolutionSpaces}). The sesquilinear form $ \tilde{ \form{ K } } : V^{\boldsymbol{ N } }\times V^{\boldsymbol{ N }} \mapsto \mathbb{ C } $ introduced in~\eqref{eq:discreteWeakForm} then follows by replacing the exact forms and maps in~\eqref{eq:stiffnessFilmMHD} with the corresponding approximate ones defined in~\eqref{eq:matKuubbUHartmannLgl} and~\eqref{eq:matKubbuBHartmannLgl}.   

\begin{rem}Our choice of precision in~\eqref{eq:lglPrecision} is motivated by Banerjee and Osborn's \cite{BanerjeeOsborn90} result that in finite-element methods for elliptical eigenvalue problems it suffices to use quadrature schemes that are exact for polynomial integrands of degree  $ 2 p - 1 $, where $ p $ is the degree of the FEM basis. Here we do not pursue a formal proof of the adequacy of~\eqref{eq:lglPrecision}, but the numerical tests in \S\ref{sec:hartmannProfile} demonstrate that eigenvalues computed using the smallest quadrature precision consistent with it converge in a virtually identical manner with those obtained via the exact quadrature scheme.\end{rem} 

For the purpose of evaluating the matrices representing $ \tilde{ \form{ K } }_{ uu}^{[U]} $, $ \tilde{ \form{ K } }_{ bb }^{[ U ]} $, $ \tilde{ \form{ K } }_{ub}^{[B]} $, and $ \tilde{ \form{ K } }_{bu}^{[B]} $, which we again denote by $ \mat{ K }_{ uu }^{[U]} $, $ \mat{ K }_{ bb }^{[U]}$, $ \mat{ K }_{ub}^{[B]} $, and $ \mat{ K }_{ bu }^{[B]} $, we introduce the differentiation matrices $ \mat{ \Delta }^{[d]}_u \in \mathbb{  R }^{ G_u \times N_u } $, $ \mat{ \Delta }^{[d]}_b \in \mathbb{ R }^{ G_b \times N_b } $, $ \mat{ \Delta }'^{[d]}_{ u } \in \mathbb{ R }^{ G_{ ub } \times N_u } $, and $ \mat{ \Delta }'^{[d]}_{ b } \in \mathbb{ R }^{ G_{ ub } \times N_b } $ with elements
\begin{subequations}
\label{eq:deltaMatrices}
\begin{gather}
\left[ \mat{ \Delta }_u^{[d]} \right]_{ k n } := 
\begin{cases}
\DDref^d \nu_n( \hat\zeta_{G_u,k} ), & \mbox{film prob.}, \\
\DDref^d \lambda^{[2]}_n( \hat\zeta_{G_u,k} ), & \mbox{channel prob.}
\end{cases} \quad
\left[ \mat{ \Delta }'^{[d]}_u \right]_{ k n } := 
\begin{cases}
\DDref^d \nu_n( \hat\zeta_{G_{ub},k} ), & \mbox{film prob.} \\
\DDref^d \lambda^{[2]}_n( \hat\zeta_{G_{ub},k} ), & \mbox{channel prob.}
\end{cases}\\
\left[ \mat{ \Delta }_b^{[d]} \right]_{ k n } := \DDref^d \mu_n( \hat\zeta_{G_b,k} ), \quad
\left[ \mat{ \Delta }'^{[d]}_b \right]_{ k n } := \DDref^d \mu_n( \hat\zeta_{G_{ub},k} ),
\end{gather}
\end{subequations}
where $ \hat\zeta_{G,k} \in [ -1, 1 ] $ for $ k \in \{ 0, 1, \ldots, G + 1 \} $ are LGL quadrature knots on the reference interval $ \Omegaref $. We also make use of the $ G \times G $ diagonal weight matrices $ \hat{\mat{ \varrho }}_G $, whose entries $ [ \hat{\mat{\varrho}}_G ]_{ kk } = \hat\varrho_{G,k} $ are equal to the quadrature weights associated with the knots $ \hat\zeta_{G,k} $ (note that $ \hat\zeta_{G,k} = Q^{-1}(\zeta_{G,k})$ and $ \hat\varrho_{G,k}=2\varrho_{G,k}/h $), and the diagonal matrices $ \mat{ U }_G^{[ d]} $ and $ \mat{ B }_G^{[d]} $, where $ \left[ \mat{ U }_G^{[d]} \right]_{ kk } := \DDref^dQ^*( U )( \hat\zeta_{G,k} ) $ and $ \left[ \mat{ B }_G^{[d]} \right]_{ kk } := \DDref^d Q^*( B )( \hat\zeta_{G,k} ) $. We then obtain
\begin{subequations}
\label{eq:matKLgl1}
\begin{align}
\nonumber
\mat{ K }_{uu}^{[U]} := \left[ \tilde{ \form{ K } }_{uu}^{[U]}( \phi_n, \phi_m ) \right] &=  - \ii \alpha \Rey j \left( j^{ - 2 } ( \mat{ \Delta }_u^{[ 1 ]} )^\mathrm{ T } \hat{\mat{ \varrho }}_{G_u} \mat{ U }_{G_u}^{[0]} \mat{ \Delta }_u^{[ 1]} + \alpha^2 ( \mat{ \Delta }_u^{[ 0 ]} )^\mathrm{ T } \hat{\mat{ \varrho }}_{G_u} \mat{ U }_{G_u}^{[0]} \mat{ \Delta }_u^{[0]} \right. \\
\label{eq:matKuuUHartmannLgl2}
& \quad \left. - j^{ -2 } ( \mat{ \Delta }_u^{[1]} )^\mathrm{ T }\hat{\mat{ \varrho }}_{G_u} \mat{ U }_{G_u}^{[ 1 ]} \mat{ \Delta }_u^{[0]} \right), \\
\label{eq:matKbbUHartmannLgl2}
\mat{ K }_{bb}^{[U]} := \left[ \tilde{ \form{ K } }_{bb}^{[U]}( \chi_n, \chi_m ) \right] &= - \ii \alpha \Reym j( \mat{ \Delta }_b^{[0]} )^\mathrm{T} \hat{\mat{ \varrho }}_{G_b} \mat{ U }^{[0 ]}_{G_b} \mat{ \Delta }_b^{[0]}, 
\end{align}
\end{subequations}
and
\begin{subequations}
\label{eq:matKLgl2}
\begin{align}
\nonumber
\mat{ K }_{ub}^{[B]} := \left[ \tilde{ \form{ K } }_{ub}^{[B]}( \chi_n, \phi_m ) \right] &=  \ii \alpha \Rey  \Harz  \Prm\sh j \left( j^{ - 2 } ( \mat{ \Delta }'^{[1]}_u )^\mathrm{ T } \hat{\mat{ \varrho }}_{G_{ub}} \mat{ B }_{G_{ub}}^{[ 0 ]} \mat{ \Delta }_b^{[1]} + \alpha^2 ( \mat{ \Delta }'^{[0]}_u )^\mathrm{T} \hat{\mat{ \varrho }}_{G_{ub}} \mat{ B }^{[0]}_{G_{ub}} \mat{ \Delta }_b^{[0]} \right. \\
& \quad \left. - j^{-2} ( \mat{ \Delta }'^{[1]}_u )^\mathrm{T} \hat{\mat{ \varrho }}_{G_{ub}} \mat{ B }^{[1]}_{G_{ub} } \mat{ \Delta }'^{[0]}_b \right),\\
\mat{ K }_{bu}^{[B]} := \left[ \tilde{ \form{ K } }_{bu}^{[B]}( \phi_n, \chi_m ) \right] & = \ii \alpha \Reym  \Harz  \Prm\sh j ( \mat{ \Delta }'^{[0]}_b )^\mathrm{T} \hat{\mat{ \varrho }}_{G_{ub}} \mat{ B }_{G_{ub}}^{[0]} \mat{ \Delta }'^{[0]}_u. 
\end{align}
\end{subequations}

\begin{rem}\label{rem:lgl}In 64-bit arithmetic the numerators and denominators in~\eqref{eq:baseHartmann} overflow at around $ \Harz = \ln( 2^{1023} ) \simeq 700 $ . This, in conjunction with the fact that neither $ U $ nor $ B $ have Taylor expansions about $ \Harz = \infty $ valid for all $ z \in [ -1, 1 ] $, renders the evaluation of the $ \mat{ U } $ and $ \mat{ B } $ matrices at large $ \Harz $ somewhat problematic. A practical workaround is to code the internal calculations for $ U $ and $ B $ in REAL*16 (128-bit) arithmetic, supported by a number of Fortran compilers (\eg the Intel and NAG compilers), pushing the occurrence of the overflow to $ \Harz = \ln( 2^{\text{16,383}} ) \simeq \text{11,000} $. Note that a similar issue arises with the exact-quadrature method, but in that case performing the internal computations with REAL*16 data types is not as straightforward (see Remark~\ref{rem:mach} in Appendix~\ref{app:hartmann}).\end{rem}

\subsubsection{\label{sec:boundaryMatrices}Boundary Terms}

Boundary terms, namely~\eqref{eq:stiffWeakBoundaryUU}, \eqref{eq:stiffWeakBoundaryBB}, \eqref{eq:stiffWeakBoundaryUB}, \eqref{eq:boundaryFormsA}, and \eqref{eq:stiffWeakUHZeroPm}, are the outcome of the natural imposition of the stress and kinematic conditions at the free surface, and the insulating boundary conditions at the wall and the free surface. One of the benefits of working in the $ \{ \mu_n \}_{n=1}^{N_b} $ and $ \{ \nu_n \}_{n=1}^{N_u} $ bases, consisting of internal and nodal shape functions, is that each of the boundary terms contributes at most two, $ p $-independent, nonzero matrix elements in the stiffness matrix $ \mat{ K }$. In consequence, its sparsity and conditioning are not affected by the boundary conditions. Specifically, $ \form{ K }_{uu}^{[\mathrm{S}]} $, $ \form{ K }_{bb}^{[\mathrm{I}]} $, and $ \form{ K }_{ ub }^{[\mathrm{S}]} $ are represented by the matrices $ \mat{ K }_{uu}^{[\mathrm{S}]} \in \mathbb{ R }^{N_u \times N_u } $, $ \mat{ K }_{bb}^{[\mathrm{I}]} \in \mathbb{ R }^{ N_b \times N_b } $, and $ \mat{ K }_{ub}^{[\mathrm{S}]} \in \mathbb{ C }^{ N_u \times N_b } $, where, as follows from~\eqref{eq:muBoundary} and~\eqref{eq:nuBoundary},
\begin{subequations}
\label{eq:boundaryMatrices1}
\begin{align}
\label{eq:matKuuS}
\mat{ K }_{ uu }^{[\mathrm{S}]} := \left[ \form{ K }_{uu}^{[\mathrm{S}]}( \phi_n, \phi_m ) \right]  & = - j^{-1} \alpha^2 ( \delta_{ m 1 } \delta_{ n 2 } + \delta_{ m 2 } \delta_{ n 1 } ), \\
\label{eq:matKbbI}
\mat{ K }_{ bb }^{[\mathrm{I}]} := \left[ \form{ K }_{bb}^{[\mathrm{I}]}( \chi_n, \chi_m ) \right] & = - \alpha( \delta_{ m 1 } \delta_{ n 1 } + \delta_{ m 2 } \delta_{ n 2 } ), \\
\label{eq:matKubS}
\mat{ K }_{ ub }^{[\mathrm{S}]} := \left[ \form{ K }_{ub}^{[\mathrm{S}]}( \chi_n, \phi_m ) \right] & = \alpha \Rey \left( \ii \alpha ( \Alfxinv + \Alfzinv \Reym B( 0 ) ) \delta_{ m 1 } - j^{ - 1 } \Alfzinv \delta_{ m 2 } \right) \delta_{ n 2 }.
\end{align}
\end{subequations}
In addition, the maps $ \form{ K }_{ua} $, $ \form{ K }_{ba} $, and $ \form{ K }_{ a u } $ respectively give rise to the column vectors $ \colvec{ K }_{ua} \in \mathbb{ C }^{ N_u } $ and $ \colvec{ K }_{ba} \in \mathbb{ C }^{ N_b } $, and the row vector $ \colvec{ K }_{au}^\mathrm{ T} \in \mathbb{ R }^{ N_u } $, where
\begin{subequations}
\label{eq:boundaryMatrices2}
\begin{align}
\label{eq:boundaryMatricesUA}
\left[ \colvec{ K }_{ua} \right]_n := \form{ K }_{ua}( \phi_n, 1 ) & = \alpha^2 \left( \frac{ -1 }{ \Prg^2 \Rey } - \frac{ \alpha^2 }{ \Ohn^2 \Rey } +  2 \ii \alpha \DD \bux( 0 ) \right) \delta_{ n 1 } + \frac{ \ii \alpha }{ j }\left( \DD^2 \bux( 0 ) + \Harz^2  \DD B( 0 ) \right) \delta_{ n 2 }, \\
\left[ \colvec{ K }_{ba} \right]_n := \form{ K }_{ba}( \chi_n, 1 ) & = \ii \alpha \Alfzinv \Reym \DD B( 0 ) \delta_{ n 2 }, \\
\left[ \colvec{ K }_{au} \right]_n := \form{ K }_{au}( 1, \phi_n ) & = \delta_{1 n }.
\end{align}
\end{subequations}
In inductionless problems the column vector corresponding to~\eqref{eq:stiffWeakUHZeroPm} is given by~\eqref{eq:boundaryMatricesUA} with $ \DD B( 0 ) $ formally set to zero.

\subsubsection{\label{sec:composingKM}Constructing the Stiffness and Mass Matrices}

Eq.~\eqref{eq:matrixWeakForm2}, according to which the stiffness and mass matrices are to be computed, has different instantiations, depending on the forms $ \form{ K } $ and $ \form{ M } $ of the variational problem at hand (Defs.~\ref{def:filmMhd}--\ref{def:channelZeroPm}), and the corresponding choice of basis functions (Def.~\ref{def:discreteBases}). The matrices introduced in \S\ref{sec:freeStreamMatrices}--\S\ref{sec:boundaryMatrices} serve as building blocks, out of which $ \mat{ K } $ and $ \mat{ M } $ can be composed in a modular manner. A number of these matrix `modules' are common among different types of problems (\eg $ \mat{ K }^{[0]}_{uu} $~\eqref{eq:matKuu0} and $ \mat{ M }_{uu} $~\eqref{eq:matMuu0} are present in all film and channel problems), which is convenient for implementation purposes. 

In film MHD problems, $ \mat{ K } $ and $ \mat{ M } $ have $ N = N_u + N_b + 1 $ rows and columns, and are given by 
\begin{equation}
\label{eq:matKFilmMhd}
\mat{ K } = 
\left(
\begin{array}{c|c|c}
\mat{ K }_{uu}^{[0]} + \mat{ K }_{uu}^{[\mathrm{S}]} & \mat{ K }_{ub}^{[0]} + \mat{ K }_{ub}^{[B]} & \colvec{ K }_{ua} \\
+ \mat{ K }_{uu}^{[U]} & + \mat{ K }_{ub}^{[\mathrm{S}]} \\
\hline
\mat{ K }_{bu}^{[0]} + \mat{ K }_{bu}^{[B]} & \mat{ K }_{bb}^{[0]} + \mat{ K }_{bb}^{[\mathrm{I}]} & \colvec{ K }_{ba} \\
& + \mat{ K }_{bb}^{[U]} \\
\hline
\colvec{ K }^\mathrm{ T }_{au} & \colvec{ 0 }^\mathrm{ T } & - \ii \alpha U( 0 ) 
\end{array}
\right), \quad
\mat{ M } =
\left(
\begin{array}{ccc}
\mat{ M }_{uu} & \mat{ 0 } & \mat{ 0 } \\
\mat{ 0 } & \mat{ M }_{bb} & \mat{ 0 } \\
\mat{ 0 } & \mat{ 0 } & 1 
\end{array}
\right).
\end{equation}
Here the submatrices with $ uu $ and $ bb $ indices are respectively dimensioned $ N_u \times N_u $ and $ N_b \times N_b $. Among them, the $ U $-independent matrices, $ \mat{ K}_{uu}^{[0]} $, $ \mat{ K }_{uu}^{[\mathrm{S}]} $, $ \mat{ K }_{bb}^{[0]} $, $ \mat{ K }_{bb}^{[\mathrm{I}]} $, $ \mat{ M }_{uu} $, and $ \mat{ M }_{bb} $, are given by Eqs.~\eqref{eq:matKuubb0}, \eqref{eq:matKuuS}, \eqref{eq:matKbbI}, and \eqref{eq:matM}. Also, the submatrix $ \mat{ K }_{ uu }^{[U]} $ is to be evaluated using either of~\eqref{eq:matKuuUPoiseuille}, \eqref{eq:matKuuUHartmannExact}, and~\eqref{eq:matKuuUHartmannLgl2}, depending on whether the velocity profile is Poiseuille, Hartmann (treated by means of the exact-quadrature method), or LGL quadrature is employed. Similarly, $ \mat{ K }_{bb}^{[U]} $ can be computed by means of either~\eqref{eq:matKbbUPoiseuille}, \eqref{eq:matKbbUHartmannExact}, or~\eqref{eq:matKbbUHartmannLgl2}. The submatrices indexed by $ ub $ and $ bu $ have dimension $ N_u \times N_b $ and $ N_b \times N_u $, respectively. The $ B $-independent ones, $ \mat{ K }_{ub}^{[0]} $, $ \mat{ K }_{ub}^{[\mathrm{S}]} $, and $ \mat{ K }_{bu}^{[0]} $, follow from~\eqref{eq:matKubbu0} and \eqref{eq:matKubS}, while for those that depend on the induced magnetic field, namely $ \mat{ K }_{ub}^{[B]} $ and $ \mat{ K }_{bu}^{[B]} $, there exist options to use exact quadrature~\eqref{eq:matKubbuBHartmannExact}, or LGL quadrature~\eqref{eq:matKLgl2}. Finally, the column vectors $ \colvec{ K }_{ua} $ and $ \colvec{ K }_{ba} $, respectively of size $ N_u $ and $ N_b $, and the row vector $ \colvec{ K }_{au}^\mathrm{T} $ of size $ N_u $ are given by~\eqref{eq:boundaryMatrices2}. In inductionless film problems (Def.~\ref{def:filmZeroPm}), the magnetic-field degrees of freedom are not present, and $ \mat{ K } $ and $ \mat{ B } $ are replaced by the $ ( N_u +1 ) \times (N_u + 1 ) $ matrices
\begin{equation}
\label{eq:matKFilmZeroPm}
\mat{ K } = 
\left(
\begin{array}{c|c}
\mat{ K }_{uu}^{[0]} + \mat{ K }_{uu}^{[\mathrm{S}]} &  \colvec{ K }_{ua} \\
+ \mat{ K }_{uu}^{[U]} + \mat{ K }_{uu}^{[\mathrm{L}]}  \\
\hline
\colvec{ K }^\mathrm{ T}_{au} & - \ii \alpha U( 0 ) 
\end{array}
\right), \quad
\mat{ M } = \diag( \mat{ M }_{uu}, 1 ),
\end{equation}
where, aside from $ \mat{ K }_{uu}^{[\mathrm{L}]} $, which is given by~\eqref{eq:matKuuL}, and $ \colvec{ K }_{ua} $ (obtained from~\eqref{eq:boundaryMatricesUA} with $ \DD B( 0 ) $ set to zero), the submatrices have the same definitions as in~\eqref{eq:matKFilmMhd}. 

In the interests of brevity, we do not write down explicit expressions for the stiffness and mass matrices in channel problems. We note, however, that they have the same structure as the corresponding film-problem matrices, but with the rows and columns representing the free-surface removed, and all boundary terms involving the velocity ($ \mat{ K }_{uu}^{[\mathrm{S}]} $, and $ \mat{ K }_{ub}^{[\mathrm{S}]} $) set to zero.

\begin{rem}The mass matrices in~\eqref{eq:matKFilmMhd} and~\eqref{eq:matKFilmZeroPm}, as well as in the corresponding channel problems, are symmetric positive definite (SPD). Rewriting~\eqref{eq:matrixWeakForm} in the form $ \gamma_M \mat{ K } \colvec{ v } = \gamma_K \mat{ M } \colvec{ v } $, where $ \gamma_K / \gamma_M = \gamma $ (the QZ algorithm \cite{MolerStewart73} actually solves this version of the problem), the non-singularity of $ \mat{ M } $ guarantees that $ \gamma_M \neq 0 $ (\ie $ \gamma $ is finite). In fact, as can be checked from~\eqref{eq:massFilmMHD}, $ \mat{ M } $ is SPD for \emph{all} choices of discrete bases. In tau methods, however, $ \mat{ M } $ can be singular. Dawkins \etal~\cite{DawkinsDunbarDouglass98} have shown that in the Legendre tau discretization of a fourth-order eigenvalue problem (structurally similar to the OS equation) $ \mat{ M } $ has a non-trivial nullspace, and, as a result, the discrete problem contains an infinite eigenvalue. Moreover, the Chebyshev tau formulation of the same problem was found to contain spurious eigenvalues, even though in that case $ \mat{ M } $ is non-singular. Treating the Legendre and Chebyshev tau methods as members of the one-parameter family of Gegenbauer tau methods, the spurious eigenvalues in the Chebyshev case were interpreted as perturbations of the infinite eigenvalues in the Legendre one. Like KMS, we found no evidence of spurious eigenvalues in any of the schemes presented here, which, in light of the analysis by Dawkins \etal, is likely due to the fact that the variational formulation described in \S\ref{sec:weakFormulation} leads to non-singular mass matrices irrespective of the choice of basis.
\end{rem}  

\begin{rem}The sparsity of $ \mat{ K } $ and $ \mat{ M } $ in problems with polynomial steady-state profiles enables the efficient use of iterative solvers. A number of implementations (\eg the ARPACK library~\cite{LehoucqSorensenYang98}, which is also available in Matlab) provide the option to specifically seek the eigenvalues with the largest real parts, which are oftentimes the ones of interest. In practice, however, we observed that these are particularly hard eigenvalues to compute, with the algorithm frequently failing to achieve convergence. Instead, we found that a more feasible strategy is to search for eigenvalues with the smallest absolute value. Due to the predominance of highly damped modes in the spectrum (\ie eigenvalues with large $ | \gamma | $ but small $ \Real( \gamma ) $), the eigenvalue with the largest real part often happens to be among the smallest absolute value ones. This approach was used to compute the eigenvalues at $ p = \ord( 10^3 ) $ in Fig.~\ref{fig:spectralConvergenceLegendreLagrange} below.\end{rem}

\section{\label{sec:numericalResults}Results and Discussion}

In this section we present a series of test calculations aiming to validate our Galerkin schemes, and illustrate the basic properties of our stability problems.  First, in \S\ref{sec:eigenvalueSpectra} we study the eigenvalue spectra of representative film and channel problems. Various aspects of numerical accuracy are examined in \S\ref{sec:spectralConvergence}. In \S\ref{sec:consistency} we test the consistency of our schemes against energy conservation in free-surface MHD, and the time evolution of small-amplitude perturbations in nonlinear simulations. The critical-parameter calculations in \S\ref{sec:criticalReynolds} is our final topic. Aside from the nonlinear simulations in \S\ref{sec:consistency}, all numerical work was carried out using a Matlab code, available upon request from the corresponding author. We remark that in order to facilitate comparison with relevant references in the literature, we oftentimes express our computed eigenvalues in terms of the complex phase velocity $ c = i \gamma / \alpha $, rather than the complex growth rate $ \gamma $. As stated in \S\ref{sec:governingEqs}, in the former representation a mode is unstable if $ \Gamma := \Imag( c ) > 0 $, while $ C := \Real( c ) $ its phase velocity.           

\subsection{\label{sec:eigenvalueSpectra}Eigenvalue Spectra of Selected Problems}

\subsubsection{\label{sec:nonMhdProblems}Non-MHD Problems}

\begin{figure}
\begin{center}
\includegraphics{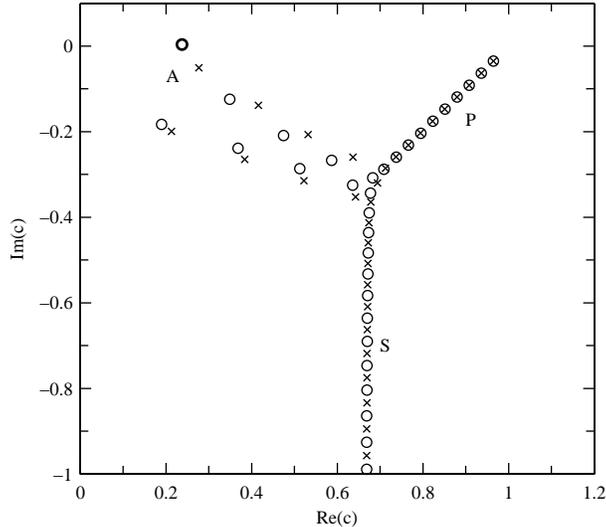}
\end{center}
\caption{\label{fig:spectrumHydroChannel}Spectrum of a non-MHD channel problem with the Poiseuille velocity profile at $ \Rey = 10^4 $, $ \alpha = 1 $, and $ p_u = 500 $, showing the A, P, and S branches. $ \circ $ and $ \times $ markers respectively correspond to even and odd modes. The even mode marked in boldface is unstable.}
\end{figure}

One of the most extensively studied problems in hydrodynamic stability is non-MHD channel flow with the Poiseuille velocity profile (\eg \cite{Kirchner00,MelenkKirchnerSchwab00,Orszag71,DongarraStraughanWalker96,DrazinReid81,Lin55}). In the high Reynolds number regime, the spectrum of the OS operator forms three branches on the complex plane, conventionally labeled A, P, and S \cite{Mack76}. The branches are shown in the numerical spectrum in Fig.~\ref{fig:spectrumHydroChannel}, obtained at $ \Rey = 10^4 $ and $ \alpha = 1 $ by solving the matrix eigenproblem \eqref{eq:matrixWeakForm} derived from Def.~\ref{def:channelZeroPm} (with $ \Harx = \Harz = 0$). According to Table~\ref{table:discreteSolutionSpaces}, $ u(z ) $ is expanded in the $ \{ \lambda_n^{[2]} \}_{n=1}^{N_u} $ basis, where for the present calculation the polynomial degree is set at $ p_u = N_u + 3 = 500 $. Due to the reflection symmetry of~\eqref{eq:orrSommerfeldZeroPm} and~\eqref{eq:noSlipChannel} with respect to $ z $, the eigenfunctions fall into even ($u(-z) = u(z)$) and odd ($u(-z)=-u(z)$) symmetry classes. The S branch contains a countably infinite set of modes, whose phase velocity is asymptotically equal (at large and negative $ \Imag( c ) $) to the mean basic flow $ \langle U \rangle = 2 / 3 $~\eqref{eq:baseUAverage}. On the other hand, the A and P branches contain a finite set of modes, respectively with $ 0 < C < \langle U \rangle $ and $ \langle U \rangle < C < 1 $. The P modes come into nearly degenerate even and odd pairs. As noted by Orszag \cite{Orszag71}, this near degeneracy is a genuine property of the spectrum, which does not disappear by increasing the spectral order. While all of the P modes are stable, the A branch contains a single unstable mode of even symmetry. This instability is of the critical-layer type \cite{DrazinReid81}: At sufficiently high Reynolds numbers, and over a suitable range of wavenumbers, the energy transfer from the basic flow to the mode (the Reynolds stress~\eqref{eq:gammaR}) exceeds the viscous dissipation, and as a result its growth rate becomes positive.  

Table~\ref{table:spectrumHydroChannel} lists in order of decreasing $ \Imag( c ) $ the first 33 eigenvalues plotted in Fig.~\ref{fig:spectrumHydroChannel}. This calculation has previously been performed by Kirchner (see Table~VII in~\cite{Kirchner00}) using the same Galerkin scheme as in the present work, so the two sets of eigenvalues should be in very close agreement. A comparison (see also the underlined digits in Table~\ref{table:spectrumHydroChannel}) reveals that for modes at the top end of the spectrum the relative agreement is of order $ 10^{-15} $, \ie close to machine precision. However, descending the spectrum, the number of decimal digits for which the calculations agree exhibits a diminishing trend, culminating to an $ \ord( 10^{-9} ) $ relative difference for Mode~33. This discrepancy is likely due to roundoff sensitivity in the computed eigenvalues close to the intersection point between the A, P and S branches, which is known to increase steeply with $ \Rey $ \cite{ReddySchmidHenningson93}. In our schemes, machine roundoff in double-precision (64 bit) arithmetic already leads to relative errors of order unity at $ \Rey \sim 5 \times 10^4 $ (see \S\ref{sec:nonNormality} below). Therefore, the observed six-digit loss in the agreement between Kirchner's eigenvalues and ours is not unreasonable at $ \Rey = 10^4 $, especially for modes like $ \mathrm{ A }_{10} $, which lies particularly close to the intersection point ($ \Imag( c ) = 0.637 \simeq 2 / 3$).
            
\begin{figure}
\begin{center}
\includegraphics{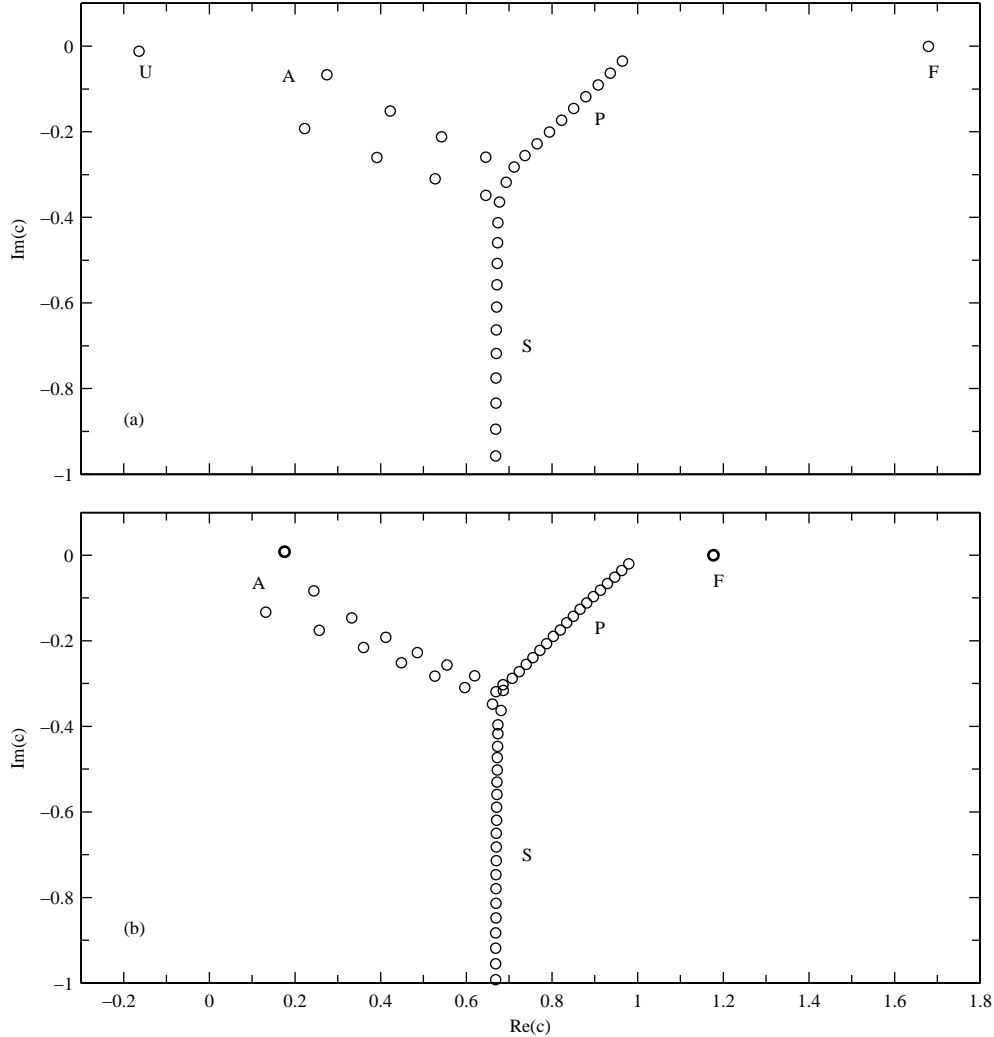}
\end{center}
\caption{\label{fig:spectrumHydroFilm}Eigenvalues of non-MHD film flow with the Poiseuille velocity profile at $ \Ohn = 3.14 \times 10^{-4} $, $ \Prg = 1.10 \times 10^{-4} $, $ \alpha = 1 $, and $ p_u = 500 $. The Reynolds numbers are $ \Rey = 10^4 $ (a) and $ \Rey = 3 \times 10^4 $ (b). In addition to the A, P, and S branches encountered in channel problems (Fig.~\ref{fig:spectrumHydroChannel}), the spectra contain downstream-propagating surface waves with $\Real( c ) > 1$, labeled F. Also, an upstream-propagating wave ($\Real( c ) < 0$), labeled U, is present in the $ \Rey = 10^4 $ spectrum. At $ \Rey = 10^4 $ all of the modes have negative growth rates, but at $ \Rey = 3 \times 10^4$ Modes $ \mathrm{A}_1 $ and F (represented by boldface markers) are unstable ($\Imag( c ) > 0 $).}
\end{figure}

In film problems, again with the Poiseuille  profile, the eigenproblem~\eqref{eq:matrixWeakForm} is derived from Def.~\ref{def:filmZeroPm} (with $ \Harx = \Harz = 0$), and, in accordance with Table~\ref{table:discreteSolutionSpaces}, the velocity eigenfunction is expressed in terms of the $ \nu_n $ polynomials. Our nominal specification of the free-surface parameters (which will also be used in the MHD calculations below) is $ \Ohn = 3.14 \times 10^{-4} $ and $ \Prg = 1.10^{-4} $, corresponding to a typical liquid-metal film of thickness $ 0.01 $~m at terrestrial gravitational fields \cite{GiannakisRosnerFischer07}. Setting $ \alpha = 1 $ and $ p_u = N_u + 1 = 500 $, we evaluate the spectra at Reynolds numbers $ \Rey = 10^4 $ and $ \Rey = 3 \times 10^4 $. The resulting eigenvalues are displayed in Fig.~\ref{fig:spectrumHydroFilm} and tabulated in Table~\ref{table:spectrumHydroFilm}, which also lists the modal free-surface energy~\eqref{eq:eA}. Like channel problems, the spectra exhibit the A, P, and S branches, and additionally contain two modes associated with the free surface, labeled U and F. Mode~F is a `fast' downstream-propagating surface wave, whose phase velocity is always greater than the basic velocity at the free surface ($ \Real( c ) > 1$). It is unstable for $ \Rey > ( 5 / 8 )\sh / \Prg $ \cite{GiannakisRosnerFischer07,Yih69}, provided that $ \alpha $ is smaller than some upper bound. This so-called \emph{soft} instability is present in Fig.~\ref{fig:spectrumHydroFilm}(b). Mode~U is an upstream-propagating mode ($ \Real( c ) < 0 $), which is present in the spectrum at sufficiently small Reynolds numbers (\eg Fig.~\ref{fig:spectrumHydroFilm}(a)). For $ \Rey \lesssim 10^3 $ (and $ \alpha = 1 $) its eigenfunction has the characteristic exponential-like profile of a surface wave. However, as $ \Rey $ grows its phase velocity increases, because the mode tends to be advected downstream by the basic flow. At the same time, its eigenfunction  develops the characteristics of an internal (shear) wave, such as well defined boundary and critical layers. Eventually, the eigenvalue crosses the $ \Real( c ) = 0 $ axis and merges with the A branch, taking over the role of the $ \mathrm{ A }_1 $ mode in channel flow (for this reason, in Table~\ref{table:spectrumHydroFilm} Mode U is also labeled $ \mathrm{ A }_1 $). In particular, provided that the Reynolds number exceeds some critical value, it experiences an instability very similar to that in channel flow, oftentimes referred to as the \emph{hard} instability. The spectra in the top and bottom panels of Fig.~\ref{fig:spectrumHydroFilm} respectively lie below and above the hard-instability threshold. As can be checked from Table~\ref{table:spectrumHydroFilm}, only a relatively small number of modes carry appreciable free-surface energy. Apart from the F mode, and the upper A and P family modes, for which $ E_a / E \sim 0.5 $, the remaining modes are internal, with $ E_a / E \lesssim 10^{-3} $. 

\subsubsection{\label{sec:zeroPmProblems}Problems in the Inductionless Limit}

The simplest version of MHD is the inductionless approximation~\eqref{eq:orrSommerfeldZeroPm}, whose weak formulation is stated in Defs.~\ref{def:filmZeroPm} and~\ref{def:channelZeroPm}, respectively for film and channel problems. Compared to the non-MHD baseline scenario, the steady-state magnetic field, parameterized by the streamwise and flow-normal Hartmann numbers $ \Harx $ and $ \Harz $, affects the stability of the flow both at the level of the basic state, as well as the perturbations. In the former case, the flow-normal component of the field leads to the establishment of the Hartmann velocity profile~\eqref{eq:baseHartmann}, which differs substantially from the Poiseuille one, even at moderate Hartmann numbers  (see Fig.~\ref{fig:baseProfile}). The departure from the parabolic profile affects the Reynolds stress, which is the main driver of critical-layer instabilities. The magnetic field also acts at the level of the perturbations by way of electrical currents induced by the perturbed fluid motion within the field. These induced currents set up Lorentz forces, which, in accordance with Lentz's law, always tend to dampen the flow. Moreover, they modify the velocity distribution of the perturbations, changing in turn the Reynolds stress and/or viscous dissipation. It is generally known, both on theoretical grounds \cite{Lock55,Drazin60}, as well as from numerical calculations \cite{PotterKutchey73,Takashima96}, that in channel problems the combined outcome of these effects is strongly stabilizing. In film problems, however, the existence of a resonance between the velocity and surface degrees of freedom may cause the F mode to deviate from that behavior \cite{GiannakisRosnerFischer07}. 

\begin{figure}
\begin{center}
\includegraphics{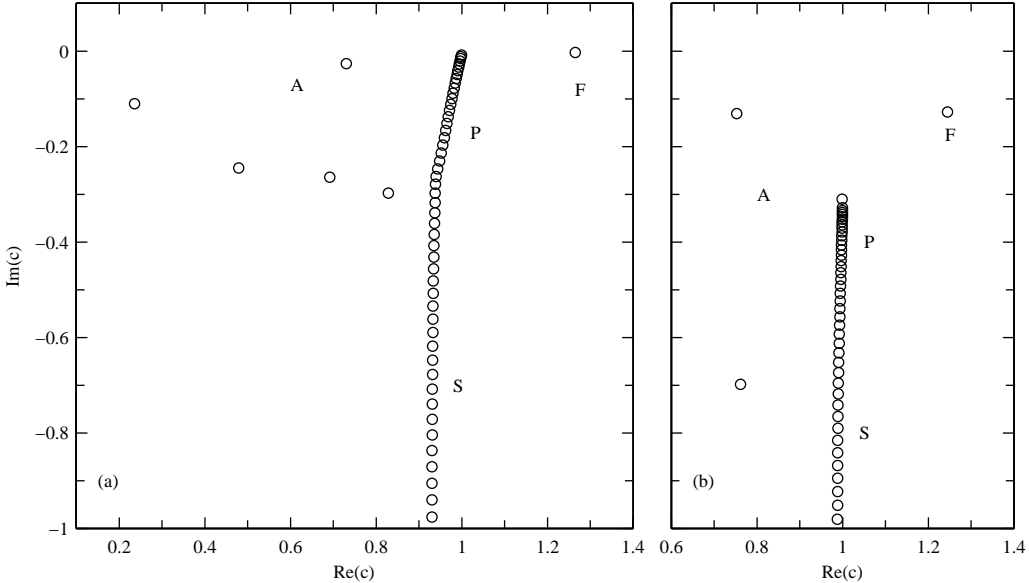}
\end{center}
\caption{\label{fig:spectrumZeroPmFilm}Eigenvalues of inductionless film problems with the Hartmann velocity profile and flow-normal background magnetic field ($ \Harx = 0$) at $ \Rey = 3 \times 10^4 $,  $ \Ohn = 3.14 \times 10^{-4} $, $ \Prg = 1.10 \times 10^{-4} $, $ \alpha = 1 $, and $ p_u = 500 $. The flow-normal Hartmann numbers are $ \Harz = 14 $ (a) and $ \Harz = 100 $ (b).}
\end{figure}

We first consider film problems with flow-normal magnetic field ($ \Harx = 0 $). Fig.~\ref{fig:spectrumZeroPmFilm} displays the eigenvalues computed at $ \Harz = 14 $ and 100, with all other spectral and flow parameters equal to those in Fig.~\ref{fig:spectrumHydroFilm}(b). Numerical results obtained using both exact and LGL quadrature for the computation of the stiffness matrix $ \mat{ K } $ (respectively \eqref{eq:matKuuUHartmannExact} and~\eqref{eq:matKuuUHartmannLgl}) are listed in Table~\ref{table:spectrumZeroPmFilm}. The maximum relative difference between the two eigenvalue sets is of order $ 10^{-11} $ (the corresponding mode is $ \mathrm{ P }_{22} $ at $ \Harz = 14$), becoming as small as $ 4.5 \times 10^{-16} $ for the $ \mathrm{ P }_3 $ mode. We note that the agreement between the lower modes does not improve by increasing $ p_u $. Like in our previous comparison of the eigenvalues of plane Poiseuille flow (Table~\ref{table:spectrumHydroChannel}) with the corresponding calculations by Kirchner \cite{Kirchner00}, the numerical convergence of the lower modes appears to be over fewer significant digits than the least stable ones. Nonetheless, our results demonstrate that the LGL quadrature scheme is a very viable alternative to the exact one, especially in light of its flexibility to treat arbitrary analytic velocity profiles  (see also \S\ref{sec:hartmannProfile} ahead).

Comparing Fig.~\ref{fig:spectrumZeroPmFilm} to Fig.~\ref{fig:spectrumHydroFilm}(b) illustrates the following basic aspects of the magnetic field's influence on the eigenvalues. First, as $ \Harz $ increases the A branch is seen to collapse. That is, the eigenvalues move towards the intersection point between the P and S branches, eventually experiencing what qualitatively appears as an inelastic collision with the S branch. In the process, Mode~$ \mathrm{A}_1$ (the hard mode) crosses the $ \Imag( c ) = 0 $ axis, \ie is stabilized. The real part of the S family eigenvalues remains (asymptotically) equal to the average value of the velocity profile, and moves from $ 2 / 3 $ towards 1, in accordance with~\eqref{eq:baseUAverage}. At the same time, the P branch becomes progressively aligned with the S branch. For sufficiently small values of $ \Harz $, including the $ \Harz = 14 $ example in Table~\ref{table:spectrumZeroPmFilm}, the P modes are somewhat less stable than in the non-MHD case (cf.~Table~\ref{table:spectrumHydroFilm}), but never cross the $ \Imag( c ) = 0 $ axis. As for the originally unstable F mode, this also becomes stabilized once $ \Harz $ exceeds some critical value (the spectra in Fig.~\ref{fig:spectrumZeroPmFilm} and Table~\ref{table:spectrumZeroPmFilm} are evaluated past that threshold). 

The behavior outlined above is encountered at moderate $ \Harz $, and is mainly due to the formation of the Hartmann velocity profile. As discussed in Ref.~\cite{GiannakisRosnerFischer07}, at sufficiently large Hartmann numbers Lorentz damping causes the decay rate $ | \Real( \gamma ) | $ of the A, P, and S modes to increase quadratically with $ \Harz $. In contrast, once $ \Harz $ crosses a threshold scaling like $ \Prg^{-1/2} $, the F mode exhibits a change in behavior, with its decay rate switching over to a \emph{decreasing} function of the Hartmann number. In the $ \Harz = 100 $ problem in Fig.~\ref{fig:spectrumZeroPmFilm} and Table~\ref{table:spectrumZeroPmFilm}, which lies close to that transition, the decay rate $ | \Gamma | = 0.12765$ of the F mode already is substantially smaller than that of the Lorentz-damped P and S modes ($ | \Gamma | \geq 0.31013$). A single A mode is present in the spectrum with comparable decay rate $ | \Gamma | = 0.13099 $, but at larger Hartmann numbers (not shown here) it also becomes suppressed. Even though the F mode remains stable, eventually it becomes the only one with small decay rate. As a result, film and channel problems differ qualitatively in that the eigenmodes of the former cannot be damped by arbitrarily large amounts solely by applying a background magnetic field.       

\begin{figure}
\begin{center}
\includegraphics{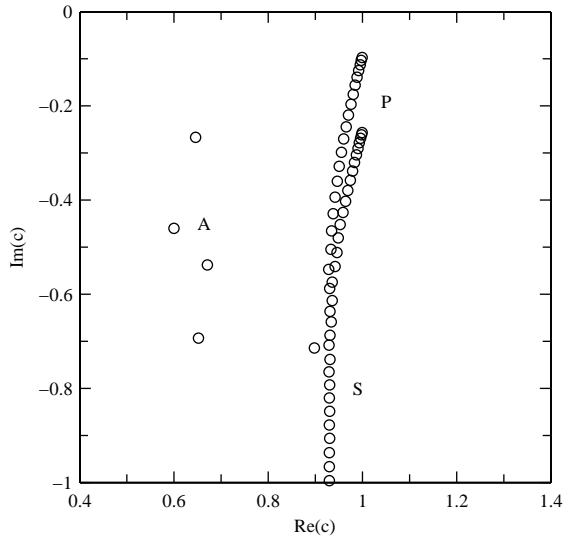}
\end{center}
\caption{\label{fig:spectrumZeroPmChannel}Spectrum of inductionless channel flow with the Hartmann velocity profile and oblique background magnetic field (orientation angle $ \phi = 1^\circ $ with respect to the streamwise direction) at $ \Rey = 10^4 $, $ \Harz = 14 $, $ \Harx = 14 / \tan( \phi ) \approx 802 $, $ \alpha = 1 $ and $ p_u = 500 $}
\end{figure}

The more general case with oblique external magnetic field (\ie $ \Harx $ and $ \Harz $ both nonzero), is especially interesting in the context of channel problems because, as can be checked from~\eqref{eq:orrSommerfeldZeroPm}, the reflection symmetry with respect to $ z $ is no longer present. As shown in Fig.~\ref{fig:spectrumZeroPmChannel}, the near-degeneracy between the even and odd P-family modes is broken, and the resulting sub-branches assume a distinctive curved shape. In general, the streamwise Hartmann number required to cause a comparable change in the eigenvalues is significantly larger than the corresponding flow-normal one. It is for this reason than in Fig.~\ref{fig:spectrumZeroPmChannel} we consider a magnetic field oriented at only $ 1^\circ $ with respect to the streamwise direction, but sufficiently strong so that $ \Harz $ is equal to the one used in Fig.~\ref{fig:spectrumZeroPmFilm}(b). In film problems, where no nearly degenerate P mode pairs exist to begin with, the oblique field still causes the P branch to adopt a qualitatively similar curved shape.

\subsubsection{\label{sec:MhdProblems}Film MHD Problems}

We now relax the inductionless approximation made in the preceding section and consider film MHD problems, defined variationally in Def.~\ref{def:filmMhd}, and discretized using the $ \{ \mu_n \}_{n=1}^{N_b} $ and $ \{ \nu_n \}_{n=1}^{N_u} $ bases for the magnetic-field and velocity eigenfunctions (see Table~\ref{table:discreteSolutionSpaces}). Throughout this section we work at polynomial degrees $ p_u = p_b = 500 $. Moreover, we compute the $ U$  and $ B $-dependent terms in the stiffness matrix $ \mat{ K } $ using the exact quadrature scheme, \ie \eqref{eq:matKuubbUHartmannExact} and~\eqref{eq:matKubbuBHartmannExact}, although accurate results can also be obtained by means of the LGL method (Eqs.~\eqref{eq:matKLgl1} and~\eqref{eq:matKLgl2}). 

\begin{figure}
\begin{center}
\includegraphics{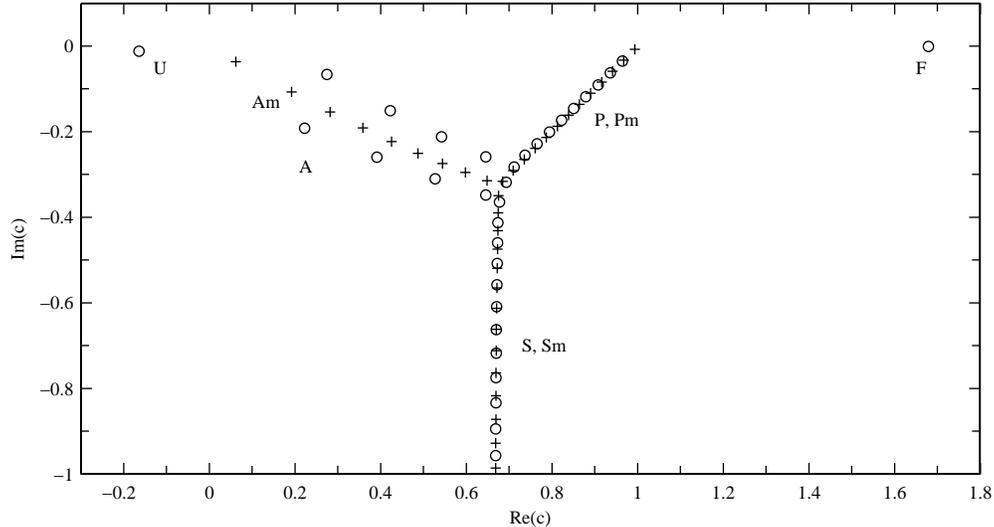}
\end{center}
\caption{\label{fig:spectrumMhdFilmNoField}Eigenvalues of film MHD flow with the Poiseuille velocity profile and vanishing steady-state magnetic field ($ \Harx = \Harz = 0$) at $ \Rey = 10^4 $, $ \Prm = 1.2 $, $ \Ohn = 3.14 \times 10^{-4} $, $ \Prg = 1.10 \times 10^{-4} $, $ \alpha = 1 $, and $ p_u = p_b= 500 $. In addition to the hydrodynamic modes, marked with $ \circ $, the spectrum contains magnetic modes, labeled by $ + $, which form the Am, Pm, and Sm branches.}
\end{figure}

Noting that the limit $ \Prm \ttz $, at which Eqs.~\eqref{eq:coupledOSInd} reduce to~\eqref{eq:orrSommerfeldZeroPm}  (under the proviso that $ \Harx $ and $ \Harz $ are non-negligible), is a singular limit of the coupled OS and induction equations, one can deduce that certain MHD modes, which we refer to as \emph{magnetic} modes, are disconnected from the inductionless spectra. These are to be distinguished from \emph{hydrodynamic} modes, that are regular as $ \Prm \ttz $. Whenever $ \Prm $ is of order unity, magnetic modes are expected to be present in the portion of the complex plane with $ \Imag( c ) > -1 $, irrespective of the background magnetic-field strength. In fact, they stand out particularly clearly in spectra evaluated at  $ \Harx = \Harz = 0$, such as the one depicted in Fig.~\ref{fig:spectrumMhdFilmNoField} for a $ \Prm = 1.2 $ problem. In this special case with zero background field, the maps $ \form{ K }_{ub} $~\eqref{eq:stiffWeakUB}, $ \form{ K }_{bu} $~\eqref{eq:stiffWeakBU}, and $ \form{ K }_{ba} $~\eqref{eq:stiffWeakBH} vanish, and the magnetic modes are independent of the hydrodynamic ones. The latter have zero magnetic-field eigenfunction and retain the same velocity eigenfunction and free-surface amplitude as in the non-MHD case, whereas for the former $ u $ and $ a $ are zero and $ b $ is non-vanishing. The magnetic modes form a three-branch structure as well, whose branches we label Am, Pm, and Sm. The magnetic S branch coincides with the hydrodynamic one, and the Pm branch lies close, but does not coincide, with P. The Am branch forms a nearly straight line that interpolates between the hydrodynamic A modes. Numerical values for the complex phase velocities of the 25 least stable magnetic modes are tabulated in Table~\ref{table:spectrumMhdFilmNoField}.

\begin{figure}
\begin{center}
\includegraphics{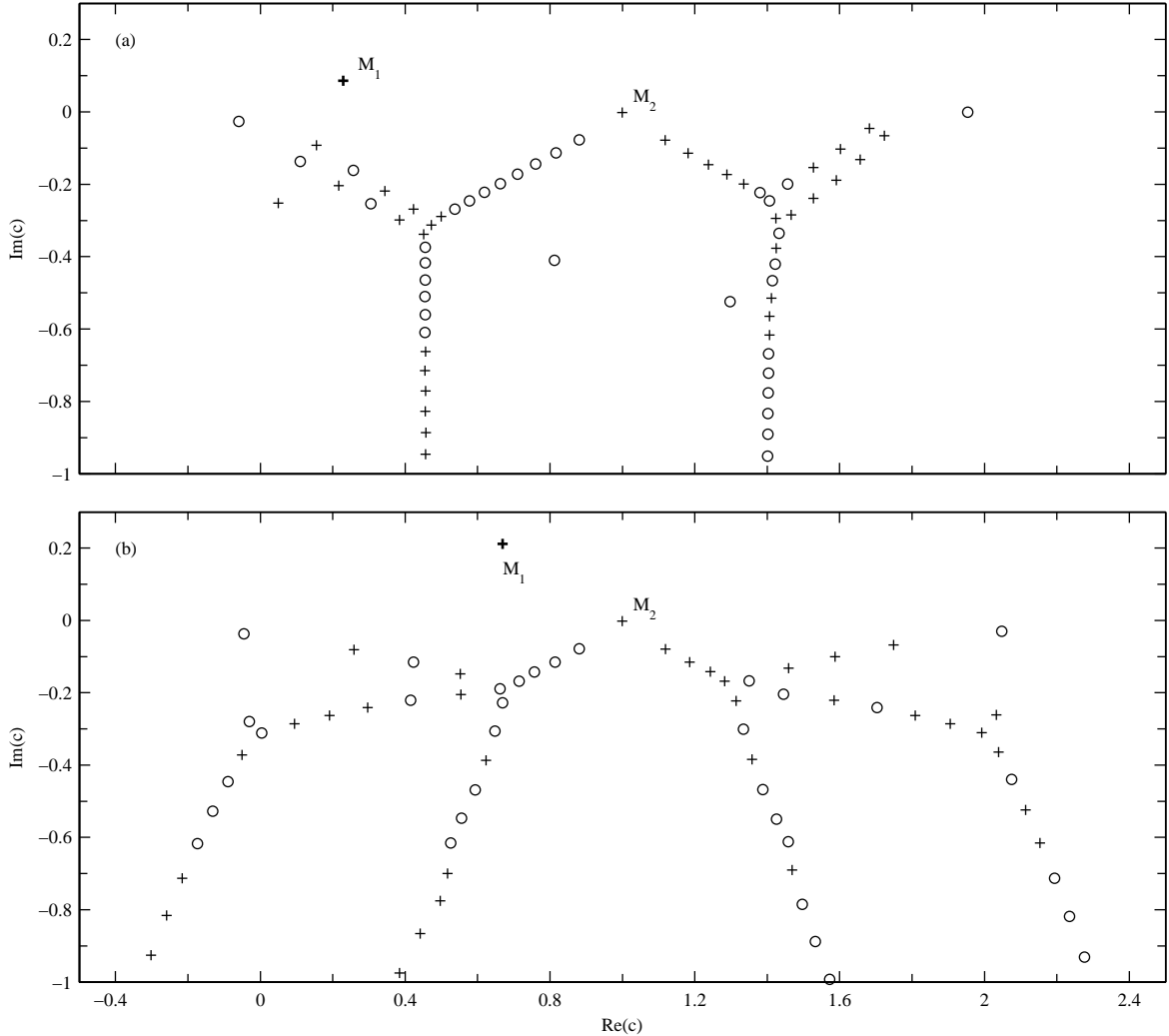}
\end{center}
\caption{\label{fig:spectrumMhdFilmFlowNormal}Eigenvalue spectra of film MHD problems at $ \Rey = 10^4 $,  $ \Prm = 1.2 $, $ \alpha = 1 $, and $ p_u = p_b= 500 $. The external magnetic field is flow-normal ($ \Harx = 0$), with $ \Harz = 14 $ (a) and $ \Harz = 100 $ (b). $ \circ $ and $ + $ markers respectively represent hydrodynamic and magnetic modes. Boldface markers correspond to unstable modes.} 
\end{figure}

\begin{figure}
\begin{center}
\includegraphics{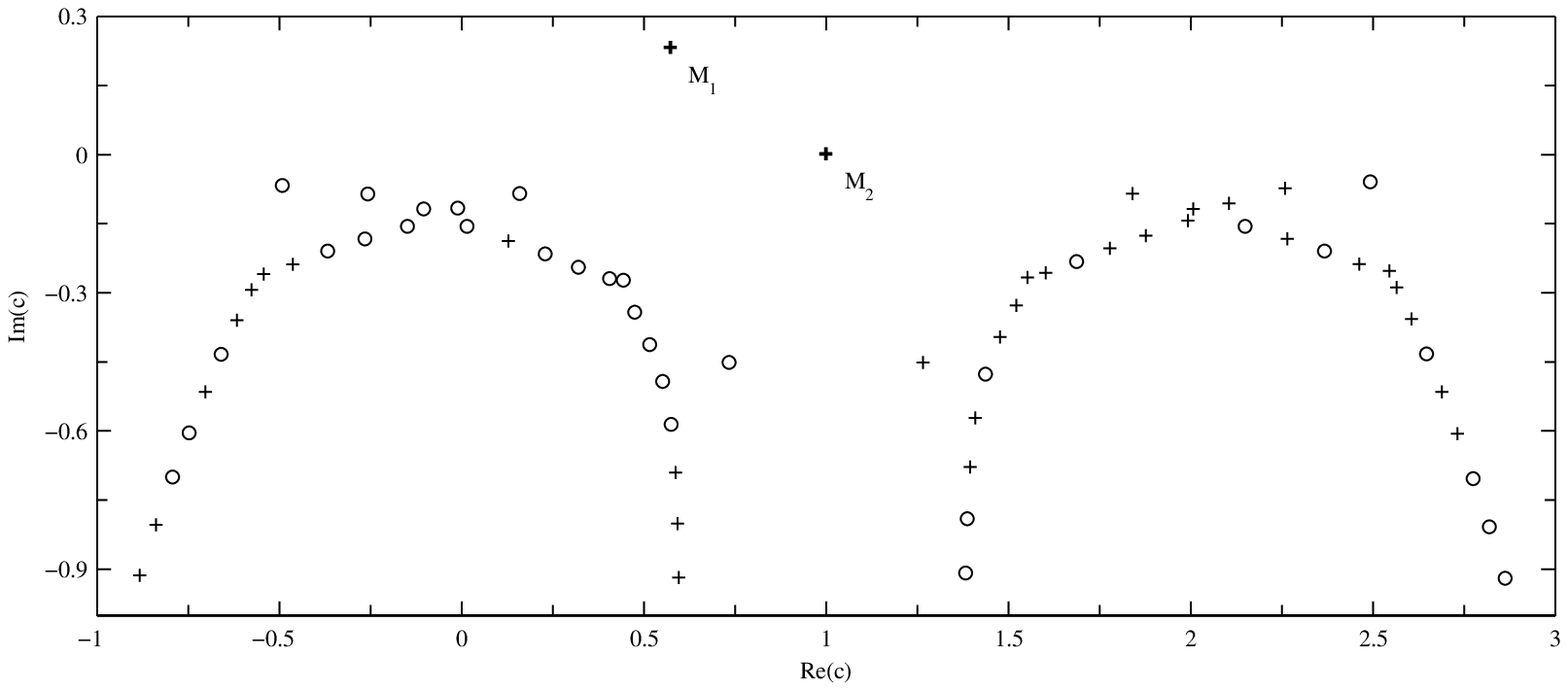}
\end{center}
\caption{\label{fig:spectrumMhdFilmOblique}Same spectrum as in Fig.~\ref{fig:spectrumMhdFilmFlowNormal} but with $ \Harx = \Harz / \tan( 1^\circ ) \approx \mbox{5,792} $}
\end{figure}

When the steady-state magnetic field is nonzero, $ \form{ K }_{bu} $, $ \form{ K }_{ub} $, and $ \form{ K }_{ba} $ couple the hydrodynamic and magnetic modes, typically resulting to the formation of multiple eigenvalue branches. This type of behavior is illustrated in Figs.~\ref{fig:spectrumMhdFilmFlowNormal} and~\ref{fig:spectrumMhdFilmOblique} for film MHD problems at $ \Prm = 1.2 $, respectively with flow-normal and oblique external magnetic field. Tables~\ref{table:spectrumMhdFilmFlowNormal} and~\ref{table:spectrumMhdFilmOblique} list the corresponding complex phase velocities and energies. As shown in Fig.~\ref{fig:spectrumMhdFilmFlowNormal}(a), instead of leading to the collapse of the A branch and alignment of the P and S branches observed in the inductionless limit (Fig.~\ref{fig:spectrumZeroPmFilm}), the magnetic field causes the nearly coincident three-branch structures at $ \Harz = 0 $ to split into two distinct ones, each of which is populated by both hydrodynamic and magnetic modes. Moreover, an unstable magnetic mode ($\mathrm{M}_1$) is now present in the spectrum. This mode, which also arises in channel problems, signifies that at sufficiently high $ \Prm $ the magnetic field can destabilize an originally stable flow. As  $ \Harz $ increases above 14, the tails of the branches split again, resulting to the intricate eigenvalue distribution observed at $ \Harz = 100 $, which, apart from Mode~$ \mathrm{ M }_1 $, is nearly symmetrical about $ \Real( c ) = 1 $. The spectrum with oblique external magnetic field (Fig.~\ref{fig:spectrumMhdFilmOblique}) exhibits multiple branches as well, and additionally contains a second unstable magnetic mode ($ \mathrm{ M }_2 $). In both examples with $ \Harz = 100 $, Mode~$ \mathrm{ M}_2 $ stands out in that its kinetic energy is significantly smaller ($E_u / E = \text{0.0023948} $ and $ E_u / E = \text{0.00091314} $ for $ \Harx = 0 $ and $ \Harx = \Harz/\tan(1^\circ) $, respectively) than the $ E_u / E = \ord( 10^{-1} ) $ values of the remaining modes.
   
\begin{figure}
\begin{center}
\includegraphics{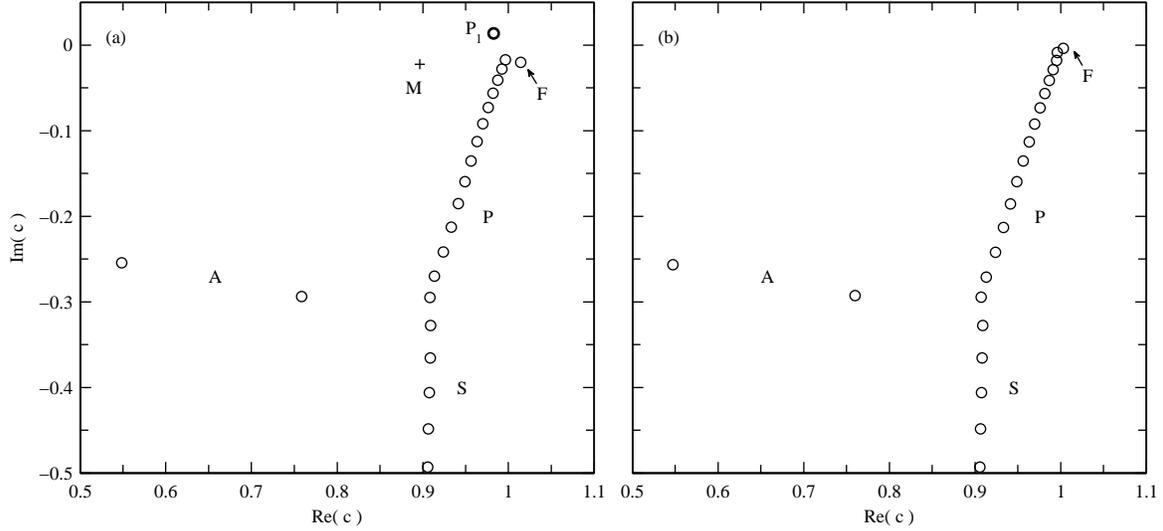}
\caption{\label{fig:spectrumMhdFilmLowPm}Spectra of film problems with flow-normal background magnetic field ($ \Harx = 0$) at $ \Rey = 10^{6} $, $ \Prg = 1.10 \times 10^{-4} $, $ \Ohn = 3.14 \times 10^{-4} $, $ \alpha = 0.01 $, $ \Prm = 10^{-4} $, $ p_u = p_b = 500 $ (a), and the corresponding inductionless problem (b)}
\end{center}
\end{figure}

At the $ \Prm < 10^{-4} $ regime of laboratory fluids, we have observed that film MHD spectra are well approximated by those in the inductionless limit, apart from the presence of (i) magnetic modes with $ \Imag( c ) > -1 $, and (ii) a diffusive interaction between the F and P modes, accompanied by an instability. These two kinds of discrepancy are shown in Fig.~\ref{fig:spectrumMhdFilmLowPm} and Table~\ref{table:spectrumMhdFilmLowPm}, where $ \Prm = 10^{-4} $ and inductionless spectra have been evaluated at $ \Rey = 10^{6 } $, $ \alpha = 0.01 $, $ ( \Harx, \Harz ) = ( 0, 10 ) $, $ \Prg = 1.10 \times 10^{-4 } $ ,and $ \Ohn = 3.14 \times 10^{-4} $. To begin, Fig.~\ref{fig:spectrumMhdFilmLowPm}(a) exhibits an isolated, damped magnetic mode (labeled M), which, due to the singular nature of the limit $ \Prm \ttz $, is entirely absent from Fig.~\ref{fig:spectrumMhdFilmLowPm}(b). Its magnetic energy $ E_b / E = 0.28628 $ is the largest of the modes with $ \Imag( c ) \geq -1 $. Mode~M, which is also present in channel problems at comparable $\{ \Rey, \Harz, \Prm, \alpha \}$, has sufficiently large decay rate so as not to affect the validity of the inductionless approximation in critical Reynolds number calculations (see \cite{Takashima96} and \S\ref{sec:criticalReynolds} ahead). The second contrastive feature is the presence of an \emph{unstable} P mode ($ \langle U \rangle < \Real( c ) < 1 $) in the $ \Prm = 10^{-4} $ spectrum, when all of the modes of the inductionless problem are stable. At the same time, the decay rate $ -\Gamma = 0.020234 \alpha $ of the F mode ($ \Imag( c ) > 1 $) in Fig.~\ref{fig:spectrumMhdFilmLowPm}(a) is about a factor of ten greater than its $ - \Gamma = -0.0035581 \alpha$ value in the inductionless limit. The magnetic energy of the $ \mathrm{ P }_1 $ and F modes at $ \Prm = 10^{-4} $, respectively amounting to 0.21534$E$ and 0.20240$E$, is somewhat smaller than that of the M mode, but still more than an order of magnitude greater than the magnetic energy $E_b / E < \text{0.0064}$ of the remaining modes with $ \Imag( c ) > - 1 $. For these latter modes, the relative error in $ c $ of the inductionless approximation is less than 0.0058. 

The unstable P mode can be continuously traced to the F mode as $ \Prm \ttz $, and likewise the F mode at $ \Prm = 10^{-4} $ originates from the $ \mathrm{ P }_1 $ mode in the inductionless limit. The relative change in $ c $ accumulated in the process is 0.027 and 0.021, respectively for $ \mathrm{ P }_1 $ and F (in the sense of the $ \Prm = 10^{-4} $ problem). This type of exchange of the modes' physical character, oftentimes accompanied by instabilities, is common in multiply diffusive systems \cite{Craik85}. Here, the velocity and magnetic-field perturbations play the role of two diffusive substances, respectively with diffusion constants $ \Rey^{-1} $ and $ \Reym^{-1} $. As discussed in more detail in~\cite{GiannakisRosnerFischer07}, the free-surface is essential to this low-$ \Prm $ instability, which does not occur in channel problems.

The markedly different types of behavior we have so far encountered are a testament that in its full generality the free-surface MHD stability problem is a complex one. One of its major aspects that we have not touched upon, and which we defer to future work, is the role of the induced magnetic field $ B $ on the instabilities, as well as the formation of multiple eigenvalue branches, at $ \Prm = \ord( 1 ) $. While we do not present these calculations here, setting $ B $ to zero while keeping all other parameters fixed yields spectra that neither contain unstable magnetic modes, nor exhibit the multiple-branch structures. Using the approach employed in \cite{GiannakisRosnerFischer07} for low-$ \Prm $ fluids, it would be interesting to investigate the energy-transfer mechanisms associated with $ B $, and the manner in which they contribute to the instability.

\subsection{\label{sec:spectralConvergence}Convergence and Stability}

The issue of convergence and stability of spectral schemes is a very broad one, and can be approached from various angles. As a minimum, certain analytical criteria must be satisfied (\eg \S2.2 in~\cite{DevilleFischerMund02}). That is, given a well-posed variational formulation of the problem at hand, the discrete solution must converge, under some suitable norm, to the exact one as the dimension $ N $ tends to infinity. Furthermore, the discretization error must be bounded by an $ N $-independent constant (stability). Among the relevant literature for eigenvalue problems (see~\cite{BabuskaOsborn91} and references therein) of particular importance to us is the work of Melenk \etal~\cite{MelenkKirchnerSchwab00}, who showed that the Galerkin method used for what we call here non-MHD channel problems is spectrally convergent. Generalizing the results in \cite{MelenkKirchnerSchwab00} to MHD is of course an essential prerequisite if our schemes are to be deemed well-posed. In what follows, however, instead of pursuing that program we adopt a less rigorous approach and limit ourselves to more practical aspects of convergence and stability. That is, implicitly assuming that our schemes convergence in the analytical sense, we perform test calculations that aim to probe their behavior in actual computing environments, emphasizing on issues related to finite arithmetic precision. 

\subsubsection{\label{sec:pConvergence}$ p $-Convergence}

In shear-flow stability problems at large Reynolds numbers both truncation and roundoff errors come into play, and in certain cases addressing them leads to self-conflicting situations. On one hand, in order to resolve the small length scales that develop (the boundary and internal friction layers) it is necessary to work at large spectral orders ($ p \gtrsim 500 $). Otherwise, the truncation error is significant. However, unless the basis polynomials are carefully chosen, the matrix representations of high-order differential operators (such as the $ \DD^4 $ operator in the OS equation) become ill conditioned as $ p $ increases, causing a growth in roundoff error to the point where it exceeds truncation error. It is precisely here that a major strength of the scheme developed by KMS for plane Poiseuille flow, and extended here to free-surface MHD problems, lies: By working in polynomial bases constructed so as to reflect the order of the Sobolev spaces of the underlying continuous problem (see Remark~\ref{rem:orthogonality}), roundoff sensitivity essentially becomes \emph{independent} of $ p $. 

As a concrete illustration, we have experimented with an alternative implementation of our Galerkin schemes for non-MHD channel flow, where instead of the $ \lambda_n^{[2]} $ polynomials prescribed in Table~\ref{table:discreteSolutionSpaces}, the basis polynomials of $ V_u^{N_u} $ are Lagrange interpolants on LGL quadrature knots of order $ p + 1 $, suitably modified to meet the essential boundary conditions~\eqref{eq:noSlipChannel}. Basis polynomials of this type, hereafter denoted by $ h_n $, are widely used in pseudospectral and spectral-element methods \cite{DevilleFischerMund02,Fornberg96}. However, they lack the orthogonality properties appropriate to $ H^2_0 $.

\begin{rem}{\label{rem:matrixGrowth}}A prominent manifestation of non-orthogonality in the $ \{ h_n \} $ basis is matrix coefficient growth with $ p $. We observed that the $ \infty $-norm of matrices with elements  $ \innerprodLtwo{ \Omega }{ \DD^{d_2}(h_n) }{ \DD^{d_1}(h_m)} $ scales as $ p^{ d_1 + d_2 } $. In contrast, all of the corresponding matrices $ \mat{ T }^{[k d_1 d_2]} $ evaluated in the $ \{ \lambda_n^{[r]} \} $, $ \{ \mu_n \} $ and $\{ \nu_n \}$ bases (see Appendix~\ref{app:innerProducts}) have $ p $-independent $ \infty $-norms. Recalling that the $ \infty $-norm of a matrix $ \mat{ A } $ is equal to $ \max_{m} \sum_n | A_{mn} | $, the latter is a direct consequence of the  orthogonality properties of the Legendre polynomials and the choice of normalization, which ensure that $ \mat{ T }^{[k d_1 d_2] } $ (i) are banded, (ii) their bandwidths are $ p $-independent, and (iii) apart from those corresponding to the nodal shape functions, the absolute values of the matrix coefficients either remain constant or decrease going down the nonzero diagonals.\end{rem} 

\begin{figure}
\begin{center}
\includegraphics{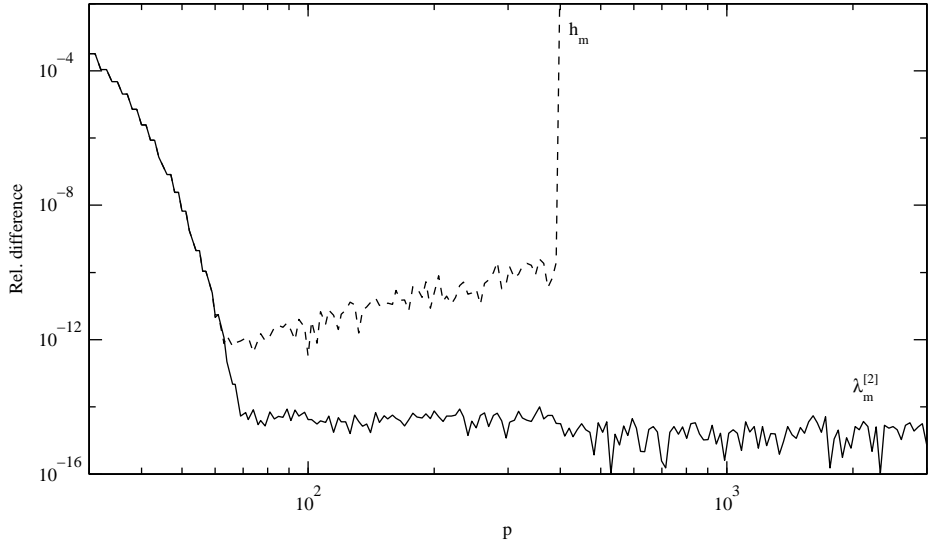}
\end{center}
\caption{\label{fig:spectralConvergenceLegendreLagrange}Spectral convergence of the least stable eigenvalue (the $\mathrm{ A }_1 $ eigenvalue in Table~\ref{table:spectrumHydroChannel}) of non-MHD channel flow with the Poiseuille velocity profile at $ \Rey = 10^4 $ and $ \alpha = 1$. The solid and dotted lines respectively correspond to eigenvalues computed using the $ \{ \lambda_n^{[2]} \} $ basis~\eqref{eq:lambda0} with polynomial degree $ p $, and the $ \{ h_n \} $ basis, consisting of Lagrange interpolants at the Legendre-Gauss-Lobatto quadrature knots of order $ p + 1 $, modified to satisfy the $ H^2_0 $ boundary conditions. Convergence is evaluated relative to a reference value computed using the $ \{ \lambda_n^{[2]} \} $ basis at $ p = \mbox{5,000} $.}
\end{figure}

The ill behaved stiffness and mass matrices arising in the Lagrange-interpolant basis lead to a rapid increase of the scheme's roundoff sensitivity with $ p $. The resulting degradation in the accuracy of the computed eigenvalues is immediately obvious in Fig.~\ref{fig:spectralConvergenceLegendreLagrange}, which shows the relative convergence of the least-stable eigenvalue at $ ( \Rey, \alpha ) = ( 10^4, 1 ) $ as a function of the polynomial degree $ p $, obtained via the $ \{ h_n \} $ and $ \{ \lambda_n^{[2]} \}$ bases. In both cases, convergence has been computed relative to a reference value obtained by means of the $ \{ \lambda_n^{[2]} \} $ basis at high polynomial degree ($p = \mbox{5,000}$). At small to moderate values of $ p$ the  results are essentially identical, and clearly display the exponential decrease in truncation error typical of spectral methods. However, in the case of the eigenvalue computed using the $ \{ h_n \} $ basis the exponential convergence trend halts abruptly at around $ p = 50 $, at which point the roundoff error caused by the ill conditioned stiffness and mass matrices becomes dominant. As $ p $ further increases, the eigenvalue is seen to progressively diverge from the reference value until, around $ p = 400 $, the algorithm for the computation of the differentiation matrices (see Appendix~C in \cite{Fornberg96}) becomes unstable and breaks down. In contrast, the eigenvalue computed using the $ \{ \lambda_n^{[2]} \} $ basis converges exponentially until close to machine precision, and although a small systematic trend can be observed for $ p \gtrsim 10^3 $, the calculation remains stable even at very large $ p $.   

\subsubsection{\label{sec:hartmannProfile}Effects of the Hartmann Profile}

\begin{figure}
\centering
\includegraphics{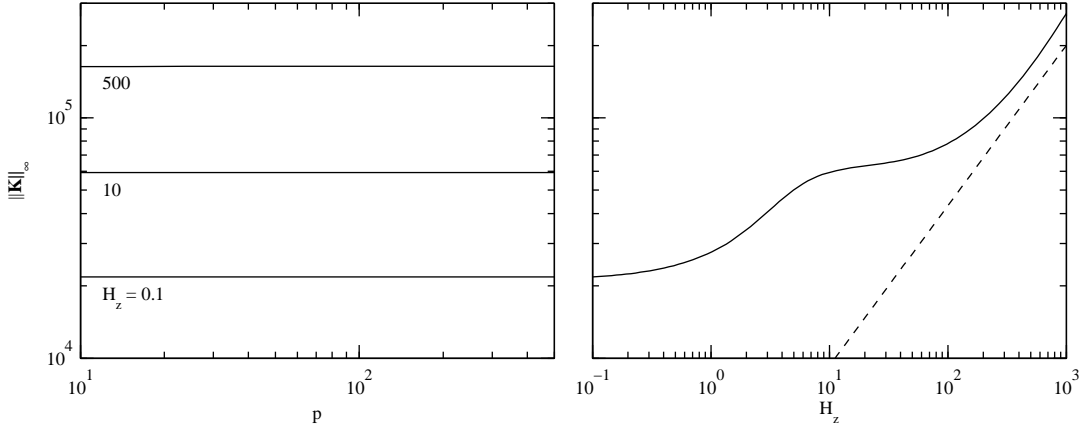}
\caption{\label{fig:infNormK}Infinity norm of the stiffness matrix $ \mat{ K } $ of film MHD problems with the Hartmann velocity and magnetic-field profiles at $ \Rey = 10^4 $, $ \Prm = 1.2 $, $ \Prg = 1.10 \times 10^{-4} $, $ \Ohn = 3.14 \times 10^{-4} $, $ \alpha = 1 $, and $ \Harx = \Harz / \tan( 1^\circ ) $. In the left-hand panel, $ || \mat{ K } ||_\infty $ is plotted as a function of $ p = p_u = p_b $ at $ \Harz = \text{0.1, 10, 500} $. The right-hand panel shows $ || \mat{ K } ||_\infty $ as a function of $ \Harz $ at fixed $ p_u = p_b = 200 $. The dashed line represents the power law $ || \mat{ K } ||_\infty \propto \Harz^{2/3} $.} 
\end{figure}

Problems with the Hartmann velocity and magnetic-field profiles~\eqref{eq:baseHartmann} differ from their counterparts with quadratic (or, more generally, polynomial) steady-state profiles in that the stiffness matrix $ \mat{ K } $ contains contributions of the form $ \int_{-1}^1 \dd \xi \, \ee^{\Harxi \xi} L_m( \xi ) L_n( \xi ) $, where $ \Harxi $ is a real parameter. These exponentially-weighted inner products are nonzero for all $( m, n )$, and, as a result, $ \mat{ K } $ is full. One immediate implication concerns memory and eigenvalue-computation costs, respectively scaling as $ N^2 $ and $ N^3 $ for a problem of dimension $ N = \dim( V^{[\boldsymbol{N}]} ) $. MHD problems are especially affected, since the spectral decompositions now have to be performed for both of the velocity and magnetic-field eigenfunctions, leading to 4-fold and 8-fold increases in size and complexity relative to inductionless or non-MHD problems. The non-sparsity of $ \mat{ K } $ also necessitates a re-evaluation of whether or not our schemes are roundoff stable at large spectral orders. For, our argument in Remark~\ref{rem:matrixGrowth} that $ || \mat{ K } ||_\infty $ is $ p $-independent relied on the number of nonzero elements in each of its rows being fixed, which no longer applies in problems with exponential profiles. Yet, as Fig.~\ref{fig:infNormK} illustrates, in practice $ || \mat{ K } ||_\infty $ is to a very good approximation $ p $-independent irrespective of the value of the Hartmann number, suggesting that our schemes are well-conditioned for the Hartmann family of steady-state profiles as well. Of course, $ || \mat{ K } ||_\infty$ does experience a growth with $ \Harz $, but that growth is due to physical parameters only.    

\begin{figure}
\centering
\includegraphics{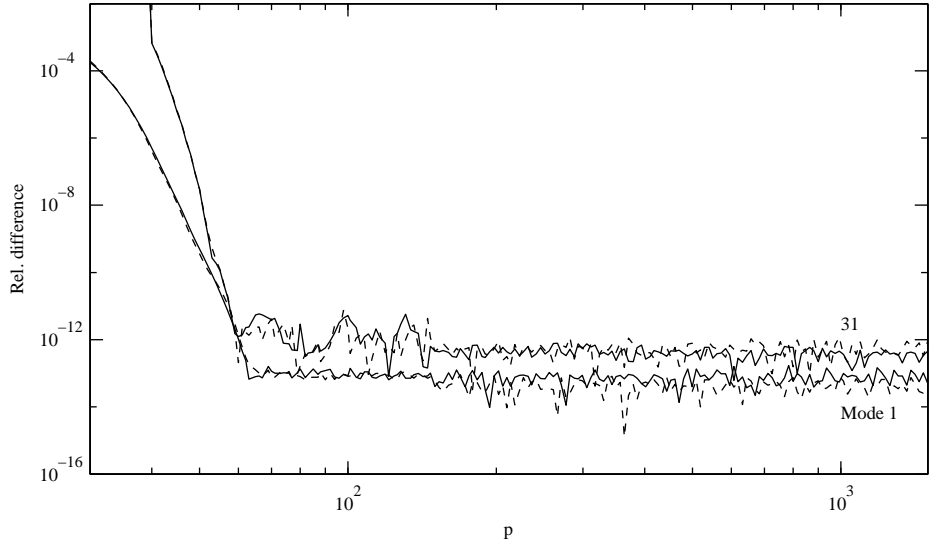}
\caption{\label{fig:compareQuadrature}$p$-convergence in film MHD problems with Hartmann velocity and magnetic-field profiles and oblique steady-state magnetic field. The flow parameters are as in Fig.~\ref{fig:spectrumMhdFilmOblique} and Table~\ref{table:spectrumMhdFilmOblique}, and the eigenvalues shown are for Modes~1, and 31. The solid and dashed lines respectively represent eigenvalues obtained via the exact and LGL quadrature schemes. Convergence is computed relative to a reference value at $ p = p_u = p_b = \mbox{2,500} $ obtained using the exact-quadrature method.}
\end{figure}

In \S\ref{sec:baseUBMatrices}, we introduced two alternative ways of evaluating the $ U $ and $ B $-dependent terms in the stiffness matrix, one of which employs suitable quadrature rules \cite{Mach84} to compute the exponentially-weighted inner products exactly, while the other is based on approximate LGL quadrature at the precision level specified in~\eqref{eq:lglPrecision}. The eigenvalue calculations in Table~\ref{table:spectrumZeroPmFilm} have already hinted at a close agreement between the two methods in inductionless flow, which we now examine in more detail, using film MHD flow with oblique magnetic field as a more challenging example. We consider a problem with the same parameters as in Fig.~\ref{fig:spectrumMhdFilmOblique}, and track the dependence of the computed eigenvalue of Mode~1 and Mode~31 (as usual, ordered in descending order of $ \Real( \gamma ) $) as $ p = p_u = p_b $ is varied from 30 to 1,500. We calculate the eigenvalues using both exact quadrature (Eqs.~\eqref{eq:matKuubbUHartmannExact} and~\eqref{eq:matKubbuBHartmannExact}), and approximate quadrature (Eqs.~\eqref{eq:matKLgl1}--\eqref{eq:matKLgl2}) at the smallest precision level consistent with~\eqref{eq:lglPrecision}. Fig.~\ref{fig:compareQuadrature} demonstrates that the eigenvalues converge exponentially towards their reference values, computed at $ p = \mbox{2,500} $ via the exact-quadrature method, in a nearly identical manner, until limited by finite arithmetic precision. Convergence for Mode~31 is about an order of magnitude less than Mode~1, but in both cases the computed eigenvalues remain stable at large $ p $. It therefore appears that a version of Banerjee and Osborn's theorem \cite{BanerjeeOsborn90} that $ 2 p - 1 $ quadrature precision is sufficient for convergence in elliptical eigenvalues problems also applies in OS-type problems. We remark that due to the aforementioned issues regarding storage and computation cost, we were not able to extend the calculation to as high values of $ p$ as we did in the non-MHD problem with the Poiseuille velocity profile (Fig.~\ref{fig:spectralConvergenceLegendreLagrange}).       

\subsubsection{\label{sec:nonNormality}Non-Normality Issues}

Despite  yielding stiffness and mass matrices that are `optimally' conditioned with $ p $, our choice of bases does comparatively little in addressing the second major source of roundoff error in our stability problems, which is due to the non-normality of the OS and induction operators \eqref{eq:coupledOSInd}. As already discussed in \S\ref{sec:statusOfTheField}, at large Reynolds numbers the OS operator is highly non-normal, and, in consequence, its spectrum contains nearly linearly dependent eigenfunctions (with respect to the $ L^2$ or energy inner products). According to Reddy \etal~\cite{ReddySchmidHenningson93}, expanding arbitrary functions of unit norm in terms of the OS eigenfunctions would require coefficients scaling as $ \exp(\gamma \Rey\sh ) $ (for $ \alpha = 1 $). At around $ \Rey = 4 \times 10^4 $, the coefficients would be as large as $ 10^{16} $, indicating that in 64-bit arithmetic (15 significant digits) expansions of arbitrary functions would be severely affected by roundoff error. Similarly, one would expect the reverse operation of decomposing the OS eigenfunctions in a basis of polynomials to be also characterized by a sharp rise in roundoff sensitivity with  $ \Rey $.    

\begin{figure}
\centering
\includegraphics{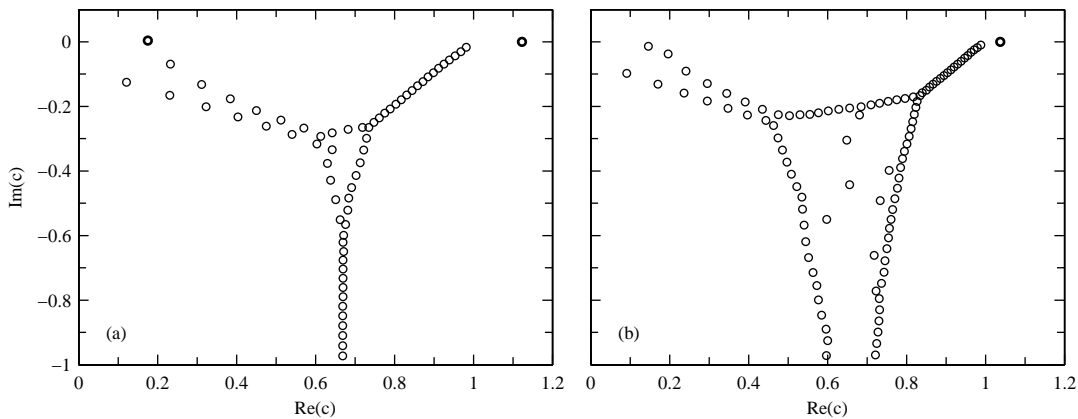}
\caption{\label{fig:spectralInstabilityHydro}Eigenvalue spectra of non-MHD film problems at $ \Prg = 1.1 \times 10^{-4} $, $ \Ohn = 3.14 \times 10^{-4} $, $ \alpha = 1 $, $ p_u = 500 $, and Reynolds numbers $ \Rey = 4 \times 10^4 $ (a) and $ 10^5 $ (b)} 
\end{figure}

Consider, for example, the spectra in Fig.~\ref{fig:spectralInstabilityHydro}, which have been computed at $ \Rey = 4 \times 10^4 $ and $ \Rey = 10^5 $ with our Matlab code, working in 64-bit arithmetic. Instead of a well-defined intersection point between the A, P, and S branches, the numerically computed spectra exhibit a diamond-shaped structure of eigenvalues, whose area on the complex plane grows with $ \Rey $. This type of spectral instability, which is entirely caused by finite-precision arithmetic, has come to be the hallmark of roundoff sensitivity due to non-normality of the OS operator \cite{Orszag71,DongarraStraughanWalker96,SchmidEtAl93,ReddySchmidHenningson93}. As expected from the analysis in \cite{ReddySchmidHenningson93}, the sensitivity increases steeply with the Reynolds number: Comparing the spectrum at $ \Rey = 4 \times 10^4 $ with the corresponding one at $ \Rey = 3 \times 10^4 $ (Fig.~\ref{fig:spectrumHydroFilm}) reveals that it only takes a factor of 0.3 increase in $ \Rey $ for a noticeable diamond-shaped pattern to form (though a small diamond is already present in Fig.~\ref{fig:spectrumHydroFilm}). 

In channel problems, the diamond-shaped pattern also emerges at around $ \Rey = 4 \times 10^4$, and at marginally smaller $ \Rey = 3 \times 10^4 $ if the calculation is performed in the Lagrange interpolant basis of \S\ref{sec:spectralConvergence}. In the latter case, decreasing $ \Rey $ to $ 2 \times 10^4 $ is sufficient for the diamond to become virtually unnoticeable by eye, despite the basis being `ill-conditioned'. A similar Reynolds number for the onset of the spectral instability ($ \Rey = 2.7 \times 10^4 $ in Fig.~2 of Ref.~\cite{DongarraStraughanWalker96}) is reported by Dongarra \etal~for their Chebyshev tau scheme. Therefore, in these examples the accuracy of the computed eigenvalues and eigenvectors appears to be limited by the physical parameters of the problem, in accordance with the estimates of Reddy \etal~\cite{ReddySchmidHenningson93}, rather than the details of the numerical scheme. If that is the case, then, as noted by Dongarra \etal~\cite{DongarraStraughanWalker96}, the only way of addressing the non-normality issue would be to increase numerical precision. Those authors have observed that working in 128-bit arithmetic does indeed remove the diamond-shaped pattern from the numerical spectra. Unfortunately, we have not been able to verify this for our Galerkin schemes, as our code was written in Matlab, which does not natively support extended-precision floating-point numbers. However, there is no reason to believe that increasing the number of significant digits would not alleviate the spectral instability in our schemes as well. 

\begin{figure}
\centering
\includegraphics{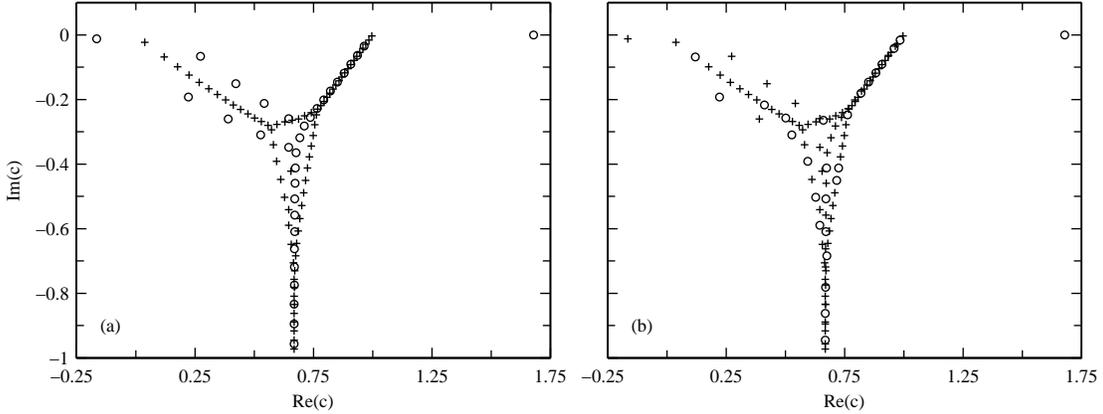}
\caption{\label{fig:spectralInstabilityMhd1}Eigenvalue spectra of film MHD flow with the Poiseuille velocity profile and zero steady-state magnetic field ($ \Harx = \Harz = 0$) at magnetic Prandtl numbers $ \Prm = 5 $ (a) and 10 (b). The remaining parameters are equal to those in Fig.~\ref{fig:spectrumMhdFilmNoField}. As $ \Prm $ increases, the magnetic modes (marked with $ + $ markers) develop the diamond-shaped pattern characteristic to roundoff errors caused by non-normality of the stability operators. The hydrodynamic modes (represented by $ \circ $ markers) are accurately computed, as they are decoupled from the magnetic ones (and do not depend on $ \Prm $).}
\end{figure}

\begin{figure}
\centering
\includegraphics{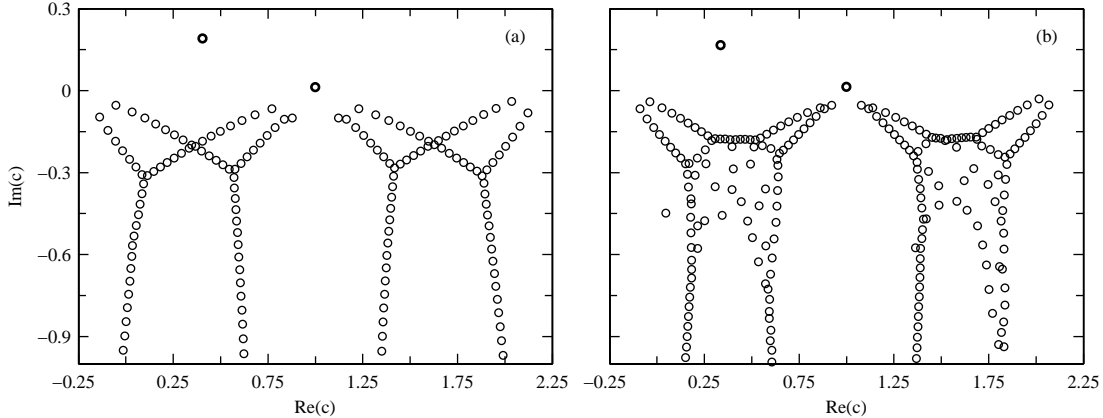}
\caption{\label{fig:spectralInstabilityMhd2}Eigenvalue spectra of film MHD flow with oblique steady-state magnetic field at Reynolds numbers $ \Rey = 5 \times 10^{4} $ (a) and $ \Rey = 10^{5} $ (b). Apart from $ \Rey $, all other flow and spectral parameters are equal to those in Fig.~\ref{fig:spectrumMhdFilmOblique}. In these plots no distinction is made between hydrodynamic and magnetic modes.} 
\end{figure}

Turning now to MHD, at a given value of the Reynolds number the effects of non-normality may be more or less severe compared to the hydrodynamic case, depending on the remaining parameters of the problem ($ \Prm $, $ \Harx $, $ \Harz $). The general rule of thumb is, however, that whenever the spectrum contains branch-intersection points, the highly non-orthogonal modes close to them will at some point experience the spectral instability if $ \Rey $ and/or $ \Reym $ are increased. The examples in Fig.~\ref{fig:spectralInstabilityMhd1} illustrate that in problems with zero background magnetic field the magnetic modes are the first to develop the diamond-shaped pattern if $ \Prm $ is greater than unity. Moreover, Fig.~\ref{fig:spectralInstabilityMhd2} shows that if $ \Rey $ is increased in the film MHD problem in Fig.~\ref{fig:spectrumMhdFilmOblique} four branch intersection points are formed, all of which are affected by roundoff errors at $ \Rey = \ord( 10^5 ) $. On the other hand, in the $ \Prm \lesssim 10^{-5}$ regime relevant to terrestrial fluids, the gradual disappearance of the three-branch structure with increasing $ \Harz $ (see \S\ref{sec:zeroPmProblems}) results to smaller regions on the complex plane being dominated by inaccurately computed eigenvalues. 

\subsection{\label{sec:consistency}Consistency Calculations}

The energy-balance relation~\eqref{eq:energyBalance} forms the basis of the following consistency check for film MHD problems: First, solve the matrix eigenproblem~\eqref{eq:matrixWeakForm} to obtain $ \gamma $ and the discrete representation $ \colvec{ v } $ of $ ( u, b, a ) $. Using $ \colvec{ v } $, the corresponding definition of the basis functions $ \psi_m $ (Def.~\ref{def:discreteBases}), and Eqs.~\eqref{eq:gammaSum}, compute the quantity $ \tilde \Gamma :=  \Gamma_\mathrm{ R } + \Gamma_\mathrm{ M } + \Gamma_J + \Gamma_\nu + \Gamma_\eta + \Gamma_{ a \nu } + \Gamma_{ a \eta } $. Then, according to~\eqref{eq:energyBalance}, the relative difference between $ \Gamma = \Real( \gamma ) $ and $ \tilde \Gamma $, given by $ \epsilon := | ( \tilde \Gamma - \Gamma ) / \Gamma | $, should be small, ideally close to machine precision. We note that the presence of the second derivatives of $ b $ in~\eqref{eq:gammaEta} and~\eqref{eq:gammaAEta}, which cannot be defined weakly for $ b \in H^1( \Omega ) $, necessitates that for the purposes of this calculation $ ( u, b, a ) $ is restricted to the strong solution space $ \mathcal{ D }_\mathcal{ K } $. Of course, in our polynomial subspaces of $ H^1 $ square integrability of second (and higher) derivatives is in principle not an issue. However, as we discuss below, practical repercussions in evaluating expressions like~\eqref{eq:gammaEta} and~\eqref{eq:gammaAEta} \emph{are} nonetheless present, since in the $ \{ \mu_n \} $ basis, which has been constructed so as to reflect $ H^1 $ regularity, the matrix representations of sesquilinear forms involving second weak derivatives are not stable with $ p $. Specifically, as can be checked either numerically or from the properties of the Legendre polynomials, matrix coefficients of the form $ \innerprodLtwo{ \Omegaref }{ \hat\DD^2 \mu_n }{ \hat\DD^2 \mu_m } $ grow like $ p^2 $, while boundary terms $\hat\DD^2\mu_n( 1 ) $ scale as $ p^{5/2} $.

\begin{figure}
\begin{center}
\includegraphics{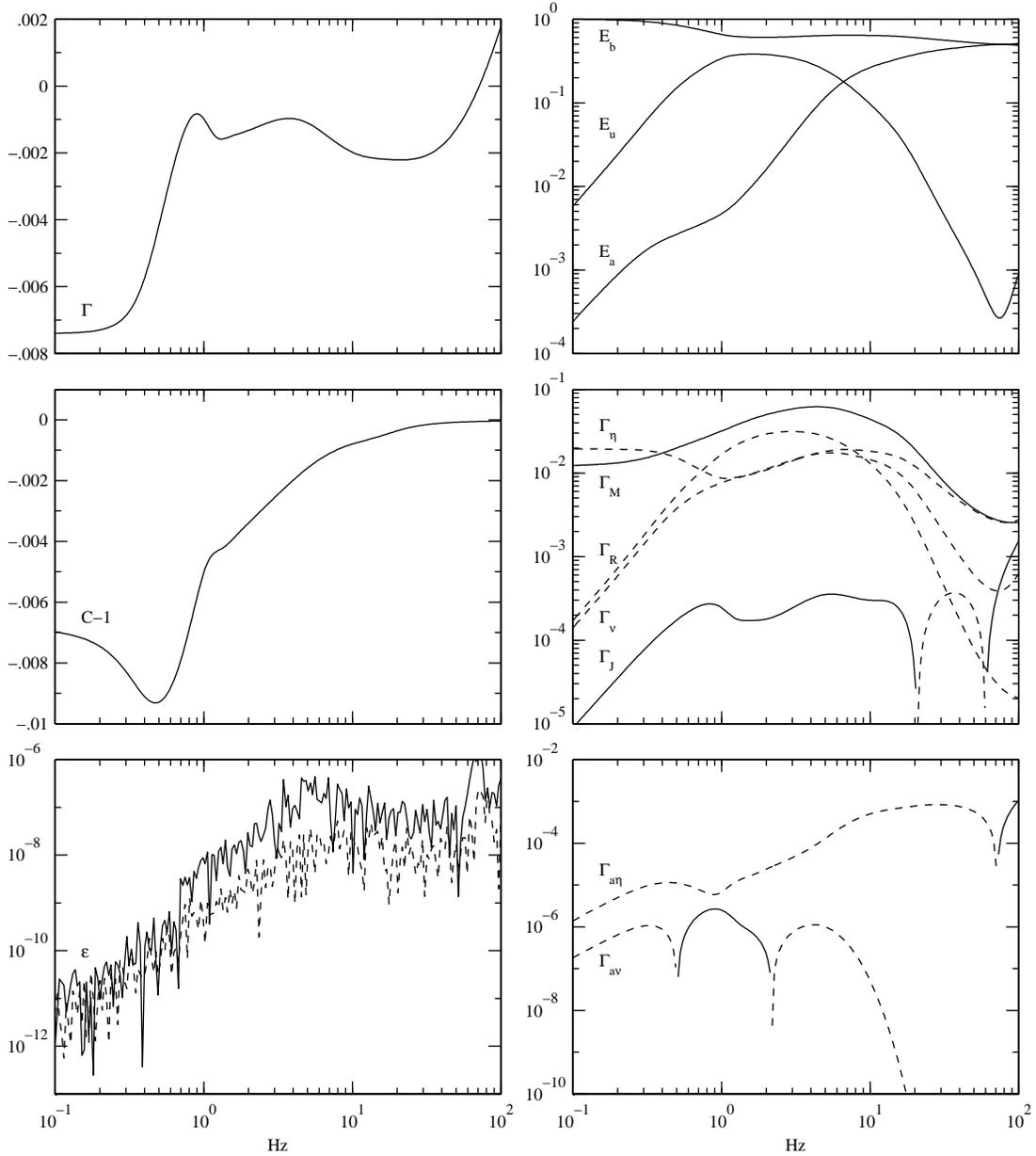}

\end{center}
\caption{\label{fig:gammaSumMhd}Energy balance for film MHD flow at $ \Rey = 10^4 $, $ \Prm = 1.2 $, $ \Prg = 1.1 \times 10^{-4} $, $ \Ohn = 3.14 \times 10^{-4} $, $ \alpha = 1 $, and $ p_u = p_b = 200 $. The flow-normal Hartmann number $ \Harz $ ranges from 0.1 to 100, with the streamwise Hartmann number given by $ \Harx = \Harz / \tan( 1^\circ ) $. The curves track the behavior of a single mode as a function of $ \Harz $, which at $ \Harz = 100 $ is Mode $ \mathrm{ M }_2 $ (see Fig.~\ref{fig:spectrumMhdFilmOblique} and Table~\ref{table:spectrumMhdFilmOblique}). The graphs in the right-hand panels show the energies~\eqref{eq:energySum}, normalized so that $ E = 1 $, and the power-transfer terms~\eqref{eq:gammaSum}. The solid (dotted) portions of the curves in the logarithmic plots correspond to positive (negative) values. The left-hand panels display the growth rate $ \Gamma $ and phase velocity $ C $, as well as the error $ \epsilon $. The latter has also been evaluated at $ p_u = p_b = 100$, and plotted as a dotted line, in order to illustrate the roundoff sensitivity in $ \Gamma_{a \eta } $.}   
\end{figure}

Fig.~\ref{fig:gammaSumMhd} shows the details of such a calculation for film MHD problems at $ \Prm = 1.2 $, with external magnetic field oriented at $ 1^\circ $ relative to the streamwise direction, and flow-normal Hartmann number $ \Harz \in [ 0.1, 100 ] $ (correspondingly, the streamwise Hartmann number $ \Harx $ ranges from approximately 5.73 to 5,730). As $ \Harz $ is varied, a single mode is continuously tracked, which corresponds to Mode~$ \mathrm{ M }_2 $ at $ \Harz = 100 $ (see Table~\ref{table:spectrumMhdFilmOblique} and Fig.~\ref{fig:spectrumMhdFilmOblique}). That mode is stable for sufficiently weak magnetic fields, but as $ \Harz $ increases it undergoes an instability in which the dominant power input is Maxwell stress (\cf the instabilities in non-MHD and low-$ \Prm $ flows caused by positive Reynolds stress). At the same time, the energy $ E $~\eqref{eq:energySum} changes from being predominantly magnetic to a nearly equal mix of magnetic and free-surface energies (at $ \Harz \sim 10 $ the energy is also seen to have a significant kinetic contribution). For all values of the Hartmann number considered, the error $ \epsilon $ remains small ($ \epsilon \lesssim 10^{-6} $), but displays  a trend with $ \Harz $ that mirrors $ \Gamma_{a \eta} $. We attribute this behavior to roundoff error in $ \Gamma_{a\eta} $ due to that term's dependence on $ \DD^2 b( 0 ) $. In fact, the reason that we chose to examine Mode~$ \mathrm{ M }_2 $, rather than, say, Mode~$ \mathrm{ M }_1 $, is that at sufficiently large Hartmann numbers the magnetic and surface energies of that mode are both appreciable, making it particularly susceptible to errors associated with $ \Gamma_{a\eta} $. Indeed, as the dotted line in the lower-left panel in Fig.~\ref{fig:gammaSumMhd} shows, decreasing $ p $ from 200 to 100 results to a noticeable change in $ \epsilon $, which diminishes roughly by an order of magnitude. On the other hand, modes with small $ | \Gamma_{a \eta } | $, are comparatively unaffected by the choice of $ p $ (\eg for Mode~$ \mathrm{ M }_1 $ $ \epsilon \sim 10^{-10} $ for both $ p = 100 $ and $ p = 200$, and for all $ \Harz \leq 100 $). It therefore appears that $ \epsilon $ is dominated by roundoff error in $ \Gamma_{a\eta} $, rather than some inconsistency in our numerical scheme and/or its implementation.     
  
As a further consistency check we have compared growth rates for the two most unstable modes of the results of Table~\ref{table:spectrumHydroFilm} at $ \Rey = 3 \times 10^{4} $ (see also Fig.~\ref{fig:spectrumHydroFilm}(b)) with the results of free-surface flows computed using a fully nonlinear Navier--Stokes solver. The Navier--Stokes code is based on the arbitrary Lagrangian--Eulerian spectral element code developed by Ho \cite{Ho89}, R{\o}nquist \cite{ronquist88}, and Fischer \cite{fischer97}.  For $\alpha=1$, the (nominal) computational domain was taken as $\Omega = [0,2\pi] \times [-1,0]$, which was tessellated with a $6 \times 10$ array of spectral elements. A uniform element distribution was used in the streamwise direction while a stretched distribution was used in the wall-normal direction. Near the wall, an element thickness of $\Delta z=\text{0.005}$ was used to resolve the boundary layer of the unstable eigenmodes. The polynomial order within each element was $N=13$ and third-order timestepping was used with $\Delta t = \text{0.00125}$.  The initial conditions corresponded to the base flow plus $\delta := 10^{-5}$ times the velocity eigenmode associated with mode $k$, with $k =1$ or 2, \ie the most and second-most unstable eigenmode for these particular flow conditions. The domain was stretched in the $z$ direction to accommodate the $\ord(\delta)$ surface displacement using transfinite interpolation \cite{gohall73}. The eigenmodes, which are defined only on $z=[-1,0]$, were mapped onto the nominal domain then displaced along with the mesh. The base flow was defined as $U(z) = 1-z^2$ over the deformed mesh. Mean growth rates were computed by monitoring the $L_2$-norm of the wall-normal velocity and defining $\tilde{\Gamma}(t) := \ln( ||w(t)||_2/||w(0)||_2 )$.  The error is again defined as $(\tilde{\Gamma}(t) - \Gamma)/\Gamma $, where $\Gamma$ is computed using the linearized code.  Aside from some initial transients, the error over $t=[\text{5,100}]$ was less than $10^{-5}$ for Mode~1 ($\Gamma =\text{0.007984943826437}$) and less than $5^{-3}$ for Mode~2 ($\Gamma=\text{0.000052447145102}$). These results provide independent confirmation of both the linear-stability and spectral-element based ALE codes.

\subsection{\label{sec:criticalReynolds}Critical-Parameter Calculations}

Our final set of calculations pertains to the critical parameters for the onset of instability in channel and film problems with flow-normal background magnetic field and Hartmann steady-state profiles. In channel problems, we seek the minimum (critical) Reynolds number $ \Reyc $, and the corresponding wavenumber $ \alphac $, for which the spectrum contains unstable modes, keeping the Hartmann number $ \Harz $ and, where applicable, the magnetic Prandtl number $ \Prm $ fixed. In film problems, we also constrain the Ohnesorge number and the gravitational Prandtl number, setting $ \Prg = 1.10 \times 10^{-4} $ and $ \Ohn = 3.14 \times 10^{-4} $. As stated in \S\ref{sec:boundaryConditions}, neither of $ \Harz $, $ \Prm $, $ \Prg $ and $ \Ohn $ depend on the velocity scale of the flow. Therefore, our formulation of the critical-parameter calculation leads, among other possibilities, to a determination of the minimum steady-state velocity at which the flow becomes linearly unstable when all of its remaining properties are held fixed. Other choices of parameter constraints may be appropriate depending on the particular application. In all cases, however, $ \Reyc $, $ \alphac $, and the corresponding modal phase velocity $ C_\mathrm{ c } $, can be obtained by solving a minimization problem for $ \Rey $, constrained by the eigenproblem~\eqref{eq:matrixWeakForm} and the normalization $ \norm{ \colvec{ v } }_{2} = 1 $. The numerical results presented in Tables~\ref{table:criticalHar} and~\ref{table:criticalPrm} were obtained in that manner, using Matlab's fmincon optimization solver to carry out the computations. 

\begin{table}
\renewcommand{\arraystretch}{1}
\caption{\label{table:criticalHar}Hartmann-number dependence of the critical Reynolds number $ \Reyc $, wavenumber $ \alphac $, and phase velocity $ C_\mathrm{ c } $ for the even mode in channel problems, and the hard and soft modes in film problems. The critical parameters have been computed in the inductionless limit and $ \Prm = 10^{-4 } $, for $ \Harx = 0 $ and $ \Harz \in [ 0, 100 ] $. $ N_u = N_b $ is the dimension of the velocity and magnetic-field solution spaces used in the calculations. The underlined digits in the results for the even channel mode indicate discrepancy from the corresponding calculations in Tables~1 and~3 of Ref.~\cite{Takashima96}. The soft mode's critical parameters in the inductionless limit were evaluated via Eqs.~\eqref{eq:criticalSoft}.}
\centering    
\begin{tabular*}{\linewidth}{@{\extracolsep{\fill}}llllllll}
\hline
& \multicolumn{3}{c}{Inductionless} & \multicolumn{3}{c}{$\Prm = 10^{-4}$}\\
\cline{2-4}\cline{5-7}
$ \Harz $ & $ \Reyc $ & $ \alphac $ & $ C_\mathrm{ c } $ & $ \Reyc $ & $ \alphac $ & $ C_\mathrm{ c } $ & $ N_u $/$ N_b $ \\
\hline
\multicolumn{7}{l}{Even channel mode}\\
$ 0 $ &$ 5.7722218\mathrm{E}+03 $ & $ 1.0205\underline{51}\mathrm{E}+00 $ & $ 2.640007\mathrm{E}-01 $ &$ 5.7722218\mathrm{E}+03 $ & $ 1.020539\mathrm{E}+00 $ & $ 2.639993\mathrm{E}-01 $ & 70 \\ 
$ 5 $ &$ 1.6408999\mathrm{E}+05 $ & $ 1.134248\mathrm{E}+00 $ & $ 1.564271\mathrm{E}-01 $ &$ 1.6372742\mathrm{E}+05 $ & $ 1.13420\underline{0}\mathrm{E}+00 $ & $ 1.565433\mathrm{E}-01 $ & 130 \\ 
$ 10 $ &$ 4.3981816\mathrm{E}+05 $ & $ 1.739136\mathrm{E}+00 $ & $ 1.547887\mathrm{E}-01 $ &$ 4.3861946\mathrm{E}+05 $ & $ 1.739025\mathrm{E}+00 $ & $ 1.549340\mathrm{E}-01 $ & 170 \\ 
$ 20 $ &$ 9.6176717\mathrm{E}+05 $ & $ 3.237635\mathrm{E}+00 $ & $ 1.550111\mathrm{E}-01 $ &$ 9.5885971\mathrm{E}+05 $ & $ 3.2373\underline{79}\mathrm{E}+00 $ & $ 1.551713\mathrm{E}-01 $ & 250 \\ 
$ 50 $ &$ 2.4155501\mathrm{E}+06 $ & $ 8.076565\mathrm{E}+00 $ & $ 1.5502\underline{95}\mathrm{E}-01 $ &$ 2.4078809\mathrm{E}+06 $ & $ 8.0759\underline{17}\mathrm{E}+00 $ & $ 1.551967\mathrm{E}-01 $ & 370 \\ 
$ 100 $ &$ 4.8311016\mathrm{E}+06 $ & $ 1.615311\mathrm{E}+01 $ & $ 1.5502\underline{95}\mathrm{E}-01 $ &$ 4.8155338\mathrm{E}+06 $ & $ 1.6151\underline{95}\mathrm{E}+01 $ & $ 1.551990\mathrm{E}-01 $ & 510 \\

% rotation is 20 degrees 
\\
\multicolumn{7}{l}{Hard Mode}\\
$ 0 $ &$ 9.8577335\mathrm{E}+03 $ & $ 2.861951\mathrm{E}+00 $ & $ 1.576040\mathrm{E}-01 $ &$ 9.8577335\mathrm{E}+03 $ & $ 2.861951\mathrm{E}+00 $ & $ 1.576040\mathrm{E}-01 $ & 70 \\ 
$ 5 $ &$ 1.6375039\mathrm{E}+05 $ & $ 1.136385\mathrm{E}+00 $ & $ 1.564194\mathrm{E}-01 $ &$ 1.6337134\mathrm{E}+05 $ & $ 1.136344\mathrm{E}+00 $ & $ 1.565384\mathrm{E}-01 $ & 130 \\ 
$ 10 $ &$ 4.3978705\mathrm{E}+05 $ & $ 1.739235\mathrm{E}+00 $ & $ 1.547884\mathrm{E}-01 $ &$ 4.3858882\mathrm{E}+05 $ & $ 1.739125\mathrm{E}+00 $ & $ 1.549336\mathrm{E}-01 $ & 170 \\ 
$ 20 $ &$ 9.6176640\mathrm{E}+05 $ & $ 3.237636\mathrm{E}+00 $ & $ 1.550111\mathrm{E}-01 $ &$ 9.5885899\mathrm{E}+05 $ & $ 3.237379\mathrm{E}+00 $ & $ 1.551713\mathrm{E}-01 $ & 250 \\ 
$ 50 $ &$ 2.4155501\mathrm{E}+06 $ & $ 8.076548\mathrm{E}+00 $ & $ 1.550295\mathrm{E}-01 $ &$ 2.4078809\mathrm{E}+06 $ & $ 8.075929\mathrm{E}+00 $ & $ 1.551967\mathrm{E}-01 $ & 370 \\ 
$ 100 $ &$4.8311016\mathrm{E}+06 $ & $ 1.615311\mathrm{E}+01 $ & $ 1.550295\mathrm{E}-01 $ &$ 4.8155338\mathrm{E}+06 $ & $ 1.615189\mathrm{E}+01 $ & $ 1.551990\mathrm{E}-01 $ & 510 \\

% rotation is 20 degrees 
\\
\multicolumn{7}{l}{Soft Mode}\\
$ 0 $ & $ 7.1869947\mathrm{E}+03 $ & 0& $ 2.000000\mathrm{E}+00 $ \\ 
$ 5 $ &$ 3.1818260\mathrm{E}+05 $ & $ 0$ & $ 1.013475\mathrm{E}+00 $ &$  6.7243983\mathrm{E}+05 $ & $ 1.447850\mathrm{E}-03 $ & $ 1.052189\mathrm{E}+00 $ & 300 \\ 
$ 10 $ &$ 3.2575755\mathrm{E}+07 $ & $ 0$ & $ 1.000091\mathrm{E}+00 $ &$ 1.0019578\mathrm{E}+05 $ & $ 3.579587\mathrm{E}-03 $ & $ 1.015528\mathrm{E}+00 $ & 300 \\ 
$ 20 $ &$ 4.9973860\mathrm{E}+11 $ & $ 0$ & $ 1.000000\mathrm{E}+00 $ &$ 1.5359746\mathrm{E}+05 $ & $ 8.979436\mathrm{E}-03 $ & $ 1.006938\mathrm{E}+00 $ & 300 \\ 
$ 50 $ &$ 3.3499906\mathrm{E}+24 $ & $ 0$ & $ 1.000000\mathrm{E}+00 $ &$ 2.7908921\mathrm{E}+05 $ & $ 3.014297\mathrm{E}-02 $ & $ 1.002347\mathrm{E}+00 $ & 300 \\ 
$ 100 $ &$ 1.2249687\mathrm{E}+46 $ & $ 0$ & $ 1.000000\mathrm{E}+00 $ &$ 4.4089365\mathrm{E}+05 $ & $ 7.382278\mathrm{E}-02 $ & $ 1.000980\mathrm{E}+00 $ & 300 \\

% rotation is 20 degrees 
\hline
\end{tabular*}
\end{table}

\begin{table}
\renewcommand{\arraystretch}{1}
\caption{\label{table:criticalPrm}Critical Reynolds number $ \Reyc $, wavenumber $ \alphac $, and phase velocity $ C_\mathrm{ c } $ for channel and film MHD problems at $ ( \Harx, \Harz ) = ( 0, 10 ) $ and  $ \Prm \in [ 10^{-8}, 10^{-4} ] $. All calculations were performed at dimension $ N_u = N_b = 300 $. The underlined digits in the channel-problem calculations differ from the corresponding ones in Table~3 of Ref.~\cite{Takashima96}.}
\centering
\begin{tabular*}{\linewidth}{@{\extracolsep{\fill}}lllllll}
\hline
& \multicolumn{3}{c}{Channel} & \multicolumn{3}{c}{Film}\\
\cline{2-4}\cline{5-7}
$ \Prm $ & $ \Reyc $ & $ \alphac $ & $ C_\mathrm{ c } $ & $ \Reyc $ & $ \alphac $ & $ C_\mathrm{ c } $ \\
\hline
$ 1.0\mathrm{E}-08 $ &$ 4.3981789\mathrm{E}+05 $ & $ 1.739135\mathrm{E}+00 $ & $ 1.547887\mathrm{E}-01 $ &$ 4.3978679\mathrm{E}+05 $ & $ 1.739235\mathrm{E}+00 $ & $ 1.547884\mathrm{E}-01 $ \\ 
$ 1.0\mathrm{E}-07 $ &$ 4.3981547\mathrm{E}+05 $ & $ 1.7391\underline{35}\mathrm{E}+00 $ & $ 1.547889\mathrm{E}-01 $ &$ 4.3978440\mathrm{E}+05 $ & $ 1.739234\mathrm{E}+00 $ & $ 1.547886\mathrm{E}-01 $ \\ 
$ 1.0\mathrm{E}-06 $ &$ 4.3979162\mathrm{E}+05 $ & $ 1.739128\mathrm{E}+00 $ & $ 1.547912\mathrm{E}-01 $ &$ 4.3976064\mathrm{E}+05 $ & $ 1.739228\mathrm{E}+00 $ & $ 1.547909\mathrm{E}-01 $ \\ 
$ 1.0\mathrm{E}-05 $ &$ 4.3958738\mathrm{E}+05 $ & $ 1.739141\mathrm{E}+00 $ & $ 1.548125\mathrm{E}-01 $ &$ 2.1862147\mathrm{E}+05 $ & $ 1.947639\mathrm{E}-03 $ & $ 1.004972\mathrm{E}+00 $ \\ 
$ 1.0\mathrm{E}-04 $ &$ 4.3861946\mathrm{E}+05 $ & $ 1.7390\underline{24}\mathrm{E}+00 $ & $ 1.549340\mathrm{E}-01 $ &$ 1.0019578\mathrm{E}+05 $ & $ 3.579587\mathrm{E}-03 $ & $ 1.015528\mathrm{E}+00 $ \\ 
$ 1.0\mathrm{E}-03 $ &$ 4.2969213\mathrm{E}+05 $ & $ 1.739870\mathrm{E}+00 $ & $ 1.559585\mathrm{E}-01 $ &$ 6.1108706\mathrm{E}+04 $ & $ 2.914764\mathrm{E}-03 $ & $ 1.020771\mathrm{E}+00 $ \\ 
$ 1.0\mathrm{E}-02 $ &$ 4.8282141\mathrm{E}+04 $ & $ 4.894029\mathrm{E}-03 $ & $ 8.973103\mathrm{E}-01 $ &$ 4.6367226\mathrm{E}+04 $ & $ 9.120983\mathrm{E}-04 $ & $ 1.089575\mathrm{E}+00 $ \\ 
$ 1.0\mathrm{E}-01 $ &$ 6.8382770\mathrm{E}+02 $ & $ 2.788195\mathrm{E}-01 $ & $ 8.899146\mathrm{E}-01 $ &$ 1.1205597\mathrm{E}+03 $ & $ 1.869298\mathrm{E}-01 $ & $ 8.803851\mathrm{E}-01 $ \\

% rotation is 20 degrees 
\hline
\end{tabular*}
\end{table}

We begin from channel and film problems in the inductionless limit, critical parameters of which are listed for $ \Harz \in [ 0, 100 ] $ in the left-hand portion of Table~\ref{table:criticalHar}. In the channel case, the critical mode is always of even symmetry and lies in the A branch of the spectrum ($ C_\mathrm{ c } < \langle U \rangle ) $. As indicated by the underlined digits in the computed parameter values, our calculations are in excellent agreement with those by Takashima~\cite{Takashima96}. Even though channel problems also exhibit an odd unstable mode, its critical Reynolds number always that of the even one \cite{Lock55}, and therefore we do not consider it here. On the other hand, in film problems either the soft or the hard mode, respectively characterized by $ C < \langle U \rangle $ and $ C > 1 $ (see \S\ref{sec:eigenvalueSpectra}), can have the smallest critical Reynolds number, depending on the Hartmann number and the free-surface parameters. The critical wavenumber of the soft mode is zero, and in light of this it is possible to derive the closed-form expressions
\begin{equation}
\label{eq:criticalSoft}
\Reyc  = \frac{ 2^{3/2}  \sinh( \Harz / 2 ) ( \Har - \tanh( \Harz ) )^{1/2} }{ \Prg ( \Harz  \coth( \Harz / 2 ) \sech^3( \Harz ) ( 2 \Harz ( 2 + \cosh( 2 \Harz ) ) - 3 \sinh( 2 \Harz ) ) )^{1/2} }, \quad
C_\mathrm{ c } = 1 + \sech( \Harz ),
\end{equation}
using regular perturbation theory about $ \alpha = 0 $ \cite{GiannakisRosnerFischer07}. In the $ \Harz \ttz $ limit, Eqs.~\eqref{eq:criticalSoft} reduce to $ \Reyc = (5/8)\sh/\Prg $ and $ C_\mathrm{c} = 2 $. At $ \Harz > 0 $ the critical Reynolds number grows  exponentially, implying that for all but small Hartmann numbers the onset of instability in inductionless film problems is governed by the hard mode. Direct numerical calculations of the soft mode's critical parameters rapidly become intractable, but we checked that the linear-stability code yields results of $ \ord( 10^{-8} ) $ accuracy for $ \Harz \leq 10 $. As for the hard mode, the inductionless results in Table~\ref{table:criticalHar} show that apart from small Hartmann numbers, where gravity and surface tension are more important than the magnetic field, its critical parameters are very close to those of the channel mode, suggesting that for sufficiently strong magnetic fields the free surface only plays a minor role in the hard instability. The critical wavenumber of the channel and hard modes increases with $ \Harz $ (\ie shorter wavelengths become unstable first), which is consistent with the fact that the decreasing Hartmann-layer thickness is the principal contributing factor in the instability suppression \cite{Lock55}.   

We now examine how nonzero magnetic Prandtl numbers modify the picture in the inductionless limit. As can be checked from Table~\ref{table:criticalHar}, the error in $ \Reyc $ incurred by making the inductionless approximation is less than $ 4 \times 10^{-3} $ for the even channel mode over all Hartmann numbers probed. Moreover, according to Table~\ref{table:criticalPrm}, at $ \Harz = 10 $ the critical Reynolds number of the even channel mode decreases by a factor smaller than $ 0.003 $ when $ \Prm $ is increased from $ 10^{-8} $ to $ 10^{-4} $. These calculations are in very good agreement  with the corresponding ones by Takashima, and are illustrative of the weak dependence of critical parameters of the channel mode on $ \Prm < 10^{-4} $ for all $ \Harz < 200 $ \cite{Takashima96}. As $ \Prm $ grows above $ \ord( 10^{-4} ) $, the accuracy of the inductionless approximation progressively deteriorates, until the critical mode undergoes a bifurcation to a magnetic mode (\ie a singular mode in the limit $ \Prm \ttz$) of odd symmetry, manifested by the sharp decrease in $ \alphac $ at $ \Prm = 10^{-2} $. 

Turning to film problems, Table~\ref{table:criticalHar} demonstrates that, as with the even channel mode, at small $ \Prm $  the inductionless approximation yields accurate results for the critical parameters of the hard mode. On the other hand, the data clearly show that a small, but nonzero, $ \Prm $ affects profoundly the critical parameters of the soft mode. In particular, the previously encountered exponential growth of $ \Reyc $ with $ \Harz $ becomes suppressed to the point that it now trails the hard mode's critical Reynolds number by a wide margin. In the right-hand portion of Table~\ref{table:criticalPrm}  the hard mode ($ C_\mathrm{c}<1$) is seen to govern the onset of instability for $ \Prm \lesssim 10^{-6} $, with the soft one, characterized by $C_\mathrm{ c } > 1 $, taking over at larger magnetic Prandtl numbers. Even though no further bifurcations occur for $ \Prm \in [ 10^{-6}, 10^{-2} ] $, the soft mode in itself appears to be sensitive to $ \Prm $. In the context of large-wavelength perturbation theory, this behavior can be traced to the coefficient $ \gamma_1 $ in the asymptotic expansion $ \gamma = \alpha \gamma_1 + \alpha^2 \gamma_2 + \ord( \alpha^3 ) $, which vanishes as $ \Prm \ttz$. In consequence, the equation $ \gamma_1 + \alpha \gamma_2 = 0 $, used to determine $ \Reyc $, becomes singularly perturbed, resulting to the observed sensitivity. From a physics standpoint, as already mentioned in \S\ref{sec:MhdProblems} and discussed in further detail in~\cite{GiannakisRosnerFischer07}, the problems with nonzero $ \Prm $ are susceptible to doubly diffusive effects, giving rise to instabilities not present in the inductionless limit. In total, over the interval $ 10^{-8} \leq \Prm \leq 10^{-4} $, which roughly coincides with the $ \Prm $ values of terrestrial incompressible fluids, the critical Reynolds number of the examined film problems decreases by almost a factor of five.

\section{\label{sec:conclusions}Conclusions}

In this paper we have presented a spectral Galerkin method for linear-stability problems in free-surface MHD. The method is essentially an extension of the scheme developed by Kirchner \cite{Kirchner00}, and Melenk, Kirchner and Schwab~\cite{MelenkKirchnerSchwab00} to solve the Orr--Sommerfeld (OS) equation for plane Poiseuille flow. Besides free-surface MHD problems, which we refer to as \emph{film MHD problems} (Def.~\ref{def:filmMhd}), our scheme provides a unified framework to solve MHD stability problems with fixed boundaries---the so-called \emph{channel MHD problems} (Def.~\ref{def:channelMhd})---and their simplified versions at vanishing magnetic Prandtl number $ \Prm $, which we refer to as \emph{inductionless film and channel problems} (Defs.~\ref{def:filmZeroPm} and~\ref{def:channelZeroPm}). We studied problems with either the Poiseuille velocity profile, or the Hartmann velocity and magnetic-field profiles, both of which are physically motivated. However, our schemes are applicable to arbitrary analytic steady-state profiles. In all cases, the Galerkin discretization results to the matrix generalized eigenvalue problem $ \mat{ K } \colvec{ v } = \gamma \mat{ M } \colvec{ v } $, where $ \mat{ K } $ and $ \mat{ M } $ are respectively the stiffness and mass matrices, $ \gamma $ is the complex growth rate, and $ \colvec{ v } $ is a column vector containing the problem's degrees of freedom. We detected no spurious eigenvalues, a fact which we attribute to the non-singularity of $ \mat{ M } $ in all bases.

We discretized the solution spaces for the velocity and magnetic-field eigenfunctions using Legendre internal shape functions and nodal shape functions, chosen according to the Sobolev spaces of the continuous problems. Separating the basis polynomials into internal and nodal ones facilitates the natural (weak) impositition of the boundary conditions for free-surface MHD, namely the stress and kinematic conditions at the free surface, and the Robin-type insulating boundary conditions for the magnetic field. The orthogonality properties of the bases guarantee that roundoff error is independent of the spectral order $ p $, allowing one to work at the large spectral orders ($ p > 500$) required to resolve the small length scales present at high Reynolds numbers $ \Rey $. Moreover, in problems with polynomial velocity and magnetic-field profiles, $ \mat{ K } $ and $ \mat{ M } $ are sparse, and iterative solvers can be used to compute $ \gamma $ and $ \colvec{ v } $ efficiently.
 
The optimal conditioning of our schemes with respect to $ p $ alleviates only marginally their roundoff sensitivity due to non-normality of the stability operators. At around $ \Rey = 4 \times 10 ^ 4 $ we observed the formation of the characteristic diamond shaped pattern on the complex eigenvalue plane caused by lack of sufficient precision in 64-bit arithmetic. An alternative discretization, performed in terms of Lagrange interpolation polynomials, was found to give rise to the pattern at only slightly smaller Reynolds numbers ($ \Rey = 3 \times 10^4 $, which is close to the value reported in~\cite{DongarraStraughanWalker96} for a Chebyshev tau scheme), despite the ill conditioning of the Lagrange interpolant basis. Roundoff errors associated to non-normality therefore appear to be governed by physical parameters, rather than the details of the discretization scheme. Working in extended-precision (\eg 128-bit) arithmetic is probably the only way to address this type of error, but, at the time of writing, that option could not be implemented with our Matlab code.

We described two ways of addressing the presence of exponentially weighted sesquilinear forms in problems with Hartmann steady-state profiles. In the first approach, the forms are evaluated without incurring quadrature error by means of the algorithm developed by Mach \cite{Mach84} to compute Gauss quadrature knots and weights for exponential weight functions on a finite interval. The second approach involves replacing the forms by approximate ones derived from Legendre--Gauss--Lobatto (LGL) quadrature rules at the $ 2 p - 1 $ precision level. The latter has been established by Banerjee and Osborn \cite{BanerjeeOsborn90} as sufficient to guarantee stability and convergence in elliptical eigenvalue problems, but, to our knowledge, no corresponding bound exists for OS problems. We found that eigenvalues computed via the LGL method agree to within roundoff error with the corresponding ones obtained using exact quadrature, indicating that a version of Banerjee and Osborn's theorem should also be applicable in eigenvalue problems of the OS type.   

As an independent consistency check, we compared modal growth rates in non-MHD free-surface flow to energy growth rates in fully nonlinear  simulations. At $ \Rey = 3 \times 10^4 $ and wavenumber $ \alpha = 1 $ we found that the error over 100 convective times is less than $ 10 ^ {-5} $ and $ 5 ^{ - 3 } $, respectively for the first and second least stable modes. We also compared modal growth rates in problems with oblique external magnetic field to the corresponding ones derived from an energy conservation law for free-surface MHD. Here the error was found to be less than $ 10^{-6 } $, with its largest portion attributed to roundoff sensitivity in the calculation of one of the energy terms, rather than inconsistencies in the numerical scheme. In channel problems, we found that our results for the critical Reynolds number, wavenumber, and phase velocity for Hartmann flow agree very well with the corresponding ones by Takashima~\cite{Takashima96}, obtained using a Chebyshev tau method.

At the magnetic Prandtl number regime of terrestrial fluids ($ \Prm \lesssim 10^{-4} $) and for $ \Harz \geq 5 $, the critical parameters of the hard instability mode in film flow were found to be close to those of the even critical mode in channel flow. Increasing $ \Prm $ from $ 10^{-8} $ to $ 10^{-4} $ at $ \Harz = 10 $ resulted to a mild, $ \ord( 10^{-3} ) $, decrease of the critical Reynolds number $ \Reyc $ for the hard and channel modes, but $ \Reyc $ dropped by more than a factor of four for the soft (surface) instability mode in film flow. A surface-wave instability at small $ \Prm $, but absent in the inductionless limit, was also observed in the spectra of film MHD problems at $ \alpha \Rey = 10^{4} $, $ \Harz = 10 $ and $ \Prm = 10^{-4} $. These results are indicative of the important role played by the working fluid's magnetic Prandtl number  in the stability of industrial and laboratory free-surface flows. In test problems at $ \Prm = 1.2 $ we observed that increasing $ \Harz $ from zero to 100 leads to the formation of multiple branches on the complex-eigenvalue plane. Unlike problems at small magnetic Prandtl numbers, the spectra at $ \Prm = \ord( 1 ) $ contain unstable magnetic modes, two of which were recorded in a film MHD problem with oblique external magnetic field.  
 
Before closing, we note a number of directions for future work. On the analytical side, it would be highly desirable to extend the convergence analysis of Melenk \etal~\cite{MelenkKirchnerSchwab00} to free-surface MHD. Even though the calculations presented in \S\ref{sec:numericalResults} provide strong numerical evidence that our schemes are stable and convergent, their well-posedness cannot be settled without a rigorous analytical backing. Similarly, our proposed method in \S\ref{sec:baseUBMatrices} of approximating weighted sesquilinear forms using LGL quadrature requires an adaptation of Banerjee and Osborn's \cite{BanerjeeOsborn90} work to OS-type eigenvalue problems. A further analytical objective would be to generalize the criterion of scale resolution \cite{MelenkKirchnerSchwab00}, which provides an estimate of the minimum spectral order required to achieve convergence at a given $ \Rey $ in non-MHD channel flow. Of course, any such criterion would have to be numerically tested. On the physics side, our discussion in \S\ref{sec:eigenvalueSpectra} and \S\ref{sec:criticalReynolds}, which is mostly phenomenological, should be supplemented by a study of the operating physical mechanisms. In \cite{GiannakisRosnerFischer07}, we pursue such a study at the low-$ \Prm $ regime, but that should be extended to cover $ \Prm = \ord( 1 ) $ flows, which have been conjectured \cite{BalbusHenri07} to be relevant in certain astrophysical accretion phenomena. 

\section*{Acknowledgments}

We thank H.~Ji and M.~Nornberg for useful conversations. This work was supported by the Mathematical, Information, and Computational Science Division subprogram of the Office of Advanced Scientific Computing Research, and by the Office of Fusion Energy Sciences (Field Work Proposal No.~25145), Office of Science, U.S. Department of Energy, under Contract DE-AC02-0611357. D.~G.~acknowledges support from the Alexander S.~Onassis Public Benefit Foundation.

\appendix

\section{\label{app:innerProducts}Matrix Representations of the Schemes' Forms and Maps}

In this Appendix we provide expressions for the matrix representations of the sesquilinear forms and maps used in the main text. In \S\ref{app:tInternal}--\S\ref{app:tCross} we consider the $ \mat{ T } $ matrices, defined in~\eqref{eq:matricesT} and~\eqref{eq:matricesTCross},  whose elements can be stably evaluated in closed form by means of the orthogonality properties of the Legendre polynomials \eqref{eq:legendreOrthogonality}. We then describe, in \S\ref{app:hartmann}, how Mach's quadrature scheme~\cite{Mach84} can be used to evaluate the matrices $ \mat{ S } $~\eqref{eq:matS} and $ \mat{ C } $. 

\subsection{\label{app:tInternal}The Matrices $ \mat{ T }_{ H^r_0 }^{[k d_1 d_2]} $}

We evaluate the matrices $ \mat{ T }_{ H^r_0 }^{[ k d_1 d_2]} \in \mathbb{ R }^{ N \times N }$~\eqref{eq:tHr0} listed in Table~\ref{table:tHr0}. In the rightmost column of that table 0 stands for the main diagonal and $ m $ ($ - m $) represents the $ m$-th upper (lower) diagonal. Also, in the third column from the left, S and A respectively identify symmetric and antisymmetric matrices. For each $ k $ and $ r $, one only needs to evaluate the cases $ ( d_1, d_2 ) = ( 0, 0 ) $ and $ ( 1, 0 ) $, since the results for the remaining values of $ d_1 $ and $ d_2 \leq r $ follow by making use of the hierarchical relation~\eqref{eq:hierarchicalT} and the property $\mat{ T }_{ H^r_0 }^{ [ k d_1 d_2 ] } = \left(  \mat{ T }_{ H^r_0 }^{ [ k d_2 d_1 ] } \right)^\mathrm{T}$. Some of the results below can also be found in the paper by Kirchner \cite{Kirchner00}. However, since that reference contains a number of typographical errors, and for the sake of completeness, we have opted to reproduce them here. 
 
\begin{table}
\centering
\caption{\label{table:tHr0}Properties of the matrices $ \mat{ T }^{[ k d_1 d_2 ]}_{H^r_0} $}
\begin{tabular*}{\linewidth}{@{\extracolsep{\fill}}lllll}
\hline
$ r $ & $ [ k d_1 d_2 ] $ & Symmetry & Bandwidth & Nonzero Diagonals \\
\hline
0 & 0 0 0 & S & 0 & 0  \\

1 & 0 0 0 & S & 2 & 0, $ \pm 2 $  \\
1 & 1 0 0 & S & 3 & $ \pm 1 $, $ \pm 3 $  \\ 
1 & 2 0 0 & S & 4 & 0, $ \pm 2 $, $ \pm 4 $  \\

2 & 0 0 0 & S & 4 & 0, $ \pm 2 $, $ \pm 4 $  \\
2 & 0 1 0 & A & 3 & $ \pm 1 $, $ \pm 3 $  \\
2 & 1 0 0 & S & 5 & $ \pm 1 $, $ \pm 3 $, $ \pm 5 $  \\
2 & 1 1 0 & N/A & 4 & 0, $ \pm 2 $, $ \pm 4 $  \\ 
2 & 2 0 0 & S & 6 & 0, $ \pm 2 $, $ \pm 4 $, $ \pm 6 $ \\
\hline
\end{tabular*}
\end{table}

Working down the rows of Table~\ref{table:tHr0}, our first result, which has already been stated in~\eqref{eq:orthogonalityLambda}, is simply $\mat{ T }_{ H^0_0}^{[ 0 0 0 ]} = \mat{ I }_{ N }$. Next, we consider the nonzero elements in the main and upper diagonals of the matrices with $ r = 1 $, all of which are symmetric. These are 
\begin{equation}
% checked for typos
\left[ \mat{ T }^{[ 0 0 0 ]}_{H^1_0} \right]_{m n } = 
\left\{ 
\begin{split}
m & = n - 2: & - \frac{ 1 }{ \sqrt{ ( 2 n - 3 ) } ( 2 n - 1 ) \sqrt{ 2 n + 1 } }, \\
m & = n: & \frac{ 1 }{ 2 n + 1 } \left( \frac{ 1 }{ 2 n - 1 } + \frac{ 1 }{ 2 n + 3 } \right),
\end{split}
\right.
\end{equation}
\begin{multline}
% Checked for typos
\left[ \mat{ T }^{[ 1 0 0 ]}_{H^1_0} \right]_{m n } = \\
\left\{
\begin{split} 
m = n - 3 : & - \frac{ n - 1 }{ \sqrt{ 2 n - 5 }( 2 n - 3 )( 2 n - 1 ) \sqrt{ 2 n + 1 } }, \\
m = n - 1 : &  \frac{ 1 }{ \sqrt{ ( 2 n - 1 )( 2 n + 1 ) } } \left( \frac{ n - 1 }{ ( 2 n - 1 )( 2 n - 3 ) } - \frac{ n }{ ( 2 n - 1 )( 2 n + 1 ) } + \frac{ n + 1 }{ ( 2 n + 1 )( 2 n + 3 ) } \right),
\end{split}
\right.
\end{multline}
and
\begin{multline}
% Checked for typos
\left[ \mat{ T }^{[ 2 0 0 ]}_{H^1_0} \right]_{m n } = \\
\left\{
\begin{split}
m = n - 4 : & - \frac{ ( n - 1 )( n - 2 ) }{ \sqrt{ 2 n - 7 } ( 2 n - 5 )( 2 n - 3 )( 2 n - 1 ) \sqrt{ 2 n + 1 } }, \\
m = n - 2 : & \frac{ 1 }{ \sqrt{ ( 2 n - 3 )( 2 n + 1 ) } } \left( \frac{ ( n - 2 )( n - 1 ) }{ ( 2 n - 5 )( 2 n - 3 )( 2 n - 1 ) } - \frac{ 1 }{ ( 2 n - 1 )( 2 n + 1 ) } \left( \frac{ 2 ( n - 1 )^2 }{ ( 2 n - 3 ) } + 1 \right) \right. \\
& \quad \left. + \frac{ n ( n + 1 ) }{ ( 2 n - 1 )( 2 n + 1 )( 2 n + 3 ) } \right), \\
m = n : & \frac{ 1 }{ ( 2 n + 1 ) } \left( \frac{ 1 }{ ( 2 n - 1 )( 2 n + 1 ) } \left( \frac{ 2 ( n - 1 )^2 }{ ( 2 n - 3 ) } + 1 \right) - \frac{ 2 n ( n + 1 ) }{ ( 2 n - 1 )( 2 n + 1 )( 2 n + 3 ) } \right. \\
& \quad \left. + \frac{ 1 }{ ( 2 n + 3 )( 2 n + 5 ) } \left( \frac{ 2 ( n + 1 )^2 }{ 2 n + 1 } + 1 \right) \right).
\end{split}
\right.
\end{multline}
Among the $ r = 2 $ matrices, $ \mat{ T }^{[ 0 0 0 ]}_{H^2_0} $, $ \mat{ T }^{[ 1 0 0 ]}_{H^2_0} $ and $ \mat{ T }^{[ 2 0 0 ]}_{H^2_0} $ are symmetric, $ \mat{ T }^{[ 0 1 0 ]}_{H^2_0} $ is antisymmetric, and $ \mat{ T }^{[ 1 1 0 ]}_{H^2_0} $ has no symmetry property. In Ref.~\cite{Kirchner00}, the matrix corresponding to $  \mat{ T }^{[ 2 0 0 ]}_{H^2_0} $ is denoted by $ T^*_6 $. An expression for $ T^*_6 $ is provided in that paper's Appendix~B.6, but contains typographical errors. In Eq.~\eqref{eq:t200} ahead we indicate the erroneous terms, and also an error in the second upper diagonal (which we have been unable to trace to individual terms), by underlines. The nonzero elements of the symmetric and antisymmetric matrices, again in their and main and upper diagonals, are
\begin{multline}
% Checked for typos
\left[ \mat{ T }^{[ 0 0 0 ]}_{H^2_0} \right]_{ m n } = \\ 
\left\{
\begin{split}
m = n - 4 : & \frac{ 1 }{ \sqrt{ 2 n - 5 }( 2 n - 3 )( 2 n - 1 )( 2 n + 1 )\sqrt{ 2 n + 3 } }, \\
m = n - 2 : & \frac{ - 1 }{ \sqrt{ 2 n - 1 } ( 2 n + 1 )\sqrt{ 2 n + 3 } } \left( \frac{ 1 }{ ( 2 n + 3 )( 2 n + 5 ) } + \frac{ 1 }{ ( 2 n + 1 )( 2 n + 3 ) } + \frac{ 1 }{ ( 2 n - 1 )( 2 n + 1 ) } \right. \\
& \quad \left. + \frac{ 1 }{ ( 2 n - 1 )( 2 n - 3 ) } \right), \\
m = n : & \frac{ 1 }{ 2 n + 3 } \left( \frac{ 1 }{ ( 2 n + 5 )^2 ( 2 n + 7 ) } + \frac{ 1 }{ ( 2 n +3 )( 2 n + 5 )^2 } + \frac{ 2 }{ ( 2 n + 1 )( 2 n + 3 )( 2 n + 5 ) } \right. \\
& \quad \left. +  \frac{ 1 }{ ( 2 n + 1 )^2( 2 n + 3 ) } + \frac{ 1 }{ ( 2 n + 1 )^2( 2 n - 1 ) } \right),
\end{split}
\right.
\end{multline}
\begin{multline}
\left[ \mat{ T }^{[ 0 1 0 ]}_{H^2_0} \right]_{ m n } = \\
\left\{
\begin{split}
m = n - 3 : & \frac{ 1 }{ \sqrt{ 2 n - 3 }( 2 n - 1 )( 2 n + 1 )\sqrt{ 2 n + 3 } }, \\
m = n - 1 : & - \frac{ 1 }{ \sqrt{ ( 2 n + 1 )( 2 n + 3 ) } } \left( \frac{ 1 }{ ( 2 n - 1 )( 2 n + 1 ) } + \frac{ 1 }{ ( 2 n + 1 )( 2 n + 3 ) } + \frac{ 1 }{ ( 2 n + 3 )( 2 n + 5 ) } \right),
\end{split}
\right.
\end{multline}
\begin{multline}
\left[ \mat{ T }^{[ 1 0 0 ]}_{H^2_0} \right]_{ m n } = \\
\left\{
\begin{split}
m = n - 5 : & \frac{n - 1}{\sqrt{2n - 7}(2n - 5)(2n - 3)(2n - 1)(2n + 1)\sqrt{2n + 3}}, \\
m = n - 3 : & \frac{ - 1 }{ \sqrt{ 2 n - 3 } ( 2 n - 1 )( 2 n + 1 )\sqrt{ 2 n + 3 } } \left( \frac{ n - 1 }{ ( 2 n -5 )( 2 n - 3 ) } + \frac{ n - 1 }{ ( 2 n - 3 )( 2 n - 1 ) } \right. \\
& \quad \left. - \frac{ n  }{ ( 2 n - 1 )( 2 n + 1 ) }  + \frac{ n + 1 }{ ( 2 n + 1 )( 2 n + 3 ) } + \frac{ n + 1 }{ ( 2 n + 3 )( 2 n + 5 ) } \right), \\
m = n - 1 : & \frac{ 1 }{ \sqrt{ ( 2 n + 1 )( 2 n + 3 ) } } \left( \frac{ n - 1 }{ ( 2 n - 3 )( 2 n - 1 )^2( 2 n + 1 ) } - \frac{ 2 n }{ ( 2 n -1 )^2 ( 2 n + 1 )( 2 n + 3 ) } \right. \\
& \quad \left. + \frac{ 4 ( n + 1 ) }{ ( 2 n - 1 )( 2 n + 1 )( 2 n + 3 )( 2 n + 5 ) } - \frac{ 2 ( n + 2 ) }{ ( 2 n + 1 )( 2 n + 3 )( 2 n + 5 )^2 } \right. \\
& \quad \left. + \frac{ n + 3 }{ ( 2 n + 3 )( 2 n + 5 )^2 ( 2 n + 7 ) } \right) ,  
\end{split}
\right.
\end{multline}
\begin{multline}
\label{eq:t200}
% checked for typos
\left[ \mat{ T }^{[ 2 0 0 ]}_{H^2_0} \right]_{ m n } = \\
\left\{
\begin{split}
m = n - 6: & \frac{ ( n - 2 )( n - 1 ) }{ \sqrt{ 2 n - 9 }( 2 n - 7 )( 2 n - 5 )( 2 n - 3 )( 2 n - 1 )( 2 n + 1 ) \sqrt{ 2 n + 3 } }, \\
m = n - 4: & \frac{ 1 }{ \sqrt{ 2 n - 5 }( 2 n - 3 )( 2 n - 1 )( 2 n + 1 ) \sqrt{ 2 n + 3 } } \left( - \frac{ n ( n + 1 ) }{ ( 2 n + 3 )( 2 n + 5 ) } - \frac{ n ( n + 1 ) }{ ( 2 n + 1 )( 2 n + 3 ) } \right. \\
& \quad \left.  + \frac{ n^2 }{ ( 2 n - 1 )( 2 n + 1 ) } + \frac{ ( n - 1 )^2 }{ ( 2 n - 3 )( 2 n - 1 ) } - \frac{ ( n - 2 )( n - 1 ) }{ ( 2 n - 5 )( 2 n - 3 ) } - \frac{ ( n - 2 )( n - 1 ) }{ ( 2 n - 7 )( 2 n - 5 ) } \right),\\
\underline{ m = n - 2 } : & \frac{ 1 }{ \sqrt{ 2 n - 1 }( 2 n + 1 ) \sqrt{ 2 n + 3 } } \left( \frac{ ( n + 2 )( n + 3 ) }{ ( 2 n + 3 )( 2 n + 5 )^2( 2 n + 7 ) } - \frac{ 2 }{ ( 2 n + 1 )( 2 n + 5 )^2 } \left( \frac{ 2 ( n + 1 )^2 }{ 2 n + 1 } + 1 \right)  \right. \\
& \quad \left. + \frac{ n ( n + 1 ) }{ ( 2n - 1 )( 2 n + 3 ) }  \left( \frac{ 1 }{ ( 2 n + 1 )( 2 n + 5 ) } + \frac{ 1 }{ ( 2 n - 3 )( 2 n + 5 ) } + \frac{ 2 }{ ( 2 n + 1 )^2 }  \right. \right. \\
& \quad \left. \left. + \frac{ 1 }{ ( 2 n + 1 )( 2 n - 3 ) } \right) - \frac{ 2 }{ ( 2 n - 3 )(2 n + 1 )^2 } \left( \frac{ 2 ( n - 1 )^2 }{ 2 n - 3 } + 1 \right) \right. \\
& \quad \left. + \frac{ ( n - 1 )( n - 2 ) }{ ( 2 n - 5 )( 2 n - 3 )^2 ( 2 n - 1 ) } \right), \\
m = n : & \frac{ 1 }{ 2 n + 3 } \left( \frac{ ( n + 4 )^2 }{ ( 2 n + 5 )^2 ( 2 n + 7 )^2 ( 2 n + 9 ) } + \frac{ ( n + 3 )^2 }{ ( 2 n + 5 )^3 ( 2 n + 7 )^2 } - \frac{ 2 ( n + 3 )( n + 2 ) }{ ( 2 n + 3 )( 2 n + 5 )^3 ( 2 n + 7 ) } \right. \\ 
& \quad \left.  -\frac{ 2 ( n + 3 )( n + 2 ) }{ ( 2 n + 1 )( 2 n + 3 )( 2 n + 5 )^2 ( 2 n + 7 ) } + \frac{ ( n + 2 )^2 }{ ( 2 n + 3 )^2 ( 2 n + 5 )^3 } + \frac{ ( n + 1 )^2 }{ ( 2 n + 1 )( 2 n + 3 )^2 ( 2 n + 5 )^2 } \right. \\
& \quad \left. + \frac{ 2 ( n + 2 )^2 }{ ( 2 n + 1 )( 2 n + 3 )^2(2 n + 5 )^2 } + \underline{ \frac{ 2 ( n + 1 )^2 }{ ( 2 n + 1 )^2 ( 2 n + 3 )^2 ( 2 n + 5 ) } } \right. \\
& \quad \left. - \frac{ 2 n ( n + 1 ) }{ ( 2 n - 1 )( 2 n + 1 )^2 ( 2 n + 3 )( 2 n + 5 ) } + \frac{ ( n + 2 )^2 }{ ( 2 n + 1 )^2( 2 n + 3 )^2( 2 n + 5 ) } + \frac{ ( n + 1 )^2 }{ ( 2 n + 1 )^3 ( 2 n + 3 )^ 2 } \right. \\
& \quad \left.  -
\frac{ 2 n ( n + 1 ) }{ ( 2 n - 1 )( 2 n + 1 )^3( 2 n + 3 ) } + \underline{ \frac{ n^2 }{ ( 2 n - 1 )^2( 2 n + 1 )^3 } } + \frac{ ( n - 1 )^2 }{ ( 2 n - 3 )( 2 n - 1 )^2 ( 2 n + 1 )^2 } \right).    
\end{split}
\right.
\end{multline}
Moreover, the nonzero elements of $ \mat{ T }^{[ 1 1 0 ]}_{H^2_0} $ are given by
\begin{multline}
%Checked for typos
\left[ \mat{ T }^{[ 1 1 0 ]}_{H^2_0} \right]_{ m n } = \\ 
\left\{
\begin{split}
m = n - 4 : & \frac{ n - 1 }{ \sqrt{ ( 2 n - 5 ) }( 2 n - 3 )( 2 n - 1 )( 2 n + 1 ) \sqrt{ ( 2 n + 3 ) } }, \\
m = n - 2 : & \frac{ 1 }{ \sqrt{ ( 2 n - 1 )( 2 n + 3 ) } } \left( - \frac{ n - 1 }{ ( 2 n - 3 )( 2 n - 1 )( 2 n + 1 ) } + \frac{ n }{ ( 2 n - 1 )( 2 n + 1 )^2 } \right. \\
& \quad \left. - \frac{ ( n + 1 ) }{ ( 2 n + 1 )^2 ( 2 n + 3) } - \frac{ ( n + 1 ) }{ ( 2 n + 1 )( 2 n + 3 )( 2 n + 5 ) } \right), \\
m = n : & \frac{ 1 }{ 2 n + 3 } \left( - \frac{ n }{ ( 2 n - 1 )( 2 n + 1 )^2 } + \frac{ n + 1 }{ ( 2 n + 1 )^2 ( 2 n + 3 ) } + \frac{ n + 1 }{ ( 2 n + 1 )( 2 n + 3 )( 2 n + 5 ) } \right. \\
& \quad \left.  - \frac{ n + 2 }{ ( 2 n + 1 )( 2 n + 3 )( 2 n + 5 ) } - \frac{ n + 2 }{ ( 2 n + 3 )( 2 n + 5 )^2 } + \frac{ n + 3 }{ ( 2 n + 5 )^2 ( 2 n + 7 ) } \right), \\
m = n + 2 : & \frac{ 1 }{ \sqrt{ ( 2 n + 3 )( 2 n + 7 ) } } \left( \frac{ n + 2 }{ ( 2 n + 1 )( 2 n + 3 )( 2 n + 5 ) }  + \frac{ n + 2 }{ ( 2 n + 3 )( 2 n + 5 )^2 } \right. \\
& \quad \left.  - \frac{ n + 3 }{ ( 2 n + 5 )^2 ( 2 n + 7 ) } + \frac{ n + 4 }{ ( 2 n + 5 )( 2 n + 7 )( 2 n + 9 ) } \right), \\
m = n + 4 : & - \frac{ ( n + 4 ) }{ \sqrt{ 2 n + 3 }( 2 n + 5 )( 2 n + 7 )( 2 n + 9 ) \sqrt{ 2 n + 11 } }.
\end{split}
\right.
\end{multline}     

\subsection{\label{app:tNodal}The Matrices $ \mat{ T }_{H^1 }^{[k d_1 d_2]} $ and $\mat{T}_{H^2_1}^{[k d_1 d_2]} $}

We now compute the matrices listed in Table~\ref{table:tH1}. In light of Remark~\ref{rem:leakage}, we explicitly consider only the elements in their first two rows and columns with indices no greater than the spectral leakage $ l$. The remaining elements can be deduced from the results in \S\ref{app:tInternal}. All of the required $ \mat{ T }_{H^1}^{[k d_1 d_2]} $ matrices are symmetric. The nonzero elements in their first two rows are given by
\begin{subequations}
\begin{gather}
\left[ \mat{ T }_{H^1}^{[0 0 0 ]} \right]_{ m n } = 
\left(
\begin{array}{llll}
 \frac{2}{3} & \frac{1}{3} & -\frac{1}{\sqrt{6}} & \frac{1}{3 \sqrt{10}} \\
 \frac{1}{3} & \frac{2}{3} & -\frac{1}{\sqrt{6}} & -\frac{1}{3 \sqrt{10}}
\end{array}
\right)
, \quad
%Checked for typos
\left[ \mat{ T }_{H^1 }^{[ 1 0 0 ]} \right]_{ m n } = 
\left(
\begin{array}{lllll}
 -\frac{1}{3} & 0 & \frac{1}{5 \sqrt{6}} & -\frac{1}{3 \sqrt{10}} & \
\frac{\sqrt{\frac{2}{7}}}{15} \\
 0 & \frac{1}{3} & -\frac{1}{5 \sqrt{6}} & -\frac{1}{3 \sqrt{10}} & \
-\frac{\sqrt{\frac{2}{7}}}{15}
\end{array}
\right)
, \\
%Checked for typos
\left[ \mat{ T }_{H^1 }^{[ 2 0 0 ]} \right]_{ m n } = 
\left(
\begin{array}{llllll}
 \frac{4}{15} & \frac{1}{15} & -\frac{1}{5 \sqrt{6}} & \frac{1}{7 \sqrt{10}} \
& -\frac{\sqrt{\frac{2}{7}}}{15} & \frac{\sqrt{2}}{105} \\
 \frac{1}{15} & \frac{4}{15} & -\frac{1}{5 \sqrt{6}} & -\frac{1}{7 \
\sqrt{10}} & -\frac{\sqrt{\frac{2}{7}}}{15} & -\frac{\sqrt{2}}{105}
\end{array}
\right)
, \quad
%Checked for typos
\left[ \mat{ T }_{H^1}^{[ 0 1 1 ]} \right]_{ m n } = 
\left(
\begin{array}{ll}
 \frac{1}{2} & -\frac{1}{2} \\
 -\frac{1}{2} & \frac{1}{2}
\end{array}
\right)
,
%Checked for typos
\end{gather}
\end{subequations}
where $ m \leq 2 $ and, in each case, $ n \leq l $. The only-non symmetric $ \mat{ T }_{H^2_1}^{[k d_1 d_2]} $ matrices are $ \mat{ T }_{H^2_1 }^{[0 1 0]} $, and $ \mat{ T }_{H^2_1 }^{[1 1 0]} $. The nonzero elements in their first two rows are
\begin{equation}
\left[ \mat{ T }_{H^2_1}^{[0 1 0]} \right]_{ m n } = 
\left(
\begin{array}{llllll}
 \frac{1}{2} & -\frac{1}{5} & \frac{3}{7 \sqrt{10}} & 0 & -\frac{1}{105 \
\sqrt{2}} & 0 \\
 \frac{1}{5} & 0 & -\frac{\sqrt{\frac{2}{5}}}{21} & \frac{1}{15 \sqrt{14}} & \
\frac{1}{105 \sqrt{2}} & 0
\end{array}
\right)
,
%Checked for typos
\end{equation}
and
\begin{equation}
\left[ \mat{ T }_{H^2_1 }^{[1 1 0]} \right]_{ m n } = 
\left(
\begin{array}{llllll}
 \frac{9}{70} & -\frac{1}{35} & 0 & \frac{1}{15 \sqrt{14}} & 0 & \
-\frac{1}{105 \sqrt{22}} \\
 \frac{5}{21} & -\frac{4}{105} & \frac{1}{21 \sqrt{10}} & 0 & \
\frac{\sqrt{2}}{315} & \frac{1}{105 \sqrt{22}}
\end{array}
\right)
,
%Checked for typos
\end{equation}
where $ n \leq l $. Moreover, for $ m \leq l $ and $ n \leq 2 $ we have $ \left[ \mat{ T }_{H^2_1 }^{[0 1 0]} \right]_{ m n } = - \left[ \mat{ T }_{H^2_1 }^{[0 1 0]} \right]_{ n m }$ and
\begin{equation}
\left[ \mat{ T }_{H^2_1 }^{[1 1 0]} \right]_{ m n } =
\left(
\begin{array}{llll}
 -\frac{1}{3 \sqrt{10}} & -\frac{\sqrt{\frac{7}{2}}}{45} & 0 & \frac{2 \
\sqrt{\frac{2}{11}}}{315} \\
 \frac{\sqrt{\frac{2}{5}}}{21} & \frac{1}{45 \sqrt{14}} & -\frac{1}{105 \
\sqrt{2}} & -\frac{2 \sqrt{\frac{2}{11}}}{315}
\end{array}
\right)^\mathrm{T}.
%Checked for typos
\end{equation}
As for the symmetric matrices, their nonzero elements are
\begin{subequations}
\begin{gather}
\left[ \mat{ T }_{H^2_1 }^{[0 0 0]} \right]_{ m n } = 
\left(
\begin{array}{llllll}
 \frac{26}{35} & -\frac{22}{105} & \frac{1}{3 \sqrt{10}} & \frac{2 \
\sqrt{\frac{2}{7}}}{45} & 0 & -\frac{1}{315 \sqrt{22}} \\
 -\frac{22}{105} & \frac{8}{105} & -\frac{1}{7 \sqrt{10}} & -\frac{1}{45 \
\sqrt{14}} & \frac{1}{315 \sqrt{2}} & \frac{1}{315 \sqrt{22}}
\end{array}
\right)
,\\
%Checked for typos
\left[ \mat{ T }_{H^2_1 }^{[0 1 1]} \right]_{ m n } = \left(
\begin{array}{llll}
 \frac{3}{5} & -\frac{1}{10} & 0 & \frac{1}{5 \sqrt{14}} \\
 -\frac{1}{10} & \frac{4}{15} & -\frac{1}{3 \sqrt{10}} & -\frac{1}{5 \
\sqrt{14}}
\end{array}
\right)
,\quad
%Checked for typos
\left[ \mat{ T }_{H^2_1 }^{[0 2 2]} \right]_{ m n } = \left(
\begin{array}{llll}
 \frac{3}{2} & -\frac{3}{2} & 0 & 0 \\
 -\frac{3}{2} & 2 & 0 & 0
\end{array}
\right)
,\\
%Checked for typos
\left[ \mat{ T }_{H^2_1 }^{[1 0 0]} \right]_{ m n } = \left(
\begin{array}{llllll}
 \frac{26}{35} & -\frac{22}{105} & \frac{1}{3 \sqrt{10}} & \frac{2 \
\sqrt{\frac{2}{7}}}{45} & 0 & -\frac{1}{315 \sqrt{22}} \\
 -\frac{22}{105} & \frac{8}{105} & -\frac{1}{7 \sqrt{10}} & -\frac{1}{45 \
\sqrt{14}} & \frac{1}{315 \sqrt{2}} & \frac{1}{315 \sqrt{22}}
\end{array}
\right)
,\\
%Checked for typos
\left[ \mat{ T }_{H^2_1 }^{[1 1 1]} \right]_{ m n } = \left(
\begin{array}{lllll}
 0 & \frac{1}{10} & -\frac{\sqrt{\frac{2}{5}}}{7} & 0 & \frac{1}{35 \
\sqrt{2}} \\
 \frac{1}{10} & \frac{2}{15} & -\frac{1}{21 \sqrt{10}} & \
-\frac{\sqrt{\frac{2}{7}}}{15} & -\frac{1}{35 \sqrt{2}}
\end{array}
\right)
,\\
%Checked for typos
\left[ \mat{ T }_{H^2_1 }^{[2 0 0]} \right]_{ m n }
= \left(
\begin{array}{llllllll}
 \frac{94}{315} & -\frac{16}{315} & \frac{1}{21 \sqrt{10}} & \
\frac{\sqrt{14}}{495} & \frac{\sqrt{2}}{315} & \frac{29}{4095 \sqrt{22}} & 0 \
& -\frac{2 \sqrt{\frac{2}{15}}}{9009} \\
 -\frac{16}{315} & \frac{4}{315} & -\frac{1}{63 \sqrt{10}} & -\frac{1}{165 \
\sqrt{14}} & -\frac{1}{693 \sqrt{2}} & -\frac{1}{1365 \sqrt{22}} & \frac{2 \
\sqrt{\frac{2}{13}}}{3465} & \frac{2 \sqrt{\frac{2}{15}}}{9009}
\end{array}
\right)
,\\
%Checked for typos
\left[ \mat{ T }_{H^2_1 }^{[2 1 1]} \right]_{ m n } = \left(
\begin{array}{llllll}
 \frac{3}{35} & \frac{1}{70} & 0 & -\frac{1}{15 \sqrt{14}} & 0 & \frac{2 \
\sqrt{\frac{2}{11}}}{105} \\
 \frac{1}{70} & \frac{16}{105} & -\frac{1}{7 \sqrt{10}} & -\frac{1}{15 \
\sqrt{14}} & -\frac{\sqrt{2}}{105} & -\frac{2 \sqrt{\frac{2}{11}}}{105}
\end{array}
\right),
%Checked for typos
\end{gather}
\end{subequations}
again for $ m \leq 2 $ and, in each case, $ n \leq l $.

\begin{table}
\centering
\caption{\label{table:tH1}Symmetry and spectral leakage $ l $~\eqref{eq:spectralLeakage} of the matrices $ \mat{ T }_{H^1 }^{[k d_1 d_2]} $ and $ \mat{ T }^{[k d_1 d_2 ]}_{H^2_1} $}
\begin{tabular*}{\linewidth}{@{\extracolsep{\fill}}llllll}
\hline
\multicolumn{3}{c}{$ \mat{ T}_{H^1}^{[k d_1 d_2]} $ } & \multicolumn{3}{c}{$ \mat{ T }_{H^2_1}^{[k d_1 d_2]} $} \\
\cline{1-3}\cline{4-6}
$ [ k d_1 d_2 ] $ & Symmetry & $ l $ & $ [ k d_1 d_2 ] $ & Symmetry & $ l $ \\
\hline
0 0 0 & S & 4 & 0 0 0 & S & 4 \\  
1 0 0 & S & 5 & 0 1 0 & N/A & 3 \\
2 0 0 & S & 6 & 0 1 1 & S & 2 \\ 
0 1 1 & S & 2 & 0 2 2 & S & 0 \\
&&& 1 0 0 & S & 5 \\
&&& 1 1 0 & N/A & 4 \\
&&& 1 1 1 & S & 3 \\
&&& 2 0 0 & S & 6 \\
&&& 2 1 1 & S & 4 \\
\hline
\end{tabular*}
\end{table}

\subsection{\label{app:tCross}The Matrices $ \mat{ T }_{H^1 H^2_0 }^{[ k d_1 d_2]}$ and $ \mat{ T }_{H^1 H^2_1 }^{[k d_1 d_2]} $}

Regarding the matrices $ \tHOHTZ^{[k d_1 d_2]} \in \mathbb{ R }^{N_b \times N_u} $~\eqref{eq:matricesTCross}, the relation $ \mu_m = \hat \DD \lambda^{[2]}_{ m - 3 } $, which follows from~\eqref{eq:lambda2} and~\eqref{eq:muNodal} for $ m \geq 4 $, leads to $ \left[ \tHOHTZ^{[k d_1 d_2]} \right]_{ m n } = \left[ \tHTZ^{[k (d_1+1) d_2]} \right]_{m-3,n }$. Therefore, given the results in~\S\ref{app:tInternal}, we only need to evaluate the elements in rows 1--3. In the main text we make use of the matrices with $ [ k d_1 d_2 ] = [000] $, $ [001] $, $ [011] $, $ [012] $, $ [100] $, and $ [111] $. Among these matrices, $ \tHOHTZ^{[012]} $ has no nonzero elements in its first three rows, and the only corresponding nonzero element of $ \tHOHTZ^{[011]} $ is $ \left[ \tHOHTZ^{[011]} \right]_{13} = - 15^{-1/2} $. Moreover, we have
\begin{subequations}
\begin{gather}
\left[ \tHOHTZ^{[000]} \right]_{ mn } = 
\left(
\begin{array}{lll}
 \frac{1}{3 \sqrt{10}} & -\frac{1}{15 \sqrt{14}} & 0 \\
 \frac{1}{3 \sqrt{10}} & \frac{1}{15 \sqrt{14}} & 0 \\
 -\frac{\sqrt{\frac{3}{5}}}{7} & 0 & \frac{1}{105 \sqrt{3}}
\end{array}
\right)
, \quad
%Checked for typos
\left[ \tHOHTZ^{[001]} \right]_{ mn } = 
\left(
\begin{array}{ll}
 \frac{1}{3 \sqrt{10}} & 0 \\
 -\frac{1}{3 \sqrt{10}} & 0 \\
 0 & -\frac{1}{5 \sqrt{21}}
\end{array}
\right)
, \\
%Checked for typos
\left[ \tHOHTZ^{[100]} \right]_{ mn } = 
\left(
\begin{array}{llll}
 -\frac{1}{21 \sqrt{10}} & \frac{1}{15 \sqrt{14}} & -\frac{\sqrt{2}}{315} & \
0 \\
 \frac{1}{21 \sqrt{10}} & \frac{1}{15 \sqrt{14}} & \frac{\sqrt{2}}{315} & 0 \
\\
 0 & -\frac{1}{15 \sqrt{21}} & 0 & \frac{1}{105 \sqrt{33}}
\end{array}
\right)
, \quad
%Checked for typos
\left[ \tHOHTZ^{[111]} \right]_{ mn } = 
\left(
\begin{array}{ll}
 \frac{1}{3 \sqrt{10}} & 0 \\
 -\frac{1}{3 \sqrt{10}} & 0 \\
 0 & -\frac{2}{5 \sqrt{21}}
\end{array}
\right)
.
%Checked for typos
\end{gather}
\end{subequations}   
As for the $ \tHOHTO^{[ k d_1 d_2]} $ matrices~\eqref{eq:matricesTCross}, one can deduce from~\eqref{eq:nuNodal} the relation $ \left[ \tHOHTO^{[k d_1 d_2]} \right]_{mn} = \left[ \tHOHTZ^{[k d_1 d_2]} \right]_{m, n-2}$, where $ n \geq 3 $. Thus, it suffices to write down the nonzero elements in their first two columns, namely
\begin{subequations}
\begin{gather}
\left[ \tHOHTO^{[000]} \right]_{ mn } =
\left(
\begin{array}{llllll}
 \frac{3}{10} & \frac{7}{10} & -\frac{1}{\sqrt{6}} & -\frac{3}{7 \sqrt{10}} \
& 0 & \frac{1}{105 \sqrt{2}} \\
 -\frac{2}{15} & -\frac{1}{5} & \frac{\sqrt{\frac{2}{3}}}{5} & \
\frac{\sqrt{\frac{2}{5}}}{21} & -\frac{1}{15 \sqrt{14}} & -\frac{1}{105 \
\sqrt{2}}
\end{array}
\right)^\mathrm{T},\\
%Checked for typos
\left[ \tHOHTO^{[001]} \right]_{ mn } =
\left(
\begin{array}{lllll}
 \frac{1}{2} & \frac{1}{2} & -\frac{\sqrt{6}}{5} & 0 & \frac{1}{5 \sqrt{14}} \
\\
 -\frac{1}{6} & \frac{1}{6} & \frac{1}{5 \sqrt{6}} & -\frac{1}{3 \sqrt{10}} \
& -\frac{1}{5 \sqrt{14}}
\end{array}
\right)^\mathrm{T},\quad
%Checked for typos
\left[ \tHOHTO^{[011]} \right]_{ mn } =
\left(
\begin{array}{llll}
 -\frac{1}{2} & \frac{1}{2} & 0 & -\frac{1}{\sqrt{10}} \\
 0 & 0 & \frac{1}{\sqrt{6}} & \frac{1}{\sqrt{10}}
\end{array}
\right)^\mathrm{T},\\
%Checked for typos
\left[ \tHOHTO^{[100]} \right]_{ mn } =
\left(
\begin{array}{lllllll}
 \frac{1}{30} & \frac{11}{30} & -\frac{3 \sqrt{\frac{3}{2}}}{35} & \
-\frac{1}{3 \sqrt{10}} & -\frac{\sqrt{\frac{7}{2}}}{45} & 0 & \frac{2 \
\sqrt{\frac{2}{11}}}{315} \\
 0 & -\frac{1}{15} & \frac{\sqrt{\frac{2}{3}}}{35} & \
\frac{\sqrt{\frac{2}{5}}}{21} & \frac{1}{45 \sqrt{14}} & -\frac{1}{105 \
\sqrt{2}} & -\frac{2 \sqrt{\frac{2}{11}}}{315}
\end{array}
\right)^\mathrm{T},\\
%Checked for typos
\left[ \tHOHTO^{[111]} \right]_{ mn } =
\left(
\begin{array}{lllll}
 0 & 0 & \frac{\sqrt{\frac{3}{2}}}{5} & 0 & -\frac{3}{5 \sqrt{14}} \\
 -\frac{1}{6} & \frac{1}{6} & \frac{\sqrt{\frac{2}{3}}}{5} & \
\frac{\sqrt{\frac{2}{5}}}{3} & \frac{3}{5 \sqrt{14}}
\end{array}
\right)^\mathrm{T}, \quad
%Checked for typos
\left[ \tHOHTO^{[012]} \right]_{ mn } =
\left(
\begin{array}{lll}
 0 & 0 & -\sqrt{\frac{3}{2}} \\
 -\frac{1}{2} & \frac{1}{2} & \sqrt{\frac{3}{2}}
\end{array}
\right)^\mathrm{T}.
%Checked for typos
\end{gather}
\end{subequations}  

\subsection{\label{app:hartmann}The $ \mat{ S } $ and $ \mat{C} $ Matrices}

Consider first the $ N \times N $ real matrices $ \hat{\mat{ S }}^{[d d_1 d_2 ]}_{ H^r_0 } $ and $ \hat{\mat{ C }}^{[ d d_1 d_2]}_{H^r_0} $, where
\begin{equation}
\label{eq:matSCHyp} 
\left[ \hat{\mat{ S }}^{[d d_1 d_2 ]}_{ H^r_0 } \right]_{ m n } = \innerprodLtwo{ \Omegaref }{ (\hat\DD^d s_{\Harxi}) \hat \DD^{ d_2 } \lambda_n^{[r]} }{ \hat\DD^{d_1} \lambda_m^{[r]} }, \quad
\left[ \hat{\mat{ C }}^{[d d_1 d_2 ]}_{ H^r_0 } \right]_{ m n } = \innerprodLtwo{ \Omegaref }{ (\hat\DD^d c_{\Harxi}) \hat \DD^{ d_2 } \lambda_n^{[r]} }{ \hat \DD^{d_1} \lambda_m^{[r]} }.
\end{equation}
Since, as can be checked from~\eqref{eq:lambda}, the polynomial degree of $ \lambda_N^{[r]} $ is $ p = N + 2 r - 1 $ and~\eqref{eq:expQuadrature} holds for polynomial integrands of degree $ 2 G - 1 $, it follows that 
\begin{equation}
\label{eq:quadG}
G \geq \lceil( 2 p + 1 - d_1 - d_2 ) / 2 \rceil 
\end{equation}
is sufficient to evaluate~\eqref{eq:matSCHyp} exactly using Mach's quadrature scheme~\eqref{eq:expQuadrature} (see Remark~\ref{rem:mach} below). Specifically, introducing the differentiation matrices $ \mat{ \Delta }^{[d]} \in \mathbb{ R }^{ G \times N } $, where $ \left[ \mat{ \Delta }^{[d]} \right]_{kn} = \hat\DD^d \lambda_n^{[r]} ( \xi_{G,k}^{[\Harxi]} ) $, the diagonal weight matrix $ \hat{\mat{ \rho }} \in \mathbb{ R }^{ G \times G } $ with $ [ \hat{\mat{ \rho }} ]_{kk} = \hat\rho^{[\Harxi]}_{G,k} $, and making use of the symmetry property $ \nu^{[r]}_m( -\xi ) = ( - 1 )^{ m + 1 } \nu^{[r]}_m( \xi ) $, leads to the expressions
\begin{subequations}
\label{eq:matSC2}
\begin{align}
  \left[ \hat{\mat{ S }}^{[d d_1 d_2 ]}_{ H^r_0 } \right]_{ m n } & = \frac{ 1 - ( - 1 )^{ m + n + d + d_1 + d_2 } }{ 2 } H^{ d}_\xi \left[ \left( \mat{ \Delta }^{[d_1]} \right)^\mathrm{T} \hat{\mat{ \rho }} \mat{ \Delta }^{[d_2 ]} \right]_{ m n }, \\
\left[ \hat{\mat{ C }}^{[d d_1 d_2 ]}_{ H^r_0 } \right]_{ m n } &= \frac{ 1 + ( - 1 )^{ m + n + d + d_1 + d_2  } }{ 2 } H^{d}_\xi \left[ \left( \mat{ \Delta }^{[d_1 ]} \right)^\mathrm{T} \hat{\mat{ \rho }} \mat{ \Delta }^{[d_2 ]} \right]_{ m n }. 
\end{align}
\end{subequations}
In order to evaluate the corresponding matrices for the $ H^1( \Omegaref ) $ and $  H^2_1( \Omegaref ) $ bases, we require, in addition to $ \mat{ \Delta }^{[d]} $ (in these cases defined in terms of the $ \mu_m $ and $ \nu_n $ polynomials), the differentiation matrices $ \tilde{ \mat{ \Delta } }^{[d]} \in \mathbb{ R }^{ G \times N } $, given by 
\begin{equation}
\left[\tilde{ \mat{ \Delta } }^{[d]} \right]_{ k n } = 
\begin{cases}
\hat\DD^d \mu_n( -\xi^{[\Harxi]}_{G,k} ), & \mbox{$ H^1( \Omegaref ) $ basis}, \\
\hat\DD^d \nu_n( -\xi^{[\Harxi]}_{G,k} ), & \mbox{$ H^2_1( \Omegaref ) $ basis},
\end{cases}
\end{equation} 
as the nodal shape functions do not have definite symmetry about $ \xi = 0 $. Note that the degree of $ \mu_N $ and $ \nu_N $ is now $ p = N - 1 $ and $ p = N + 1 $, respectively (see Propositions~\ref{prop:mu} and~\ref{prop:nu}), and the quadrature order $ G $~\eqref{eq:quadG} must be modified accordingly. Introducing $ \hat{\mat{ S }}^{[d d_1 d_2 ]}_{ H^1 } $ and $ \hat{\mat{ C }}^{[d d_1 d_2 ]}_{ H^1 } $, where
\begin{equation}
\label{eq:matSCNodal}
\left[ \hat{\mat{ S }}^{[ d d_1 d_2 ]}_{ H^1 } \right]_{ m n } = \innerprodLtwo{ \Omegaref }{ (\hat\DD^d s_H ) \hat \DD^{ d_2 } \mu_n }{ \hat \DD^{d_1} \mu_m }, \quad
\left[ \hat{\mat{ C }}^{[d d_1 d_2 ]}_{ H^1 } \right]_{ m n } = \innerprodLtwo{ \Omegaref }{ ( \hat\DD^d c_{\Harxi} ) \hat\DD^{ d_2 } \mu_n }{ \hat\DD^{d_1} \mu_m },
\end{equation}
we obtain
\begin{subequations}
\label{eq:matSC3}
\begin{align}
\hat{\mat{ S }}^{[d d_1 d_2 ]}_{ H^1 } & = \left( \left( \mat{ \Delta }^{[d_1]} \right)^\mathrm{ T }\mat{ \rho } \mat{ \Delta }^{[d_2 ]} - (-1)^d ( \tilde{ \mat{ \Delta } }^{[d_1]} )^\mathrm{ T }\mat{ \rho } \tilde{ \mat{ \Delta } }^{[d_2 ]} \right) H^d_\xi / 2, \\
\hat{\mat{ C }}^{[d d_1 d_2 ]}_{ H^1 } & = \left( \left( \mat{ \Delta }^{[d_1]} \right)^\mathrm{ T } \mat{ \rho } \mat{ \Delta }^{[d_2 ]} + (-1)^d ( \tilde{ \mat{ \Delta } }^{[d_1]} )^\mathrm{ T } \mat{ \rho } \tilde{ \mat{ \Delta } }^{[d_2 ]} \right) H^d_\xi / 2,
\end{align}
\end{subequations}
and analogous expressions for $ \hat{\mat{ S }}^{[d d_1 d_2 ]}_{ H^2_1 } $ and $ \hat{\mat{ C }}^{[d d_1 d_2 ]}_{ H^2_1 } $. We remark that relations similar to~\eqref{eq:tH2MinusSubmatrix} also apply for the matrices in~\eqref{eq:matSCNodal}, and can be used to economize on computational and coding effort. Taking into account~\eqref{eq:weightHartmannUB}, the matrices defined in~\eqref{eq:matSVu} and~\eqref{eq:matSVb} follow from
\begin{equation}
\label{eq:matVuVb}
\frac{ X }{ \sinh( \Harz  z_0 ) } \mat{ S }_{uu}^{[d d_1 d_2]} = 
\begin{cases}
\mat{ S }_{H^2_0}^{[d d_1 d_2]}, & \mbox{channel problems}, \\
\mat{ S }_{H^2_1}^{[d d_1 d_2]}, & \mbox{film problems},
\end{cases}
\quad
\frac{ X }{ \sinh( \Harz  z_0 ) } \mat{ S }_{bb}^{[d d_1 d_2]} = \mat{ S }_{H^1}^{[ d d_1 d_2]},
\end{equation}
where the matrix dimensions are respectively set to $ N_u \times N_u $ and $ N_b \times N_b $, and the quadrature order $ G $satisfies~\eqref{eq:quadG} for the given $ N_u $ and $ N_b $ (see Table~\ref{table:discreteSolutionSpaces}). The matrices $ \mat{ C }_{uu}^{[d d_1 d_2]} $ and $ \mat{ C }_{bb}^{[d d_1 d_2]} $ can be obtained in a similar manner. 

The $ N_b \times N_u $ matrices $ \mat{ S }_{bu}^{[d d_1 d_2]} $ in~\eqref{eq:matSVuVb}, and the corresponding $ \mat{ C }_{bu}^{[d d_1 d_2]} $, are evaluated by means of a small modification of the method described above. Specifically,  setting $ G \geq \lceil ( p_u + p_b + 1 - d_1 - d_2 ) / 2 \rceil $, where $ p_u $ and $ p_b $ are respectively the polynomial degrees of the velocity and magnetic-field bases, we compute the $ G \times N_u $ differentiation matrices 
\begin{equation}
\left[ \mat{ \Delta }^{[d]}_u \right]_{ k n } = 
\begin{cases}
\hat\DD^{d_2} \lambda_n^{[2]}( \xi^{[ \Harxi ]}_{G, k} ), & \mbox{channel prob.}\\
\hat\DD^{d_2} \nu_n( \xi^{[\Harxi]}_{G, k} ), & \mbox{film prob.}
\end{cases}
\quad
\left[ \tilde{\mat{ \Delta }}^{[d]}_u \right]_{ k n }  = 
\begin{cases}
\hat\DD^{d_2} \lambda_n^{[2]}( -\xi^{[ \Harxi ]}_{G, k} ), & \mbox{channel prob.}\\
\hat\DD^{d_2} \nu_n( -\xi^{[\Harxi]}_{G, k} ), & \mbox{film prob.}
\end{cases}
\end{equation}
and the $ G \times N_b $ matrices
\begin{equation}
\left[ \mat{ \Delta }^{[d]}_b \right]_{ k n }= \hat\DD^{d_1} \mu_n( \xi^{[\Harxi]}_{G,k} ), \quad \left[ \tilde{\mat{ \Delta }}^{[d]}_2 \right]_{ k n }= \hat\DD^d \mu_n( -\xi^{[\Harxi]}_{G,k} ).
\end{equation}
Then, using~\eqref{eq:weightHartmannUB}, we obtain
\begin{align}
\mat{S}_{bu}^{[d d_1 d_2]} & = \frac{ \cosh( \Harz  z_0 ) }{ \Harz ( \cosh( \Harz ) - 1) } \left( \left( \mat{ \Delta}_b^{[d_1]} \right)^\mathrm{ T} \hat{\mat{ \rho }} \mat{ \Delta }_u^{[d_2]} - ( - 1 )^d \left( \tilde{\mat{ \Delta }}_b^{[d_1]} \right)^\mathrm{ T } \hat{\mat{\rho}} \tilde{ \mat{ \Delta }}_u^{[d_2]} \right), \\
\mat{C}_{bu}^{[d d_1 d_2]} & = \frac{ \sinh( \Harz  z_0 ) }{ \Harz ( \cosh( \Harz ) - 1) } \left( \left( \mat{ \Delta}_b^{[d_1]} \right)^\mathrm{ T} \hat{\mat{ \rho }} \mat{ \Delta }_u^{[d_2]} + ( - 1 )^d \left( \tilde{\mat{ \Delta }}_b^{[d_1]} \right)^\mathrm{ T } \hat{\mat{\rho}} \tilde{ \mat{ \Delta }}_u^{[d_2]} \right).
\end{align}

\begin{rem}
\label{rem:mach}Mach's algorithm \cite{Mach84} for the $ a_n $ and $ b_n $ coefficients (with $ n \in \{ 0, 1, 2, \ldots, G + 1 \} ) $ of polynomials orthogonal with respect to the weight function $ e^{H_\xi \xi}$ consists of two parts. For $ n \leq G_0 := \min\{ [ \Harxi ], G + 1 \} $ the coefficients are evaluated algebraically, while if $ G + 1 > [ \Harxi ] $ an iterative procedure is used for $ n > G_0 $. We observed that for the typical $ \Harxi $ and $ G $ used in our linear-stability schemes (both of which are significantly larger than the ones considered in Mach's paper), the quadrature knots $ \xi_{G,k}^{[\Harxi]} $ and weights $ \hat\rho_{G,k}^{[\Harxi]} $, which follow from the eigenvalues and eigenvectors of the Jacobian matrix $ \mat{ J } $ constructed from $\{ a_n \}$ and $\{ b_n \}$ \cite{DavisRabinowitz07}, are more accurately computed if the iterative procedure is employed for all $ n $. Moreover, using a specialized solver for symmetric tridiagonal matrices (\eg the LAPACK routine DSTEV \cite{AndersonEtAl99}), rather than a generic one, enhances the stability of the computation at large $ G $. Regarding the algorithm's large-$ \Harxi $ behavior, in 64-bit arithmetic the weight calculation overflows at around $ \Harxi = 700 $. This limitation can be mitigated by increasing the arithmetic precision, but doing so is significantly more complicated than in the case of the LGL method (see Remark~\ref{rem:lgl}), as it involves porting the routines for the $ \mat{ J } $ eigenproblem.
\end{rem}     

\section{\label{app:eigenvalueSpectra}Eigenvalues  of Selected Film and Channel Problems}

This appendix contains tables of eigenvalues for the stability problems studied in \S\ref{sec:eigenvalueSpectra}. In each case, the eigenproblem~\eqref{eq:matrixWeakForm} has been solved using the QZ algorithm, and the resulting complex phase velocity $ c = i \gamma / \alpha $ is listed in order of decreasing $ \Imag( c ) $. In the examples where the spectrum exhibits the A, P, and S branches (Tables~\ref{table:spectrumHydroChannel}--\ref{table:spectrumMhdFilmNoField} and~\ref{table:spectrumMhdFilmLowPm}) the modes are also labeled in order of decreasing $ \Imag( c ) $ within their respective families. In Tables~\ref{table:spectrumMhdFilmFlowNormal} and~\ref{table:spectrumMhdFilmOblique}, hydrodynamic and magnetic modes are respectively labeled H and M. In addition to $ c $, Tables~\ref{table:spectrumHydroFilm} and~\ref{table:spectrumMhdFilmFlowNormal}--\ref{table:spectrumMhdFilmLowPm} also display the modal energies~\eqref{eq:energySum}. All problems with $ \Harz > 0 $  (Tables~\ref{table:spectrumZeroPmFilm}, \ref{table:spectrumZeroPmChannel} and~\ref{table:spectrumMhdFilmFlowNormal}--\ref{table:spectrumMhdFilmLowPm}) have the Hartmann profiles~\eqref{eq:baseHartmann}. In these cases, the stiffness matrix $ \mat{ K } $ has been computed by means of the exact-quadrature method (Eqs.~\eqref{eq:matKuubbUHartmannExact} and~\eqref{eq:matKubbuBHartmannExact}), aside from the inductionless problems in Table~\ref{table:spectrumZeroPmFilm}, where LGL quadrature~\eqref{eq:matKuuUHartmannLgl} has also been used. The free-surface parameters are $ \Prg = 1.10 \times 10^{-4} $ and $ \Ohn = 3.14 \times 10^{-4} $ for all film problems.

\begin{table}
\renewcommand{\arraystretch}{1}
\centering
\caption{\label{table:spectrumHydroChannel}Complex phase velocity of the 33 least stable modes of non-MHD channel flow at $ \Rey = 10^4 $, $ \alpha = 1 $, and $ p_u = 500 $. E and O respectively denote even and odd modes. The underlined digits differ from Table~VII in Kirchner~\cite{Kirchner00}.}
\begin{tabular*}{\linewidth}{@{\extracolsep{\fill}}llll}
\hline
& & \multicolumn{1}{l}{Symmetry} & \multicolumn{1}{c}{$c$}\\
\cline{3-4}
1 & $ \mathrm{ A }_{ 1 } $ &  E &  $ 2.37526488820470\underline{8}\EE-01 + 3.739670622977\underline{800}\EE-03 \, \ii $ \\ 
2 & $ \mathrm{ P }_{ 1 } $ &  O &  $ 9.64630915450\underline{5980}\EE-01 - 3.5167277631027\underline{88}\EE-02 \, \ii $ \\ 
3 & $ \mathrm{ P }_{ 2 } $ &  E &  $ 9.646425100392\underline{813}\EE-01 - 3.5186583792443\underline{06}\EE-02 \, \ii $ \\ 
4 & $ \mathrm{ A }_{ 2 } $ &  O &  $ 2.7720434380880\underline{44}\EE-01 - 5.089872725696\underline{847}\EE-02 \, \ii $ \\ 
5 & $ \mathrm{ P }_{ 3 } $ &  O &  $ 9.363165358813\underline{226}\EE-01 - 6.320149583992\underline{000}\EE-02 \, \ii $ \\ 
6 & $ \mathrm{ P }_{ 4 } $ &  E &  $ 9.363517811647\underline{262}\EE-01 - 6.325156907426\underline{749}\EE-02 \, \ii $ \\ 
7 & $ \mathrm{ P }_{ 5 } $ &  O &  $ 9.079830546294\underline{242}\EE-01 - 9.122273543365\underline{197}\EE-02 \, \ii $ \\
\hline
\end{tabular*}
\end{table}

\begin{table}
\renewcommand{\arraystretch}{1}
\caption{\label{table:spectrumHydroFilm}Complex phase velocity and free-surface energy of the 25 least stable modes of non-MHD film problems at $ \alpha = 1 $, $ p_u = 500 $, and $ \Rey = 10^4$, $ 3 \times 10^4$}
\begin{tabular*}{\linewidth}{@{\extracolsep{\fill}}llrl}
\hline
& & \multicolumn{1}{c}{$c$} & \multicolumn{1}{c}{$E_a/E$}\\
\cline{3-4}
\multicolumn{4}{l}{$ \Rey = 10^4$}\\
1 &  $ \mbox{F} $ &  $ 1.679075867077238\mathrm{E}+00 - 5.309992786818418\mathrm{E}-04 \, \ii $ & $ 5.91709761\mathrm{E}-01 $ \\ 
2 &  $ \mbox{U}/\mathrm{A}_{ 1 } $ &  $ -1.643176086760401\mathrm{E}-01 - 1.185929268008322\mathrm{E}-02 \, \ii $ & $ 3.23193182\mathrm{E}-01 $ \\ 
3 &  $ \mbox{P}_{ 1 } $ &  $ 9.645730769959389\mathrm{E}-01 - 3.500080786980365\mathrm{E}-02 \, \ii $ & $ 7.72375828\mathrm{E}-03 $ \\ 
4 &  $ \mbox{P}_{ 2 } $ &  $ 9.361509487313044\mathrm{E}-01 - 6.277440588217570\mathrm{E}-02 \, \ii $ & $ 7.42572688\mathrm{E}-03 $ \\ 
5 &  $ \mbox{A}_{ 2 } $ &  $ 2.741497879597380\mathrm{E}-01 - 6.654001933197072\mathrm{E}-02 \, \ii $ & $ 1.87107321\mathrm{E}-01 $ \\ 
6 &  $ \mbox{P}_{ 3 } $ &  $ 9.077684797573511\mathrm{E}-01 - 9.050782178609683\mathrm{E}-02 \, \ii $ & $ 4.22349138\mathrm{E}-03 $ \\
\hline
\end{tabular*}
\end{table}

\begin{table}
\renewcommand{\arraystretch}{1}
\caption{\label{table:spectrumZeroPmFilm}Complex phase velocity of the 25 least stable modes of inductionless film problems at $ \Rey = 3 \times 10^4 $, $ \Harx = 0 $, $ \alpha = 1 $, and $ p_u = 500 $ and $ \Harz = 14 $, 100. The mean basic velocities~\eqref{eq:baseUAverage} are $ \langle U \rangle = 0.92857$ ($\Harz = 14$) and $ \langle U \rangle = 0.99000 $ ($ \Harz = 100$)}
\centering
\begin{tabular*}{\linewidth}{@{\extracolsep{\fill}}lllr}
\hline
& & \multicolumn{1}{c}{$ c $ (exact)} & \multicolumn{1}{c}{$c$ (LGL)} \\
\cline{3-4}
\multicolumn{4}{l}{$ \Harz = 14 $}\\
1 & $ \mathrm{ F } $ &  $ 1.264889076407188\mathrm{E}+00 - 2.709480082539037\mathrm{E}-03 \, \ii $ &  $ 1.264889076407183\mathrm{E}+00 - 2.709480082550979\mathrm{E}-03 \, \ii $ \\ 
2 & $ \mathrm{ P }_{ 1 } $ &  $ 9.992558947033491\mathrm{E}-01 - 8.278625506899678\mathrm{E}-03 \, \ii $ &  $ 9.992558947033513\mathrm{E}-01 - 8.278625506898540\mathrm{E}-03 \, \ii $ \\ 
3 & $ \mathrm{ P }_{ 2 } $ &  $ 9.983412661330004\mathrm{E}-01 - 1.153953543381368\mathrm{E}-02 \, \ii $ &  $ 9.983412661330017\mathrm{E}-01 - 1.153953543381040\mathrm{E}-02 \, \ii $ \\ 
4 & $ \mathrm{ P }_{ 3 } $ &  $ 9.971872338663629\mathrm{E}-01 - 1.574118837369919\mathrm{E}-02 \, \ii $ &  $ 9.971872338663625\mathrm{E}-01 - 1.574118837369889\mathrm{E}-02 \, \ii $ \\ 
5 & $ \mathrm{ P }_{ 4 } $ &  $ 9.958459023889774\mathrm{E}-01 - 2.082236225171166\mathrm{E}-02 \, \ii $ &  $ 9.958459023889762\mathrm{E}-01 - 2.082236225171246\mathrm{E}-02 \, \ii $ \\ 
6 & $ \mathrm{ A }_{ 1 } $ &  $ 7.299947887448923\mathrm{E}-01 - 2.612159002726035\mathrm{E}-02 \, \ii $ &  $ 7.299947887448834\mathrm{E}-01 - 2.612159002730650\mathrm{E}-02 \, \ii $ \\
\hline
\end{tabular*}
\end{table}

\begin{table}
\renewcommand{\arraystretch}{1}
\caption{\label{table:spectrumZeroPmChannel}Complex phase velocity $ c $ of the 33 least stable modes of an inductionless channel problem at $ \Rey = 10^4 $, $ \Harz = 14 $, $ \Harx = \Harz / \tan( 1^\circ ) = \text{802.06} $, $ \alpha = 1 $, and $ p_u = 500 $}
\centering
\begin{tabular*}{\linewidth}{@{\extracolsep{\fill}}lll}
\hline
1 & $ \mathrm{ P }_{ 1 } $ &  $ 9.993700195546175\mathrm{E}-01 - 9.721004530793138\mathrm{E}-02 \, \ii $ \\ 
2 & $ \mathrm{ P }_{ 2 } $ &  $ 9.976612096433158\mathrm{E}-01 - 1.030917967244671\mathrm{E}-01 \, \ii $ \\ 
3 & $ \mathrm{ P }_{ 3 } $ &  $ 9.951774765278338\mathrm{E}-01 - 1.123185821184923\mathrm{E}-01 \, \ii $ \\ 
4 & $ \mathrm{ P }_{ 4 } $ &  $ 9.921089021026304\mathrm{E}-01 - 1.244427518042686\mathrm{E}-01 \, \ii $ \\ 
5 & $ \mathrm{ P }_{ 5 } $ &  $ 9.885576846065738\mathrm{E}-01 - 1.391394166517699\mathrm{E}-01 \, \ii $ \\ 
6 & $ \mathrm{ P }_{ 6 } $ &  $ 9.845858924007105\mathrm{E}-01 - 1.561699942143518\mathrm{E}-01 \, \ii $ \\ 
7 & $ \mathrm{ P }_{ 7 } $ &  $ 9.802375881163239\mathrm{E}-01 - 1.753455921597468\mathrm{E}-01 \, \ii $ \\ 
8 & $ \mathrm{ P }_{ 8 } $ &  $ 9.755530144308208\mathrm{E}-01 - 1.965030475951849\mathrm{E}-01 \, \ii $ \\
\hline
\end{tabular*}
\end{table}

\begin{table}
\renewcommand{\arraystretch}{1}
\caption{\label{table:spectrumMhdFilmNoField}Complex phase velocity $ c$ of the 25 least stable magnetic modes of film MHD flow with zero background magnetic field ($ \Harx = \Harz = 0 $) at $ \Rey = 10^4 $, $ \Prm = 1.2 $, $ \alpha = 1 $, and $ p_u = p_b = 500 $. The numbering in the left-hand column takes into account the hydrodynamic part of the spectrum, where the eigenvalues are identical to those listed in the $ \Rey = 10^4 $ portion of Table~\ref{table:spectrumHydroFilm}.}
\centering
\begin{tabular*}{\linewidth}{@{\extracolsep{\fill}}llr}
\hline
2 & $ \mbox{Pm}_{ 1 } $ &  $ 9.931687361486334\mathrm{E}-01 - 7.411597582310019\mathrm{E}-03 \, \ii $ \\ 
4 & $ \mbox{Pm}_{ 2 } $ &  $ 9.675367306486845\mathrm{E}-01 - 3.281002191260387\mathrm{E}-02 \, \ii $ \\ 
6 & $ \mbox{Am}_{ 1 } $ &  $ 6.190480900456202\mathrm{E}-02 - 3.658789628263018\mathrm{E}-02 \, \ii $ \\ 
7 & $ \mbox{Pm}_{ 3 } $ &  $ 9.417639706299372\mathrm{E}-01 - 5.851812430456833\mathrm{E}-02 \, \ii $ \\ 
10 & $ \mbox{Pm}_{ 4 } $ &  $ 9.159676389192429\mathrm{E}-01 - 8.428165278355801\mathrm{E}-02 \, \ii $ \\ 
12 & $ \mbox{Am}_{ 2 } $ &  $ 1.920049933116885\mathrm{E}-01 - 1.071612998031651\mathrm{E}-01 \, \ii $ \\ 
13 & $ \mbox{Pm}_{ 5 } $ &  $ 8.901624704645013\mathrm{E}-01 - 1.100662088133551\mathrm{E}-01 \, \ii $ \\ 
15 & $ \mbox{Pm}_{ 6 } $ &  $ 8.643528846481575\mathrm{E}-01 - 1.358613270394366\mathrm{E}-01 \, \ii $ \\
\hline
\end{tabular*}
\end{table}

\begin{table}
\renewcommand{\arraystretch}{1}
\caption{\label{table:spectrumMhdFilmFlowNormal}Complex phase velocity, free-surface energy and magnetic energy of the 25 least stable modes of film MHD problems at $ \Rey = 10^4 $, $ \Prm = 1.2 $, $ \Harx = 0 $, $ \alpha = 1 $, $ p_u = p_b = 500 $ and $ \Harz = 14 $, 100}
\begin{tabular*}{\linewidth}{@{\extracolsep{\fill}}llrll}
\hline
& & \multicolumn{1}{c}{$ c $} & \multicolumn{1}{c}{$E_a / E$} & \multicolumn{1}{c}{$E_b / E$} \\
\cline{3-5}
\multicolumn{5}{l}{$ \Harz = 14$}\\
1 &  $ \mbox{M}_{ 1 } $ &  $ 2.284864655115033\mathrm{E}-01 + 8.576128126042017\mathrm{E}-02 \, \ii $ & $ 3.58332998\mathrm{E}-02 $ & $ 5.03962296\mathrm{E}-01 $ \\ 
2 &  $ \mbox{H}_{ 1 } $ &  $ 1.953613740443688\mathrm{E}+00 - 9.982305603345799\mathrm{E}-04 \, \ii $ & $ 3.37400672\mathrm{E}-01 $ & $ 1.86792551\mathrm{E}-01 $ \\ 
3 &  $ \mbox{M}_{ 2 } $ &  $ 9.995800306958438\mathrm{E}-01 - 1.692125148642362\mathrm{E}-03 \, \ii $ & $ 2.81931497\mathrm{E}-01 $ & $ 6.62062506\mathrm{E}-01 $ \\ 
4 &  $ \mbox{H}_{ 2 } $ &  $ -5.953484440515415\mathrm{E}-02 - 2.669748683162519\mathrm{E}-02 \, \ii $ & $ 2.14898729\mathrm{E}-01 $ & $ 2.76511889\mathrm{E}-01 $ \\ 
5 &  $ \mbox{M}_{ 3 } $ &  $ 1.682655825737726\mathrm{E}+00 - 4.600693501346862\mathrm{E}-02 \, \ii $ & $ 1.14097814\mathrm{E}-02 $ & $ 5.03786816\mathrm{E}-01 $ \\ 
6 &  $ \mbox{M}_{ 4 } $ &  $ 1.723325890582706\mathrm{E}+00 - 6.575509902272228\mathrm{E}-02 \, \ii $ & $ 2.46965409\mathrm{E}-03 $ & $ 5.02168713\mathrm{E}-01 $ \\
\hline
\end{tabular*}
\end{table}

\begin{table}
\renewcommand{\arraystretch}{1}
\caption{\label{table:spectrumMhdFilmOblique}Complex phase velocity, free-surface energy, and magnetic energy of the 50 least stable modes of film MHD flow at $ \Rey = 10^4 $, $ \Prm = 1.2 $, $ \Harz = 100 $, $ \Harx = 100 / \tan( 1^\circ ) = \text{5,729.0} $, $ \alpha = 1 $, and $ p_u = p_b = 500 $}
\begin{tabular*}{\linewidth}{@{\extracolsep{\fill}}llrll}
\hline
& & \multicolumn{1}{c}{$ c $} & \multicolumn{1}{c}{$E_a / E$} & \multicolumn{1}{c}{$E_b / E$} \\
\cline{3-5}
1 &  $ \mbox{M}_{ 1 } $ &  $ 5.730871002785992\mathrm{E}-01 + 2.323881719088570\mathrm{E}-01 \, \ii $ & $ 1.93633161\mathrm{E}-03 $ & $ 6.27669637\mathrm{E}-01 $ \\ 
2 &  $ \mbox{M}_{ 2 } $ &  $ 9.999617714618909\mathrm{E}-01 + 1.793345085085649\mathrm{E}-03 \, \ii $ & $ 4.83249270\mathrm{E}-01 $ & $ 5.15837595\mathrm{E}-01 $ \\ 
3 &  $ \mbox{H}_{ 1 } $ &  $ 2.493142559767902\mathrm{E}+00 - 5.926747205627726\mathrm{E}-02 \, \ii $ & $ 4.49457980\mathrm{E}-03 $ & $ 4.88316363\mathrm{E}-01 $ \\ 
4 &  $ \mbox{H}_{ 2 } $ &  $ -4.913953871374847\mathrm{E}-01 - 6.680787479130106\mathrm{E}-02 \, \ii $ & $ 2.77518243\mathrm{E}-02 $ & $ 4.28717357\mathrm{E}-01 $ \\ 
5 &  $ \mbox{M}_{ 3 } $ &  $ 2.259088416138722\mathrm{E}+00 - 7.319745692046410\mathrm{E}-02 \, \ii $ & $ 1.65453101\mathrm{E}-03 $ & $ 4.95707415\mathrm{E}-01 $ \\ 
6 &  $ \mbox{M}_{ 4 } $ &  $ 1.840029866609624\mathrm{E}+00 - 8.474865062968835\mathrm{E}-02 \, \ii $ & $ 1.87189078\mathrm{E}-04 $ & $ 4.99790274\mathrm{E}-01 $ \\ 
7 &  $ \mbox{H}_{ 3 } $ &  $ 1.598392105033923\mathrm{E}-01 - 8.476466722677443\mathrm{E}-02 \, \ii $ & $ 5.83406131\mathrm{E}-04 $ & $ 4.98491557\mathrm{E}-01 $ \\
\hline
\end{tabular*}
\end{table}

\begin{table}
\renewcommand{\arraystretch}{1}
\caption{\label{table:spectrumMhdFilmLowPm}Complex phase velocity, free-surface energy, and magnetic energy of the 25 least stable modes of $ Pm = 10^{-4} $ and inductionless film problems at $ \Rey = 10^6 $, $ \alpha = 0.01 $, $ \Harx = 0 $, $ \Harz = 10 $, and $ p_u = p_b = 500 $}
\centering
\begin{tabular*}{\linewidth}{@{\extracolsep{\fill}}lllll}
\hline
& & \multicolumn{1}{c}{$ c $} &  \multicolumn{1}{c}{$ E_a / E $} & \multicolumn{1}{c}{$E_b/ E $} \\
\cline{3-5}
\multicolumn{5}{l}{$\Prm = 10^{-4}$}\\
1 &  $ \mbox{P}_{ 1 } $ &  $ 9.827762904552182\mathrm{E}-01 + 1.379303998591685\mathrm{E}-02 \, \ii $ & $ 7.21784034\mathrm{E}-02 $ & $ 2.15340701\mathrm{E}-01 $ \\ 
2 &  $ \mbox{P}_{ 2 } $ &  $ 9.966319464666803\mathrm{E}-01 - 1.729497204355026\mathrm{E}-02 \, \ii $ & $ 1.59127542\mathrm{E}-03 $ & $ 6.39646788\mathrm{E}-03 $ \\ 
3 &  $ \mbox{F} $ &  $ 1.014164030326762\mathrm{E}+00 - 2.023492335823250\mathrm{E}-02 \, \ii $ & $ 5.55459542\mathrm{E}-02 $ & $ 2.02404403\mathrm{E}-01 $ \\ 
4 &  $ \mbox{M} $ &  $ 8.963943583270458\mathrm{E}-01 - 2.230611416635310\mathrm{E}-02 \, \ii $ & $ 2.62501396\mathrm{E}-03 $ & $ 2.81923599\mathrm{E}-01 $ \\ 
5 &  $ \mbox{P}_{ 3 } $ &  $ 9.924107060542195\mathrm{E}-01 - 2.791684540382556\mathrm{E}-02 \, \ii $ & $ 1.37991443\mathrm{E}-03 $ & $ 5.42646365\mathrm{E}-03 $ \\ 
6 &  $ \mbox{P}_{ 4 } $ &  $ 9.875153580671349\mathrm{E}-01 - 4.091279696616719\mathrm{E}-02 \, \ii $ & $ 5.85319839\mathrm{E}-04 $ & $ 2.43710490\mathrm{E}-03 $ \\
\hline
\end{tabular*}
\end{table}

%\bibliographystyle{elsart-num}
%\bibliography{bibliography}


% notes:

% \bibitem{label2}
% If there is a note, it should come last:
% \bibitem{label3} \note
% \end{subbibitems}

%\bibitem{}

%\end{thebibliography}

\end{document}